\documentclass[twocolumn]{aastex631}

\newcommand{\numsample}{39 }

\let\tablenum\relax
\usepackage{siunitx}
\DeclareSIUnit\angstrom{\text {Å}}
\usepackage{rotating}

\received{\today}

\shorttitle{CL-AGN in Prospector}
\shortauthors{Verrico et al.}

\graphicspath{{./}{figures/}}

\begin{document}

\title{Modeling Star Formation Histories of Changing-Look AGN Host Galaxies with \texttt{Prospector}}
\author[0000-0003-1535-4277]{Margaret E. Verrico}
\affiliation{University of Illinois Urbana-Champaign Department of Astronomy, University of Illinois, 1002 W. Green St., Urbana, IL 61801, USA}
\affiliation{Center for AstroPhysical Surveys, National Center for Supercomputing Applications, 1205 West Clark Street, Urbana, IL 61801, USA}
\author[0000-0002-4235-7337]{K. Decker French}
\affiliation{University of Illinois Urbana-Champaign Department of Astronomy, University of Illinois, 1002 W. Green St., Urbana, IL 61801, USA}
\author[0000-0002-1714-1905]{Katherine A. Suess}
\affiliation{Department for Astrophysical \& Planetary Science, University of Colorado, Boulder, CO 80309, USA}
\author[0009-0000-9992-3072]{Tanay Agrawal}
\affiliation{University of Illinois Urbana-Champaign Department of Astronomy, University of Illinois, 1002 W. Green St., Urbana, IL 61801, USA}
\author[0000-0002-3249-8224]{Lauranne Lanz}
\affiliation{Department of Physics, The College of New Jersey, Ewing, NJ 08628, USA}
\author[0000-0002-0696-6952]{Yuanze Luo}
\affiliation{Department of Physics and Astronomy and George P. and Cynthia Woods Mitchell Institute for Fundamental Physics and Astronomy, Texas A\&M University, 4242 TAMU, College Station, TX 77843-4242, US}
\author[0000-0002-9471-8499]{Pallavi Patil}
\affiliation{William H. Miller III Department of Physics and Astronomy, Johns Hopkins University, Baltimore, MD 21218, USA}
\author[0000-0001-7883-8434]{Kate Rowlands}
\affiliation{William H. Miller III Department of Physics and Astronomy, Johns Hopkins University, Baltimore, MD 21218, USA}
\affiliation{AURA for ESA, Space Telescope Science Institute, 3700 San Martin Drive, Baltimore, MD 21218, USA}
\author[0009-0005-1158-1896]{Margaret Shepherd}
\affiliation{University of Illinois Urbana-Champaign Department of Astronomy, University of Illinois, 1002 W. Green St., Urbana, IL 61801, USA}
\author[0009-0004-0844-0657]{Maya Skarbinski}
\affiliation{William H. Miller III Department of Physics and Astronomy, Johns Hopkins University, Baltimore, MD 21218, USA}


\date{\today}


\correspondingauthor{Margaret E.  Verrico}
\email{verrico2@illinois.edu}

\begin{abstract}

Changing-look active galactic nuclei, or CL-AGN, are AGN which appear to transition between Seyfert Type 1 and 2 over periods of months to years.  Several mechanisms to trigger these transitions have been proposed, but we have yet to conclusively determine their cause.  Recent studies suggest CL-AGN are hosted primarily in galaxies which are shutting down star formation \citep{dodd2021,liu2021,wang2023}, which may indicate a link between galaxy quenching and changing look events.  We use \texttt{Prospector} stellar population synthesis software \citep{leja17,johnson17,johnson2021} to model non-parametric star formation histories for \numsample CL-AGN host galaxies.  We find that $43^{+13}_{-12}\%$ of our gold sample CL-AGN at z $<$ 0.15 are star forming, while $29^{+13}_{-10}\%$ fall in the Green Valley of the stellar mass-sSFR diagram.  At z $>$ 0.15, $57^{+13}_{-18}\%$ of CL-AGN in the gold sample are star-forming and $29^{+19}_{-14}\%$ are in the Green Valley.  CL-AGN hosts have similar star formation properties to the host galaxies of Seyfert 1 and 2 AGN at z $<$ 0.15 and to Seyfert 2 AGN at z $>$ 0.15.  We find no statistically significant differences in the star formation properties of turn-on and turn-off CL-AGN.  We also find no evidence for rapid quenching in the Green Valley CL-AGN. We conclude that CL-AGN state transitions are not associated with the formation history of CL-AGN host galaxies on large spatial scales, implying CL-AGN state transitions may instead result from nuclear-scale or accretion disk effects.



\end{abstract}
\section{Introduction} \label{sec:intro}

\begin{figure}[t]
    \centering
    \includegraphics[width=\columnwidth]{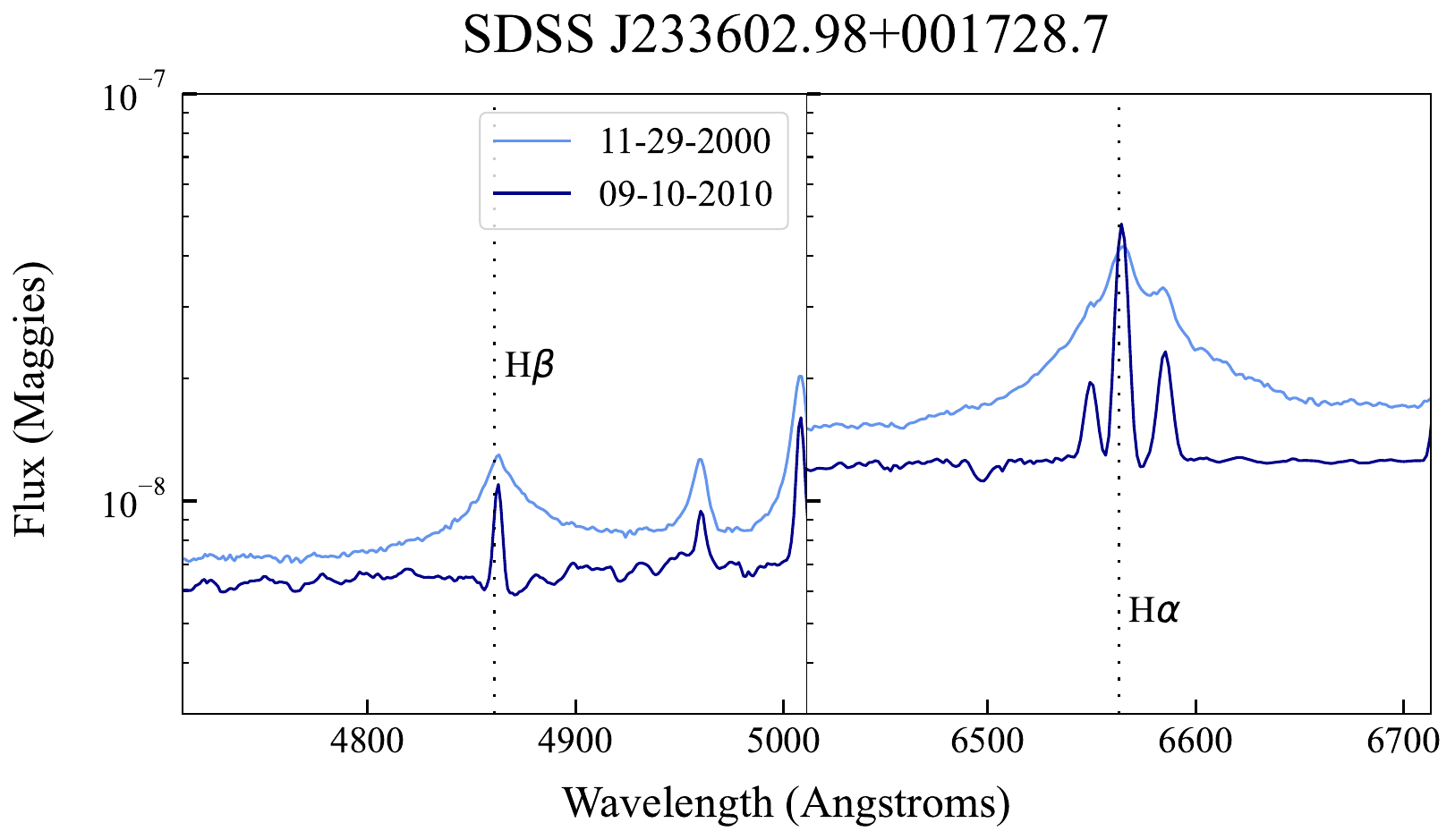}
    \caption{SDSS spectra at two epochs for a changing-look AGN from \cite{ruan2016}.  The continuum emission dims and the broad H$\alpha$ and H$\beta$ emission lines disappear between the two epochs as the AGN transitions from a Seyfert Type 1 to Type 2, making this a ``turn-off" CL-AGN.}
    \label{fig:clagn_def}
\end{figure}

 Active galactic nuclei, or AGN, are the luminous centers of massive galaxies powered by supermassive black hole accretion.  In optical observations, AGN can be divided into Seyfert Type 1 (containing both narrow and broad emission lines) or Type 2 (lacking broad emission lines). Our current understanding of AGN physics attributes these spectral differences \textit{either} to the relative orientation of an AGN accretion disk and the observer (with Type 2 Seyferts being oriented such that the dusty torus obscures the broad line emitting region of the AGN, e.g. \citet{Antonucci1993,urry1995}), \textit{and/or} to intrinsic changes in molecular gas and dust in the nucleus, either impacting accretion rates \citep{elitzur2014} or obscuration due to a merger 
 \citep{alexander2012}.  A certain amount of stochastic variation in AGN luminosity and spectra is to be expected; however, in recent years, some AGN have been observed to transition between different AGN types with either the appearance or disappearance of broad emission lines over the course of months or years, directly contrasting the unification models referenced above \citep[][see Figure \ref{fig:clagn_def} for an example]{matt2003, ricci2022}. These objects, also known as ``changing-look" AGN (CL-AGN), can be further divided into ``turn-on" CL-AGN, which gain broad-line emission, and ``turn-off," which lose their broad lines.  While some CL-AGN can be explained by external gas or dust obscuring or revealing the broad-line region of the AGN accretion disk \citep[e.g.][]{Malizia1997,Risaliti2007,kokubo2019}, other objects, like the ones studied in e.g. \cite{lamassa2015,Trakhtenbrot2019}, and \cite{green2022}, cannot be explained by changes in obscuration alone.  Instead, it is likely that the spectra of these objects are changing due to intrinsic changes in the accretion disk.  
 
  Accretion changes in AGN disks are thought to occur on timescales either too short (e.g. dynamical timescales of $\sim$days or thermal timescales of $\sim$1 month) or too long (viscous timescale $\sim10^5$ years) to explain CL-AGN \citep[see][and references therein]{ricci2022}.  Current constraints on the timescales for CL-AGN state changes are generally limited by the gap between repeat spectra for individual objects.  Recent work has attempted to further constrain state change timescales in a subset of CL-AGN using a combination of repeat spectroscopy and photometric analysis.  A few CL-AGN with high-cadence spectroscopy transitioned within a few months \citep{Trakhtenbrot2019,zeltyn2022}, while photometric studies of CL-AGN have measured transition timescales as short as three months and as long as five years \citep{hon2022,lopeznavas2022,wang2024,zhu2024}.  The largest spectroscopic searches to date find upper limits on transition timescales between 244-5762 days ($\sim15$ years) between SDSS and DESI \citep{guo2024a} and around 1-19 years between SDSS and SDSS-V \citep{zeltyn2024}.  However, these timescales are again limited by the low cadence of repeat spectroscopy for AGN in large surveys.  For this reason, the exact timescales and spectral changes associated with CL-AGN transitions remain limited to single-object studies, and the mechanisms powering CL-AGN are still a mystery.


One potential avenue for population studies of CL-AGN lies in their host galaxies.  The host galaxy environment around an AGN can impact its accretion properties; a gas-rich galaxy may fuel luminous AGN and quasars with high accretion rates, while a gas-poor galactic nucleus could starve an AGN and lead to lower luminosities and accretion rates \citep{kauffmann2009}.  These transitions in nuclear gas content may be associated with galaxy-wide gas processes which also fuel star formation, leaving imprints on the stellar population of an AGN host galaxy.  Few studies of the stellar populations of CL-AGN host galaxies currently exist, but those that do find that CL-AGN host galaxies have a higher proportion of intermediate-age stars than the host galaxies of other Seyfert AGN \citep{jin2021} and lie in either the Green Valley \citep{dodd2021,liu2021,wang2023}, a brief stage of galaxy evolution in which the galaxy is shutting off or ``quenching" star formation \citep{salim2014}, or along the star forming main sequence \citep{yu2020}.  Galaxies can move through the Green Valley quickly or slowly depending on the quenching mechanism \citep{schawinski2014}. Fast quenching mechanisms include major mergers which are thought to remove gas through tidal forces, trigger a rapid starburst which consumes much of the remaining gas, and trigger AGN activity which may induce feedback, further suppressing star formation \citep{hopkins2005a}. Nuclear transients like tidal disruption events (TDEs) and quasi-periodic eruptions are overrepresented in post-starburst galaxies, and it is thought that galaxy mergers create dynamical conditions at galaxy centers which are more likely to trigger these nuclear transients \citep{stone2020,french2016,french2020,wevers2024}.  Slower quenching mechanisms, like halo quenching or morphological quenching \citep{Martig2009,Martig2013}, would not be associated with the same dynamical effects that induce TDEs.  However, they all involve the availability of gas to fuel both star formation and AGN activity.  If CL-AGN are related to other nuclear transients like TDEs, we may expect their host galaxies to be similar to those of TDEs (namely, to have experienced recent bursts of star formation). If, on the other hand, CL-AGN are caused by distinct gas dynamics in the nuclei of their host galaxies, as has been proposed by \cite{liu2021}, their host galaxies may not be post-starburst and may or may not be distinct from the hosts of other AGN types. If CL-AGN are caused by instabilities in the AGN accretion disk, their host galaxies would likely reflect the underlying AGN host galaxy population.  We can therefore use the host galaxies of CL-AGN to distinguish between different proposed physical mechanisms for changing look behavior.

 While initial CL-AGN host galaxy studies are promising, they only fit a small number of emission lines or use galaxy colors to probe quenching, both of which may be contaminated by AGN emission. No study currently exists which models a detailed star formation history of CL-AGN host galaxies, and we are thus unable to distinguish between quenching mechanisms which may provide a clue to the origin of CL-AGN.  This requires a Stellar Population Synthesis (SPS) model, or a model which uses the present-day stellar population of the galaxy to infer the history of star formation \citep[for a review of modern SPS procedures, see][]{conroy2013}.


In this work, we use \texttt{Prospector} \citep{leja17,johnson17,johnson2021}, an SPS modeling tool, to model the star formation histories of \numsample CL-AGN host galaxies using SDSS spectra.  We compare our results to the star formation properties and histories of non-AGN host galaxies and Seyfert 1 and 2 host galaxies to determine whether CL-AGN host galaxies are statistically distinct from other types of AGN host galaxies.  In Section \ref{sec:data}, we describe the data and methods used throughout this paper.  In Section \ref{sec:prospectorfitting}, we describe  our \texttt{Prospector} model and its ability to fit host galaxy properties for AGN.  In Section \ref{sec:sfr}, we analyze the current star formation properties of these galaxy samples to determine whether CL-AGN host galaxies are truly more likely to fall in the Green Valley than comparison galaxies.  We then analyze the star formation histories of CL-AGN (Section \ref{sec:sfh}) to understand how their host galaxy star formation histories compare to those of other galaxies.  In doing so, we seek to determine whether a specific type of star formation history is associated with changing look behavior.  We use the changing nature of CL-AGN to investigate the impact of broad-line emission on SPS fitting in Section \ref{sec:sps_agn}. In Section \ref{sec:discussion}, we explore the possible implications of the CL-AGN host galaxy on the possible mechanisms to produce CL-AGN. 

Throughout this paper, we use AB magnitudes and assume a flat $\Lambda$CDM cosmology with $H_0 = 70$ km/s/Mpc and $\Omega_m = 0.3$, consistent with SDSS values \citep{sdssqso}.

\section{Data and samples} \label{sec:data}

\begin{figure}[t!]
    \centering
    \includegraphics[width=\columnwidth]{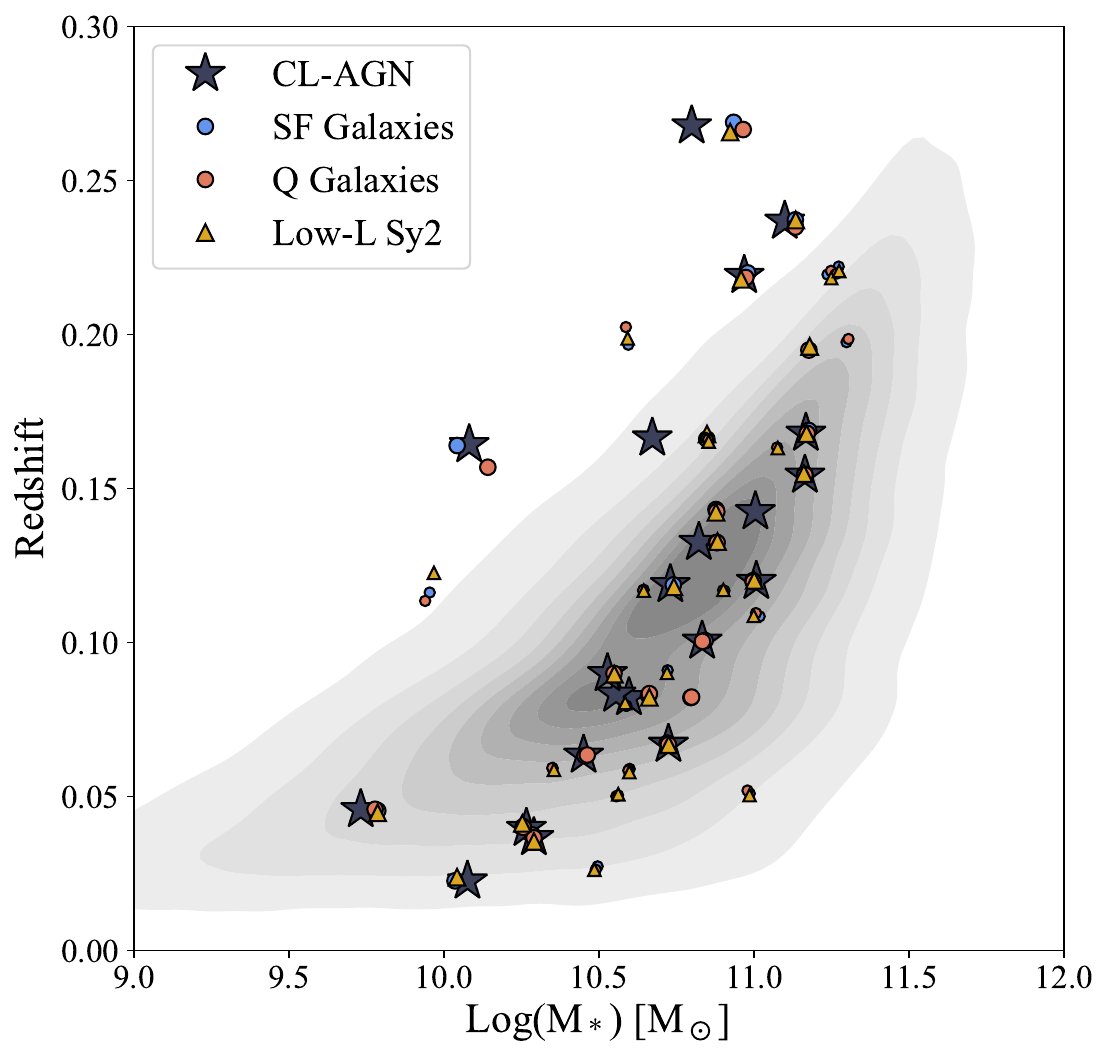}
    \includegraphics[width=\columnwidth]{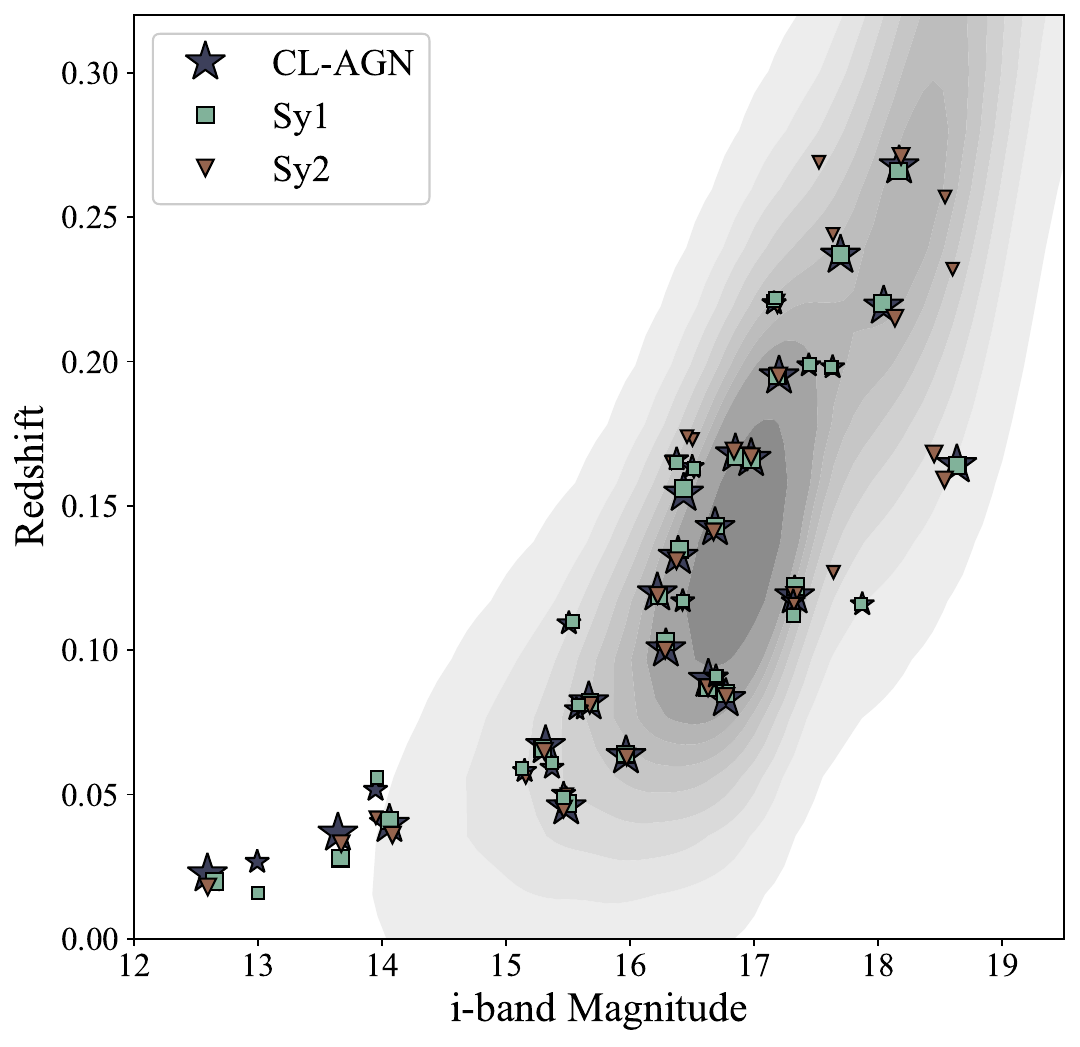}
    \caption{Comparison sample selection for our CL-AGN host galaxy sample.  Smaller markers indicate the ``silver" sample as described in Section \ref{sec:data}. (Top) Comparison low-luminosity (``Low-L") Seyfert 2- and non-AGN hosts are selected from the MPA-JHU value-added catalog from SDSS Data Release 7 \citep[][shown here as gray contours]{sdss8,Brinchmann2004}. We assign each CL-AGN with comparison galaxies with the minimum Euclidean distance in stellar mass-redshift space.  (Bottom) Seyfert 1 and 2 hosts used as the comparison sample in this paper are from \cite{veroncetty2010}  (shown in gray contours).  We perform the same Euclidean distance minimization to select comparison galaxies, but now in \textit{i}-band magnitude and redshift space.}
    \label{fig:compare}
\end{figure}


 
 We are interested in modeling the CL-AGN host galaxy parameters, but we must contend with the AGN contribution to the galaxy spectrum. To that end, we select sources from the literature which have a narrow-line spectrum which is (1) publicly available from the SDSS and (2) not classified as `QSO' by SDSS, limiting the luminosity and emission line widths of the CL-AGN in the off state to M$_i \leq -22$ and FWHM $\leq 1000 \ \si{km} \ \si{s}^{-1}$ \citep{sdssqso}.  We start with CL-AGN hosts from \cite{ruan2016, gezari2017, yang2018, Frederick2019, yu2020, green2022, tozzi2022, wang2023, dong2024, guo2024a}, and \cite{zeltyn2024}, which contain a collective 281 unique spectroscopically-confirmed CL-AGN. 
 
 These parent samples use a variety of selection methods to identify CL-AGN. \cite{ruan2016} and \cite{yang2018} look for a change in SDSS object classification (‘GALAXY’ or ‘QSO’) between SDSS epochs. \cite{gezari2017} and \cite{Frederick2019} select objects previously identified as LINERs \citep[thought to be the hosts of fading- or extremely low-luminosity AGN; see][]{heckman1980} which have brightened in optical photometry from the Zwicky Transient Facility. \cite{wang2023} also looks for photometric changes, this time in MIR, which is thought to change due to reprocessing of optical changes by the dusty torus. Another common method to identify CL-AGN is looking for changes in measured Balmer line width or flux between observations \citep[e.g.][]{yu2020, tozzi2022, dong2024, zeltyn2024}, sometimes in combination with or confirmed by visual inspection \citep{green2022, guo2024a}. We elect to combine these comparison samples to try to reduce the influence of selection choices on our final results.
 
 We restrict our analysis to the redshift range $0.03 \leq z \leq 0.40$ to reduce aperture bias and to ensure both the H$\alpha$ and H$\beta$ emission lines are within our optical spectra.  Our redshift cut removes 89 CL-AGN.  We require that at least one of H$\alpha$ or H$\beta$ completely appear or disappear for each object; this removes 20 CL-AGN.  We remove a further 133 objects which are `QSO' or have broad lines in all SDSS spectra. This leaves us with 39 CL-AGN hosts. We list the parent sample, coordinates, and fit results for each CL-AGN host in Appendix \ref{sec:fittable}. 

We use publicly-available SDSS spectra for all of our analysis. In the majority of our CL-AGN, these spectra are from the Legacy SDSS spectrograph, which covers a wavelength range of 3800–9200 \si{\angstrom} with a resolving power of 1500 at 3800 \si{\angstrom} and 2500 at 9000\si{\angstrom} and a $3^{\prime\prime}$ fiber \citep{sdss8,sdss18}. For three objects in the gold sample (J2210+2459, J1610+5436, and J0845-0027) we instead use spectra taken with the BOSS spectrograph, which has the same spectral resolution but a slightly wider wavelength range ($3650–10400$ \si{\angstrom}) and smaller fiber ($2^{\prime\prime}$). We use the narrow-line version of the spectrum for all fitting (see further discussion in Section 3). CL-AGN spectra across our sample have signal-to-noise ratios ranging from 6.1 to 40.7, with a mean value of 19.5; we mask all pixels with a signal-to-noise ratio less than five in our fitting.




\subsection{Gold and Silver Samples}

A subset of CL-AGN hosts in our sample do not change fully from a Seyfert Type 1 to 2 or vice versa.  Some objects retain small amounts of broad-line emission when in the narrow-line state (e.g. changing between Type 1 and Type 1.8). For others, especially in the upper redshift end of our sample, follow-up spectra observed only H$\alpha$ or H$\beta$ due to the wavelength ranges of follow-up telescopes.  Since the physical mechanism behind CL-AGN is unclear, we cannot determine whether objects which have only changed one line are caused by the same physical mechanism as those with changes in both. 

We therefore define a purer  ``gold" sample and a  potentially more contaminated ``silver" sample of CL-AGN hosts.  The gold sample totally lacks broad emission in H$\alpha$ and H$\beta$ in the narrow-line state and contains broad emission in both lines in the broad-line state (in other words, transitions fully between Seyfert types 1 and 2).  We have 21 CL-AGN hosts in the gold sample.  The  silver sample  includes the 18 objects for which only one broad Balmer line fully appears or disappears (those which transition to intermediate Seyfert types or with limited coverage in at least one spectrum).  We therefore have \numsample CL-AGN in our ``combined" sample.

For all analysis in this work, we state results for both the gold sample alone and the combined gold and silver samples. In doing so, we take advantage of the larger number of galaxies available when including the silver sample without loss of purity.  We match separate comparison samples for the gold and silver samples and ensure there is no overlap.

\subsection{Comparison Sample Selection} \label{sec:comparisonsample}


We use the MPA-JHU value-added catalog from SDSS Data Release 7 as our non-AGN galaxy and low-luminosity Seyfert 2 comparison samples \citep{Kauffmann2003a,Brinchmann2004,tremonti2004,sdss8}. We select Seyfert 2 hosts as any object with a Baldwin-Phillips-Terlevich classification ``AGN" \citep{baldwin1981}; we use objects not classified as ``AGN," ``composite," or ``low S/N LINER" for our parent galaxy sample. We then use the redshift-dependent star forming main sequence relations given in \cite{Whitaker2012b} along with the MPA-JHU redshift, mass, and star formation rate of each individual object, to compute its location relative to the star forming main sequence. To select objects which are robustly star forming or quiescent, we classify galaxies falling above the star forming main sequence as ``star forming" and objects at least 1.5 dex below the star forming main sequence as ``quiescent."  To select individual comparison galaxies and Seyferts, we  compute a Euclidean distance in mass-redshift space between each CL-AGN and each galaxy or Seyfert.  We then select the star-forming galaxy, quiescent galaxy, and low-luminosity Seyfert 2 host with the minimum distance to each CL-AGN host.  We restrict this distance to 0.05 in both logarithmic mass and redshift.  The results of this matching can be seen in the top panel of Figure \ref{fig:compare}.  We note that the masses shown for CL-AGN hosts are the result of the \texttt{Prospector} fitting described in Section \ref{sec:prospectorfitting}, while the masses for the galaxy and Seyfert 2 comparison samples are the MPA-JHU catalog values (\texttt{lgm\_tot\_p50}). For our matching, we do not correct the outputted CL-AGN as described in Appendix \ref{sec:sfrcalib}, as the outputted masses from nonparametric \texttt{Prospector} templates are on average $\sim0.2$ dex higher than those of other mass tracers \citep[e.g. UV/IR mass tracers; see][]{leja2021_prospectormasses}.  As a result, the uncorrected \texttt{Prospector} surviving stellar mass is in good agreement with the masses in the MPA-JHU catalog.  The masses stated throughout the rest of this paper have been corrected as discussed in Section \ref{sec:sfrcalib}.

\begin{deluxetable*}{cccc}[t]
\tablecaption{Parameters for \texttt{Prospector} fits \label{tab:priors}}
\tablewidth{0pt}
\tablehead{
\colhead{Name} & \colhead{Description} & \colhead{Prior} & \colhead{(Initial/Central)$^{(a)}$ Value} 
}
\startdata
zred & Redshift & Normal & SDSS Value \\
dust\_type & Dust Attenuation Model & Fixed & Power-Law \\ 
dust2 & Dust attenuation & Uniform [0.0,0.2] & 0.05 \\
imftype & IMF & Fixed & \citet{chabrier2003} \\
logmass & Mass formed & Uniform [7,12] & From \citet{bell2003} Table 7\\
logzsol & Metallicity & Normal & From \cite{gallazzi2005} Table 2 \\
sfrratio\_old & SFR in old bins to the first flex bin & Student T & [0.0, 0.0, 0.0] \\
sfrratio\_young & SFR in the youngest bin to the last flex bin & Student T & 0.0 \\
sfrratio & SFR ratios in flex bins & Student T & [0.0, 0.0, 0.0, 0.0] \\
tflex & The length of time containing all flex bins & Fixed & 4 Gyr \\
tlast & The length of the last time bin & Uniform [0.01, 3.9] & 1 Gyr \\
f\_outlier\_spec & Fraction of outlier pixels & Uniform [0.0, 0.1] & 0.001 \\
spec\_jitter & Spectroscopic white noise term & Uniform [0.5, 15.0] & 1.0
\enddata
\tablecomments{This table contains all priors, initial values, and fixed parameters for the CL-AGN \texttt{Prospector} fits. $(a)$ Values listed are the fixed value for the fit if the prior is "Fixed." Otherwise, they are the initial guess (for Uniform prior) or center of the prior (for Gaussian or Student T priors).}
\end{deluxetable*}

To select comparison Type 1 Seyferts, we use the catalog of \cite{veroncetty2010}, a catalog of all known AGN at time of publication, for our Seyfert 1 comparison sample based on their classification of Seyfert types. For these objects, no stellar mass estimates are available, so we repeat the above analysis using i-band magnitude as a proxy for stellar mass.  For \textit{i}-band magnitude, we restrict the Euclidean distance to 0.50 mag (bottom panel of Figure \ref{fig:compare}).  As the \cite{veroncetty2010} sample contains an overall more luminous sample of AGN than \cite{Brinchmann2004}, we want to ensure any differences in results between the two catalogs are due to real differences in AGN populations, not selection; to this end, we also match a Seyfert 2 comparison sample from \cite{veroncetty2010} to the CL-AGN hosts in \textit{i}-band magnitude and redshift.  To differentiate between our two Seyfert 2 comparison samples, we refer to the sample from the MPA-JHU catalog as ``Low-L Sy2" in all plots and ``low-luminosity Seyfert 2" in the text of this paper; the sample from \cite{veroncetty2010} is simply referred to as ``Sy2" in plot legends, and would be labeled ``high-luminosity Seyfert 2" in the text. In general, our results are similar for both Seyfert 2 samples, so we use ``Seyfert 2 hosts" to refer to both samples when appropriate.



\section{Prospector Fitting} \label{sec:prospectorfitting}

\subsection{Prospector Assumptions and Use} \label{sec:prospectorassumptions}

We fit all galaxies in this work with \texttt{Prospector} Stellar Population Software \citep{leja17, johnson17, johnson2021}. \texttt{Prospector} uses a spectral library built by the Flexible Stellar Population Synthesis (FSPS) code \citep{conroy2009,conroy2010} and accessed through \texttt{py-fsps} \citep{pyfsps} to infer physical galaxy parameters from observed photometry and spectroscopy. We use the MILES spectral library \citep{sanchez-blazquez2006} and the MIST isochrones \citep{dotter2016, choi2016, paxton2011, paxton2013, paxton2015}. The output uncertainties from our \texttt{Prospector} fits are generally symmetric, though the systematic errors discussed in Section \ref{sec:AGNimpact} sometimes are not.  To account for this, we add the 16th and 84th percentiles of the outputted chains in quadrature with upper and lower systematic errors to create the error bars shown throughout this work. 



In our fits, we use a Gaussian prior for redshift centered on the reported SDSS spectroscopic redshift with a width of 0.005.  We convert the SDSS \textit{g} and \textit{r} band magnitudes to an initial guess for stellar mass using the mass-to-light ratios from \cite{bell2003} Table 7.  Based on this initial guess, we use the data from \cite{gallazzi2005} Table 2 to inform our metallicity prior, a Gaussian centered at the median metallicity for a galaxy of the initial stellar mass guess with a width encompassing the 16th and 84th percentile metallicity values for that stellar mass. We utilize the \texttt{outlier\_model} template and a term which can inflate the spectral white noise with a likelihood penalty term (\texttt{spec\_jitter}) from \texttt{Prospector} to fit for observational noise.  We assume a Chabrier IMF \citep{chabrier2003} and assume a fixed power-law dust absorption model with an index of -0.7.  We do not account for additional dust attenuation around young stars. All priors used in our fits are listed in Table \ref{tab:priors}.

We fit only the spectrum, since photometry may change due to changing contribution from the CL-AGN. This introduces the possibility for offsets in recovered parameters due to spectrophotometric calibration and aperture losses. To address this, we use our non-AGN host galaxy samples to calibrate output SFRs based on the difference in recovered SFR when applying a spectrophotometric calibration during fitting; for further details, see Appendix \ref{sec:sfrcalib}.




\begin{deluxetable}{cccc}
\tablecaption{Dynesty settings \label{tab:dynesty}}
\tablewidth{0pt}
\tablehead{
\colhead{Name} & \colhead{Setting} 
}
\startdata
Bounding method & Multi \\
Sampling method & Random walk \\
Nested walks & 32 \\
Initial nlive & 400 \\
Nlive batch & 400 \\
Pfrac & 1 \\
Max batch & none \\
Max calls & none \\
Nested bootstrap & 20 \\
Dlogz init & 0.01 \\
Target n effective & 20,000 \\
\enddata
\tablecomments{This table contains the Dynesty settings used in our \texttt{Prospector} fits.}
\end{deluxetable}



We use Dynesty \citep{speagle2020} Dynamic Nested Sampling \citep{higson2019}, a method which speeds up sampling by sampling coarsely over the entire parameter space, then taking finer and finer ``slices" of higher likelihood and sampling more finely to trace potentially complicated posterior distributions. Stellar population synthesis often entails disentangling correlated parameters (e.g. dust reddening versus an older stellar population) with many local maxima in likelihood.  Dynesty allows us to efficiently explore the overall posterior distribution and does well with these complicated parameter spaces. Dynesty settings are listed in Table \ref{tab:dynesty}.  We use a multiple ellipsoid-bounded \citep{feroz2009} random walk sampling method \citep{Skilling2006} with 32 walks. 


Because limited analysis has been performed on the stellar populations of CL-AGN host galaxies, we elect to use a non-parametric model that allows for, but does not force, events like multiple past starbursts and other irregularities in the star formation rate. Non-parametric fits are flexible enough to encompass irregular star formation histories, which is crucial for our attempt to constrain the path to quenching if it is occurring \citep{leja2019}. The \texttt{continuity\_psb\_sfh} template developed in \cite{suess2022c} is ideal for this type of analysis,  as it is publicly available, designed to fit a flexible star formation history, and has been tested on a variety of mock galaxy types, including (but not limited to) galaxies undergoing quenching. 


The \texttt{continuity\_psb\_sfh} template is a non-parametric star formation history template with three fixed-time, variable-mass star formation rate (SFR) bins at the beginning of the universe followed by five flexible-time, fixed-mass bins and a final star-formation bin which can vary in both stellar mass formed and duration. The \texttt{continuity\_psb\_sfh} template is parameterized by three sets of ratios: \texttt{logsfr\_ratio\_old}, a three-dimensional array of the log(SFR) ratio between subsequent fixed-width time bins to the first flex bin; \texttt{logsfr\_ratio}, a four-dimensional array of log(SFR) ratios between adjacent flex-width bins; and \texttt{logsfr\_ratio\_young}, a one-dimensional array of the log(SFR) ratio between the final flex bin and the youngest bin, which can vary in both time and star formation rate. We set each of these initial values to 0. \citet{suess2022b} sets each of these values using simulated galaxies from UniverseMachine, a physically-motivated catalog of star formation histories based on galaxy halo masses \citep{behroozi2019}, to better accommodate early-Universe star formation, which can be underestimated in SED modeling due to outshining from younger stellar populations. In our testing, we do not find any significant differences in recovered star formation history or galaxy properties when using the UniverseMachine prior or setting all initial values to 0 (see further discussion in Section \ref{sec:modelchoices}.) We use a Student T prior for all log(SFR) ratios. We set \texttt{tflex}, or the total time containing all six of the flexible-width bins (including the final time bin), to 4 Gyr.  We use a uniform prior for the length of the last time bin, which will be well-constrained if a starburst has occurred and unconstrained if no starburst has occurred.

The \texttt{continuity\_psb\_sfh} template recovers basic parameters (e.g. stellar mass, metallicity, dust content, and SFR at time of observation) comparably well to other non-parametric star formation history templates while allowing for increased flexibility in the shape of the star formation history \citep{suess2022b}. In particular, a flex-width final time bin distinguishes the \texttt{continuity\_psb\_sfh} template from other non-parametric star formation history templates \citep[e.g. those described in][]{leja2019}. As discussed in \citet{suess2022b}, allowing the width of the final star formation bin to vary is necessary to recover the time since a starburst ended. If CL-AGN are caused by TDEs in AGN disks, they should be sensitive to the time since a starburst occurred \citep[][Shepherd et al. in prep]{arcavi2014,French2017,stone2018,melchor2024,wangm2024,wangyihan2024a,teboul2025}. At the same time, if CL-AGN have entered the Green Valley without experiencing a starburst (as is suggested in \citet{dodd2021} and \citet{wang2023}), we do not want to artificially fit a starburst which did not occur. The \texttt{continuity\_psb\_sfh} template has been extensively tested by \citet{suess2022b} in galaxies which have quenched without a starburst or which have ongoing star formation. Again, this template is able to capture  the galaxy's SFR \citep[in the cases where there is ongoing star formation; see][and Section \ref{sec:AGNimpact}]{suess2022b} and stellar mass, as well as the shape of the star formation history. The \texttt{continuity\_psb\_sfh} template is the only star formation history template available which can accurately recover the time since a starburst ended without forcing a particular star formation history shape; as such, it is the ideal template for our analysis.

AGN have a characteristic power-law spectrum and can contribute significantly to the blue continuum of their host galaxies.  They also contribute Balmer emission lines which can infill absorption from older stellar populations.  Both this blue light and Balmer emission can easily be mistaken for ongoing star formation in an AGN host galaxy.  This contamination can be minimized in CL-AGN hosts by choosing the narrow-line spectrum, as the lines are easy to mask and the spectrum in this epoch has minimal blue contamination from the AGN.  For the purposes of our fitting, we use only the narrow-line spectrum and mask significant emission lines that are likely from the AGN: H$\alpha$; H$\beta$; H$\gamma$; H$\delta$; [OI] $\lambda 6300$; [OII] $\lambda 3727$; [OIII] $\lambda 5007$, $\lambda 4969$, and $\lambda 4363$; [NII] $\lambda 6583$; [SII] $\lambda 6717$ and $\lambda 6737$; and [NeIII] $\lambda 3869$. Although we select the narrow-line spectrum for our fits, it is possible that a small amount of broad-line contamination remains; to that end we apply a 50 \si{\angstrom} mask to the H$\alpha$ emission line and a 30 \si{\angstrom} mask to the H$\beta$ and [OIII] $\lambda 5007$ lines. We apply a 20 \si{\angstrom} mask to all other lines. We also mask any individual pixels with a signal-to-noise ratio less than five.


\subsection{Evaluating Prospector with Mock Data for AGN and non-AGN} \label{sec:AGNimpact}

We test the performance of the \texttt{continuity\_psb\_sfh} template by generating 105 mock spectra using \texttt{Prospector}. We use the built-in \texttt{build\_model} functionality in Prospector, which is normally used to forward-model spectra in fitting, to generate a suite of mock galaxy spectra using the \texttt{continuity\_psb\_sfh} template. For each spectrum, we vary stellar mass, AGN fraction, and the ratio of star formation rates between bins.  We generate galaxies with a log uniform stellar mass distribution between $10^{9}$ and $10^{12} \text{ M}_\odot$, roughly spanning the masses probed by our CL-AGN sample. We fix dust  and log metallicity to $\tau_{5500 \si{\angstrom}} = 0.6$ and $\text{Log}(\text{Z}/\text{Z}_{\odot}) = -0.5$, respectively, as these are the default values for \texttt{Prospector} and fall well within the recovered values for our samples. We then randomly sample the prior distribution of SFR ratios from the \texttt{continuity\_psb\_sfh} template to generate a star formation history to input to the model (see priors in Table \ref{tab:priors}). We add varying levels of AGN contamination from a quasar template taken from \citet{shen2011} to approximately half of the model spectra. The AGN fraction varies from zero to 0.9 to allow us to test parameter recovery even in cases when the spectrum has a significant AGN contribution, both from broad line emission and continuum. We extend the quasar continuum with a simple power law to cover the wavelength range of SDSS for an object at $z=0.3$. We redshift each spectrum to $z = 0.3$ so that we can evaluate the performance of our model for our most distant CL-AGN spectra.  Rather than generating errors of our own, we use the wavelength-dependent uncertainties from a real SDSS spectrum at a redshift of 0.3 and rescale it to fit the overall flux of our mock spectra, as varying the mock galaxy mass will influence the overall flux of the spectrum.  We do the same re-scaling to add the quasar template spectrum to construct our AGN host galaxy spectra. We then re-fit these spectra using the same pipeline we use for our real CL-AGN spectra.

\begin{figure}[t!]
    \centering
    \includegraphics[width=.5\textwidth]{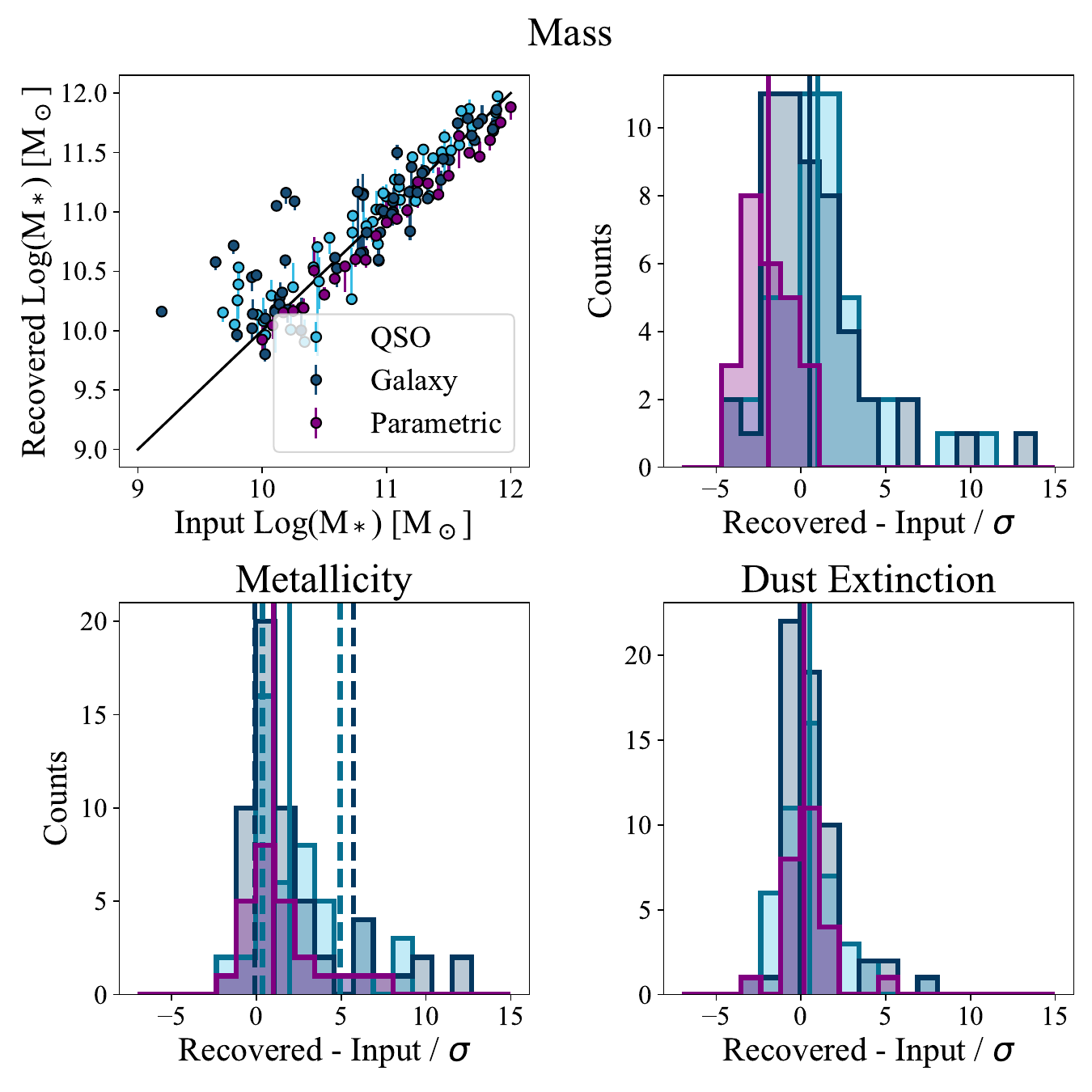}
    \caption{Comparison between the input and recovered mass, metallicity, and dust extinction of our mock galaxy spectra.  Solid lines indicate the median difference in recovered versus input values, while dashed lines indicate the 16th and 84th percentile values. ``QSO" denotes mock spectra with added AGN emission, and ``Parametric" denotes mock spectra generated with a delayed-tau plus exponential burst star formation history template. These parameters are well-recovered for spectra generated with the \texttt{continuity\_psb\_sfh} template, even with AGN contamination. While dust extinction is well-recovered for mock spectra generated with a parametric star formation history template, our pipeline underestimates that mass and overestimates the metallicity for these spectra.}
    \label{fig:mockspecdustandmass}
\end{figure}

\begin{figure*}[ht!]
    \centering
    \includegraphics[width=\textwidth]{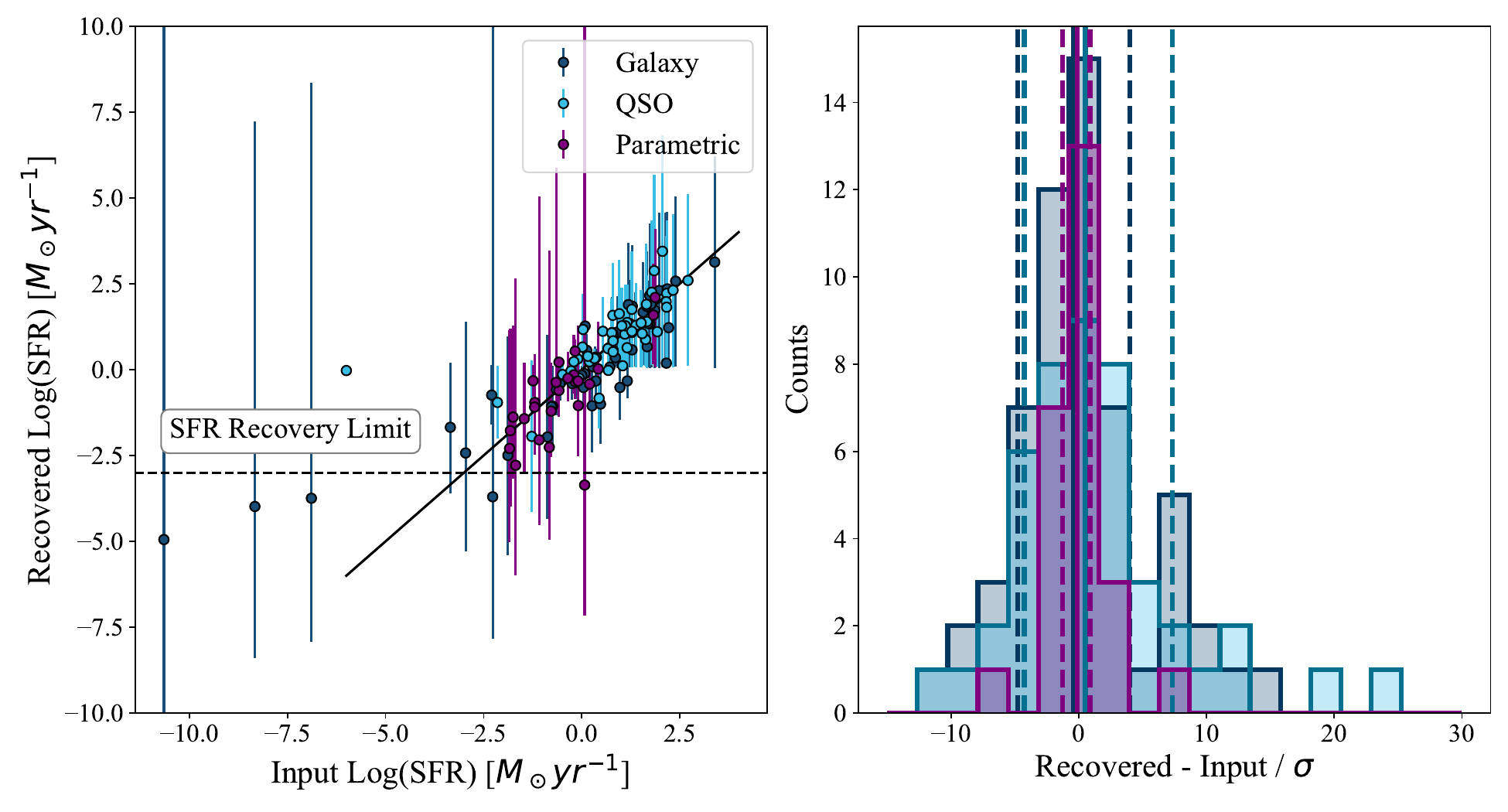}
    \caption{Comparison between the input and recovered star formation rate from our non-parametric \texttt{Prospector} fits. Solid lines indicate the median difference in recovered versus input values, while dashed lines indicate the 16th and 84th percentile values. ``QSO" denotes mock spectra with added AGN emission. We recover star formation rate well regardless of input star formation history shape} and do not see a significant overestimation of recent star formation even for spectra with AGN contamination.  Due to the difficulty of recovering very low SFRs, we take $\text{Log(SFR)} = -3$ as our lowest recoverable value and use it as an upper limit on some SFR values in this work.
    \label{fig:sfr_recovery}
\end{figure*}

 We find that our pipeline reliably recovers the mass and SFR for each mock galaxy spectrum as shown in Figures \ref{fig:mockspecdustandmass} and \ref{fig:sfr_recovery}. When we do not include AGN contamination, the recovered log galaxy masses have a median offset of 0.13 dex from the input masses with a 16th and 84th percentile offset of 0.20 and 0.02 dex, respectively. The uncertainties on the recovered stellar mass, which is the only directly recovered parameter from \texttt{Prospector} which is relevant to this study, overlap the input values. The recovered log star formation rates are $0.047^{+0.48}_{-0.66}$ dex lower than the input rates, again overlapping the true value. Importantly, we find that the recovered SFR becomes increasingly uncertain as the SFR dips below $10^{-3} \text{ M}_\odot~\text{yr}^{-1}$ (Figure \ref{fig:sfr_recovery}).  We take this as our detection limit for ongoing star formation in any given galaxy.  For any object with a recovered SFR below this value, we assign an upper limit on star formation of $10^{-3} \text{ M}_\odot~\text{yr}^{-1}$. \texttt{Prospector} also reliably recovers the dust extinction and metallicity for each model spectrum. The recovered dust extinctions are $0.0054^{+0.31}_{-0.069}$ $\tau_{5500\si{\angstrom}}$ higher than the input extinctions. The recovered metallicities are $0.17^{+0.52}_{-0.19}$ dex higher than the input value.  
 
We use the scatter in recovered parameters to calculate systematic errors for all parameters from the \texttt{continuity\_psb\_sfh} template by taking the 16$^\text{th}$ and 84$^\text{th}$ percentile of the recovered minus input parameter values.  These systematic errors are added in quadrature to the error derived from \texttt{Prospector} itself to create the error bars throughout this work.

\begin{figure}
    \centering
    \includegraphics[width=\columnwidth]{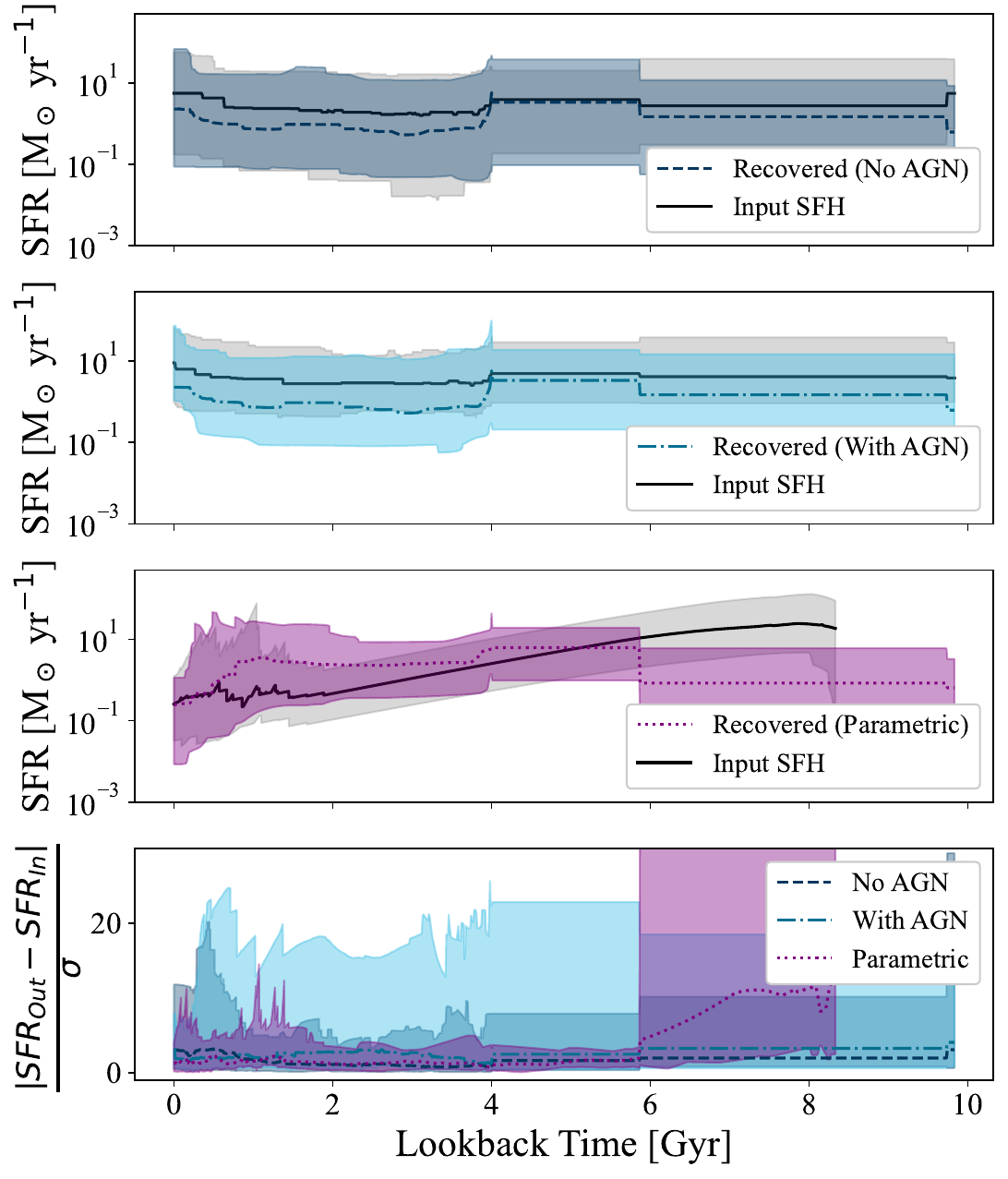}
        \caption{Average input and recovered star formation histories for the mock spectra described in Section \ref{sec:AGNimpact}. The top two panels represent spectra generated with the \texttt{continuity\_psb\_sfh} template, without (top) and with (second from top) added AGN emission. The second panel from the bottom represents spectra generated with a delayed-tau plus exponential burst star formation history. Shaded regions represent the 16th and 84th percentile input or recovered star formation histories across the sample. The \texttt{continuity\_psb\_sfh} template can reliably recover star formation histories for model spectra with no AGN contamination.  (Bottom) Recovered minus input star formation histories for the mock spectra normalized by the output uncertainties from \texttt{Prospector}.}
    \label{fig:sfhrecovery}
\end{figure}


We do not see a significant offset in mass or SFR recovery when fitting for AGN host galaxy spectra (Figure \ref{fig:mockspecdustandmass}).  With AGN contamination, the recovered log stellar masses are $0.042^{+0.13}_{-0.17}$ dex lower than the input masses, and the recovered log SFRs are $0.036^{+0.58}_{-0.48}$ dex higher than the input SFR. We also do not see a significant offset in the shape of most of the recovered star formation history (Figure \ref{fig:sfhrecovery}).  We do see a slight tendency towards overestimation of the star formation rate in the last $\sim0.5$ Gyr for both the non-contaminated and AGN contaminated mock spectra.  This overestimation is accompanied by an increase in the uncertainties outputted by \texttt{Prospector}, which is incorporated into our systematic uncertainty calculations for the recovered SFR.

\subsection{Evaluating the \texttt{continuity\_psb\_sfh} template on Post-Starburst Star Formation Histories} \label{sec:parametrictemplate}

We next test the ability of the \texttt{continuity\_psb\_sfh} to recover galaxy properties for mock spectra generated using a parametric star formation history template consisting of a delayed-tau function plus an exponential starburst; this is the star formation history shape expected for post-starburst galaxies, which may indicate a TDE origin for CL-AGN. The \texttt{continuity\_psb\_sfh} template has already been extensively tested for use with post-starburst spectra \citep{suess2022b}; here, we simply test our own implementation to ensure we are able to capture, but do not force, a starburst in our fits. We generate 25 additional mock spectra using this star formation history template. We again randomly draw masses log-uniformly from 10$^{10}$ to 10$^{12}$ M$_\odot$. We randomly draw ages from a uniform distribution between 8 and 10 Gyr old. We randomly draw burst mass fractions from a log-uniform distribution between 0.01 and 1. The time since the burst beginning in this model is parameterized as a fraction of the galaxy’s age measured from the time of the galaxy’s formation; we draw this fraction from a uniform distribution between 0.8 and 0.99, spanning a potential range from .08 to 2 Gyr ago. We again use the noise from a real SDSS spectrum scaled to the mock spectrum. We perform our normal fitting routine on the mock spectra as described above and with the priors described in Section \ref{sec:prospectorfitting}.

We find that the \texttt{}{continuity\_psb\_sfh} template somewhat underestimates the stellar mass formed over the course of the delayed tau plus exponential starburst star formation history by a median value of 0.25$^{+0.17}_{0.09}$ dex, or about 2 M$_\odot$ yr$^{-1}$. This is consistent with the star formation history shown in Figure \ref{fig:sfhrecovery}, where the recent star formation is recovered better than the older period of star formation. This is expected, as older stellar populations have lower light fractions and longer lifetimes which can obscure the exact shape of the earliest period of star formation. For the purposes of this study, we care mostly about star formation trends within the past few Gyr, and we therefore are not concerned about this less constrained fit on the earliest epoch of star formation. We find that the SFR is well-recovered with a median recovered offset of 0.02$^{+0.31}_{-0.94}$ dex. Dust attenuation is also well-recovered within 0.23 dex, and the metallicity is overestimated with a median offset of 0.19$^{+0.32}_{-0.19}$ dex. We explore the impacts of this offset in recovered mass in Section \ref{sec:modelchoices}.

 

\section{Star formation in CL-AGN hosts} \label{sec:sfr}

\subsection{Color-magnitude analysis}

\begin{figure}
    \centering
    \includegraphics[width=\columnwidth]{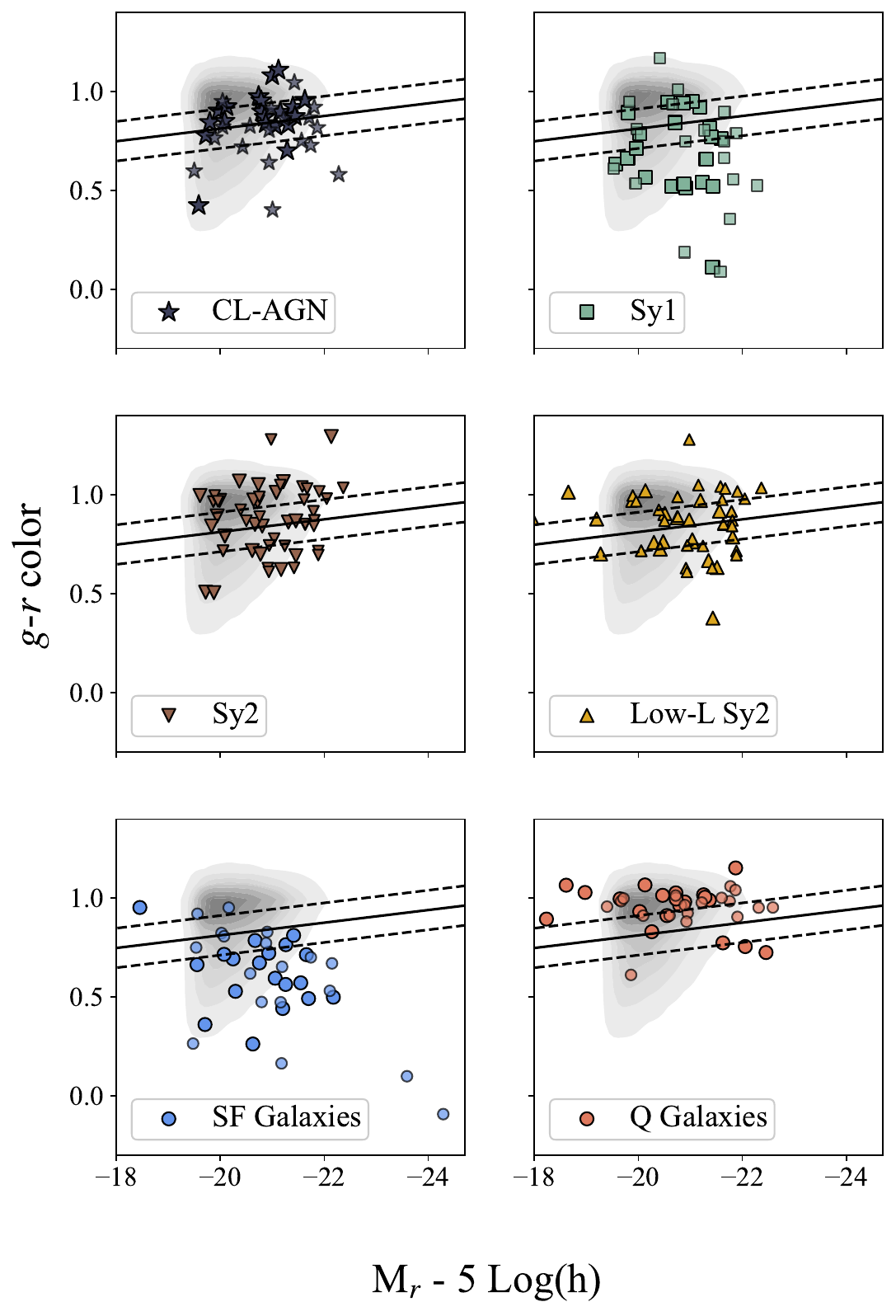}
    \caption{Color-magnitude diagrams for the CL-AGN and comparison samples. Objects from the silver samples are shown with smaller markers and lower opacities.  The lines dividing star-forming, Green Valley, and quiescent galaxies comes from \cite{Weinmann2006} with a width of 0.2 magnitudes.  Underlying gray contours are the colors and magnitudes of MPA-JHU galaxies at a redshift of 0.1.  We find that CL-AGN host galaxies fall between star-forming and quiescent galaxies in $g$-$r$ color and are similarly distributed to Seyfert 2 hosts.  Seyfert 1 hosts may be bluer due to higher AGN luminosity or due to ongoing star formation.}
    \label{fig:colormag}
\end{figure}

To facilitate comparison to previous work, we first investigate the \textit{g}-\textit{r} color of CL-AGN host galaxies as a proxy for current star formation. We use the reported SDSS model magnitudes for our color analysis. We use the \cite{Weinmann2006} distinctions for star-forming versus quiescent galaxies in \textit{g}-\textit{r} color-magnitude space.  We define the Green Valley as the 0.2 magnitude region centered on the Weinmann dividing line.  We use the Python package \textit{kcorrect} \citep[][version 5.1.5]{kcorrect} to correct all magnitudes to a redshift of 0.1.  In Figure \ref{fig:colormag}, we present the colors and magnitudes of all samples studied in this paper, as well as those of the MPA-JHU sample at $z=0.1$ (gray contours). 

The star-forming main sequence changes with redshift. Over a redshift range of 0.03 $<$ z $<$ 0.4, the median SFR of a $10^{10.5}$ M$_\odot$ star-forming galaxy is expected to change by $\sim0.4$ dex, slightly larger than the width of the star forming main sequence \citep{Whitaker2012b}. As such, a galaxy on the upper end of the Green Valley in the local Universe could be considered star-forming at a redshift of 0.4. To account for this evolution, we divide our sample into a lower- and higher- redshift bin. We elect a redshift of 0.15 for this division to facilitate comparison to other works. For example, \cite{dodd2021} finds that CL-AGN hosts are in the Green Valley at $z < 0.15$ while they are on the star-forming main sequence at $z>0.15$; meanwhile, the works of \cite{liu2021} and \cite{wang2023} examine CL-AGN at z $<$ 0.15 and z $\lesssim$ 0.1 respectively.

$60^{+9}_{-10}\%$ of z $<$ 0.15 and $50^{+13}_{-13}\%$ of of z $>$ 0.15 CL-AGN hosts fall in the color-magnitude Green Valley. CL-AGN hosts tend to be redder than star-forming galaxies and bluer than quiescent galaxies.  Under a Kolmogorov-Smirnov test, CL-AGN hosts are statistically distinct in $g-r$ color from both galaxy comparison samples and the Seyfert 1 host galaxy sample, which follows a similar distribution to the star-forming galaxy sample.  This suggests CL-AGN hosts, like Seyfert 2 hosts, have intermediate colors reminiscent of the color-magnitude Green Valley.

The colors of AGN host galaxies may be significantly contaminated by either the blue contamination from the AGN or dust contamination from the AGN torus.  For example, it is difficult to distinguish whether Seyfert 1 hosts are more star forming than Seyfert 2 hosts or whether this color difference is due to increased dust obscuration in Seyfert 2 hosts, which may both redden the galaxy and obscure the broad-line region of the AGN.  Similarly, the CL-AGN hosts may fall in the Green Valley of the color-magnitude diagram due at least in part to AGN contamination.  We therefore turn to stellar population synthesis to distinguish between these star formation and AGN contributions to the host galaxy's light.



\subsection{Current star formation} \label{sec:current_sfr}

\begin{figure}
    \centering
    \includegraphics[width=\columnwidth]{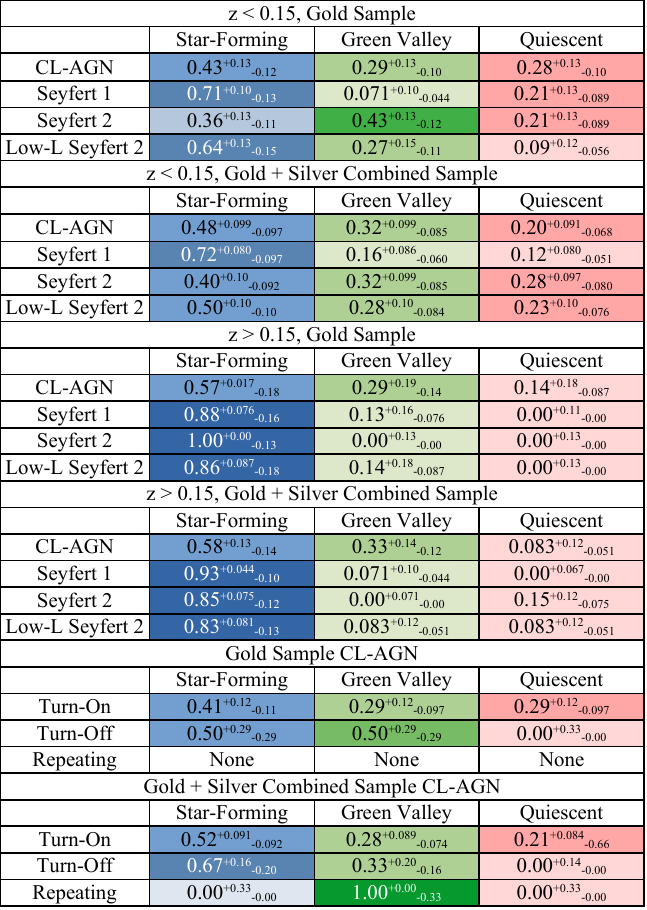}
    \caption{The fraction of each sample which is classified as star forming, Green Valley, or quiescent based on star formation and stellar mass.  Classifications are based on the \cite{Whitaker2012b} star forming main sequence for the redshift of each individual object. Fractions are rounded to two decimal places and may not sum exactly to 1. We report one-sigma binomial errors for each classification. Bins are shaded by fraction of galaxies in each sample, with darker bins indicating more galaxies in that sample fall in that bin. Like other AGN host galaxies, a plurality of CL-AGN hosts reside on the star forming main sequence across all redshift bins and in both the gold and combined samples.}
    \label{fig:samplefractions}
\end{figure}

We use our \texttt{Prospector} model star formation history to compute a final mass and specific star formation rate (or sSFR, a mass-normalized measure of star formation) for each object in our sample of interest. \texttt{Prospector} returns a stellar mass for each objects in units of stellar masses formed over the lifetime of the galaxy.  To account for stellar deaths, we convert this mass to a surviving stellar mass using the output \texttt{mfrac} parameter, which quantifies the fraction of stellar mass surviving in the last time bin.  Masses reported throughout this paper for all samples are in units of surviving mass. The \texttt{continuity\_psb\_sfh} model does not directly output an sSFR, so we define the star formation rate as the median star formation rate of the last 10 Myr \citep[consistent with observational constraints, see][]{kennicutt2012} and normalize by the surviving mass. The results of our \texttt{Prospector} fits can be found in Appendix \ref{sec:fittable}.

This type of analysis is highly dependent on the choice of star-forming main sequence \citep{Cristello2024}, a property which evolves with redshift \citep[see review by][]{madau2014}.  Our CL-AGN sample extends to a maximum redshift $z=0.39$.  To account for redshift evolution of the star forming main sequence, we use the redshift of each individual object and the redshift-dependent star formation main sequence from \cite{Whitaker2012b} (formulas 1\textendash3) to determine whether objects are star-forming, Green Valley, or quiescent for the purposes of this study.  Any galaxy which falls above or within the 0.5 dex scatter of the star-forming main sequence is considered star-forming; galaxies falling 0.5\textendash1.5 dex below the star-forming main sequence are in the Green Valley; galaxies falling more than 1.5 dex below the main sequence are quiescent.  


 We classify each sample into star forming, Green Valley, or quiescent to determine whether CL-AGN are overrepresented in any of these categories.  We find that a plurality of CL-AGN reside in star-forming galaxies.  $43^{+13}_{-12}\%$ of gold CL-AGN hosts at $z<0.15$ are star-forming, while $29^{+13}_{-10}\%$ are in the Green Valley and $28^{+13}_{-10}\%$ are quiescent.  When we include both the gold and silver samples, $48^{+9.9}_{-9.7}\%$ of CL-AGN hosts fall in the star-forming main sequence at $z<0.15$. This is similar to the other gold AGN samples at low redshift.  The fraction of each sample which is star-forming, Green Valley, or quiescent is shown in Figure \ref{fig:samplefractions} along with one-sigma binomial confidence intervals for each value.  

At $z>0.15$, we see a similar trend, though all samples are predominantly on the star forming main sequence at this redshift.
At this redshift, we find that $57^{+13}_{-18}\%$ of CL-AGN in the gold sample are star-forming, $29^{+19}_{-14}\%$ are in the Green Valley, and $14^{+18}_{-8.7}\%$ are quiescent.

We evaluate whether CL-AGN hosts are statistically distinct from other types of AGN host galaxies in star forming classification using a chi-squared test of homogeneity.  Under this test, CL-AGN hosts in both redshift bins are indistinguishable from other AGN host galaxies, regardless of whether we include the silver sample.

\begin{figure*}
    \centering
    \includegraphics[width=\textwidth]{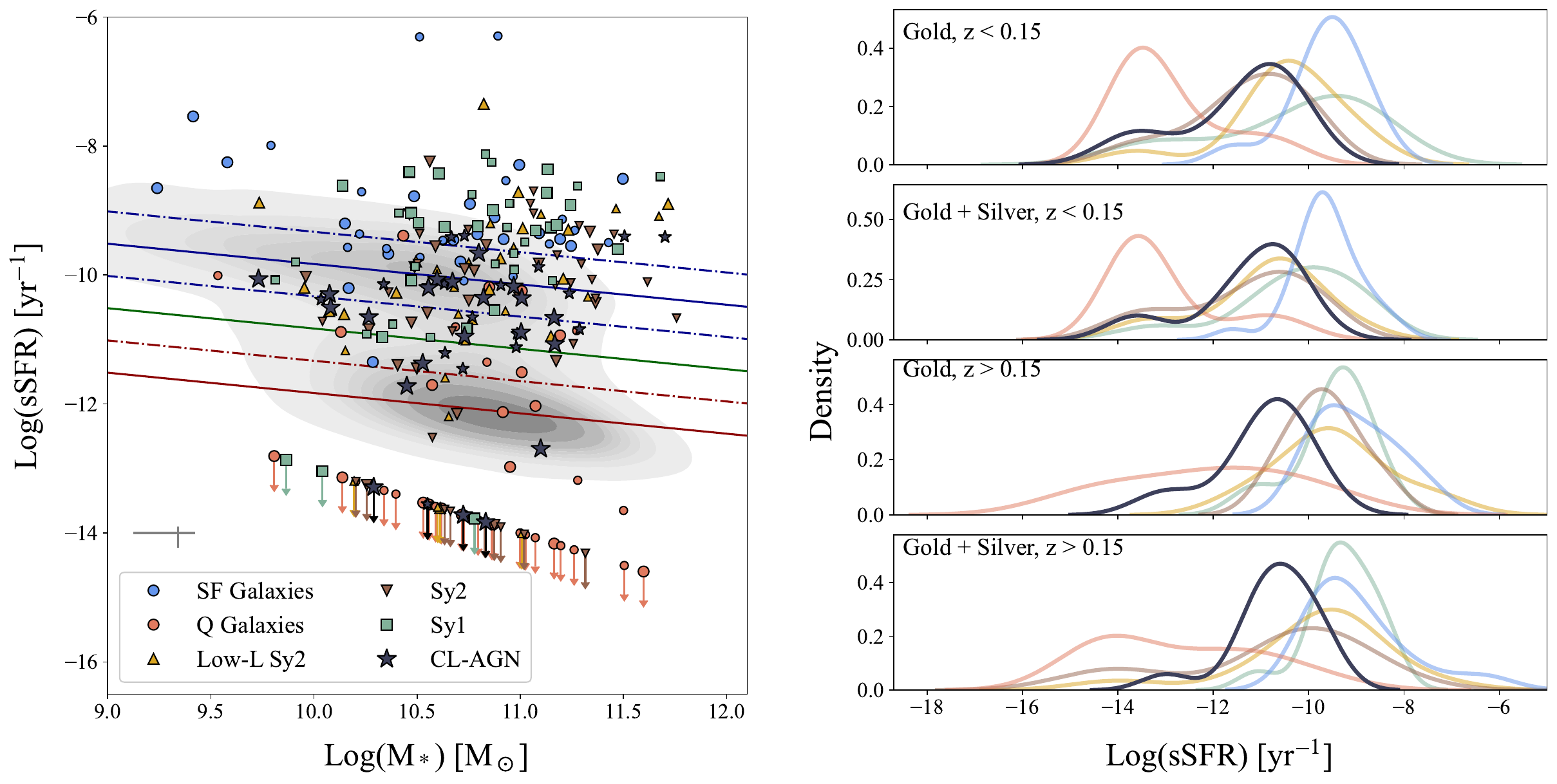}
    \caption{(Left) Mass-sSFR relationship for all samples discussed in this paper. The relation shown is the redshift-dependent mass-sSFR relation from \cite{Whitaker2012b} at z = 0.126, the median redshift of our sample.  Large marker size indicates the Gold sample, while small markers represent the Silver sample.  A characteristic errorbar is shown in gray.  Gray contours represent the total MPA-JHU galaxy sample.  The galaxies studied in this paper are higher in mass on average than the MPA-JHU sample.  CL-AGN hosts tend to reside in the upper end of the Green Valley or the lower end of the star-forming main sequence, though a few have quenched star formation.  (Right) Density of each population in sSFR.  CL-AGN hosts are similarly distributed in sSFR to Seyfert 2 hosts at all redshifts. CL-AGN hosts are similar to Seyfert 1 hosts at $z<0.15$ and are less star forming for their mass than Seyfert 1 hosts at $z>0.15$.  }
    \label{fig:mass-sfr}
\end{figure*}

We next examine the distribution of CL-AGN host galaxies in mass-sSFR space. Distributions in mass-sSFR for all samples are shown in Figure \ref{fig:mass-sfr}. For visualization of the sample relative to the star forming main sequence, we use star forming main sequence at the median redshift of the gold sample, $0.126$.  CL-AGN hosts are most similarly distributed in sSFR to Seyfert 2 host galaxies.  To investigate whether any differences in sSFR distributions are significant between samples, we perform a series of Kolmogorov-Smirnov tests.  We account for errors on our sSFR and mass measurements by bootstrapping over the posterior distributions of sSFR output from \texttt{Prospector}. We perform 1000 Kolmogorov-Smirnov tests for each bootstrapped sample and take the median p-value of these tests. If the p-value is below 0.05, we reject the null hypothesis that the two samples are drawn from the same distribution.

Under the Kolmogorov-Smirnov tests we performed, we find CL-AGN hosts are indistinguishable from other AGN host galaxies at $z<0.15$, regardless of whether we include the silver sample in our distributions.  They are significantly less star forming than mass- and redshift-matched star forming galaxies at this redshift and significantly more star-forming than quiescent galaxies at this redshift. Due to the significant number of non-detections for star formation in our quiescent galaxy sample, we perform the above Kolmogorov-Smirnov test twice: once where we sample over the full posterior, and once where we assume that the posterior for all non-detections is a gaussian centered at Log(sSFR) of -3 with a width of 0.5 dex. In both cases, our results are qualitatively the same, though the p value is systematically lower in the latter case.

At $z>0.15$, we find that CL-AGN hosts are indistinguishable from Seyfert 2 host galaxies, regardless of whether we include the silver sample. We do find that CL-AGN hosts are significantly less star forming than Seyfert 1 hosts at this redshift when we include the silver sample, though they are not significantly different when we only include gold-sample CL-AGN hosts and their Seyfert 1 counterparts. CL-AGN are significantly less star forming than the star forming galaxy sample at this redshift, regardless of whether we include the silver sample. The gold sample of CL-AGN at $z>0.15$ is statistically indistinct from quiescent galaxies, but the combined gold and silver sample is significantly more star forming than the quiescent galaxy sample at this redshift. 

\subsubsection{Turn-on versus turn-off CL-AGN}

\begin{figure}
    \centering
    \includegraphics[width=\columnwidth]{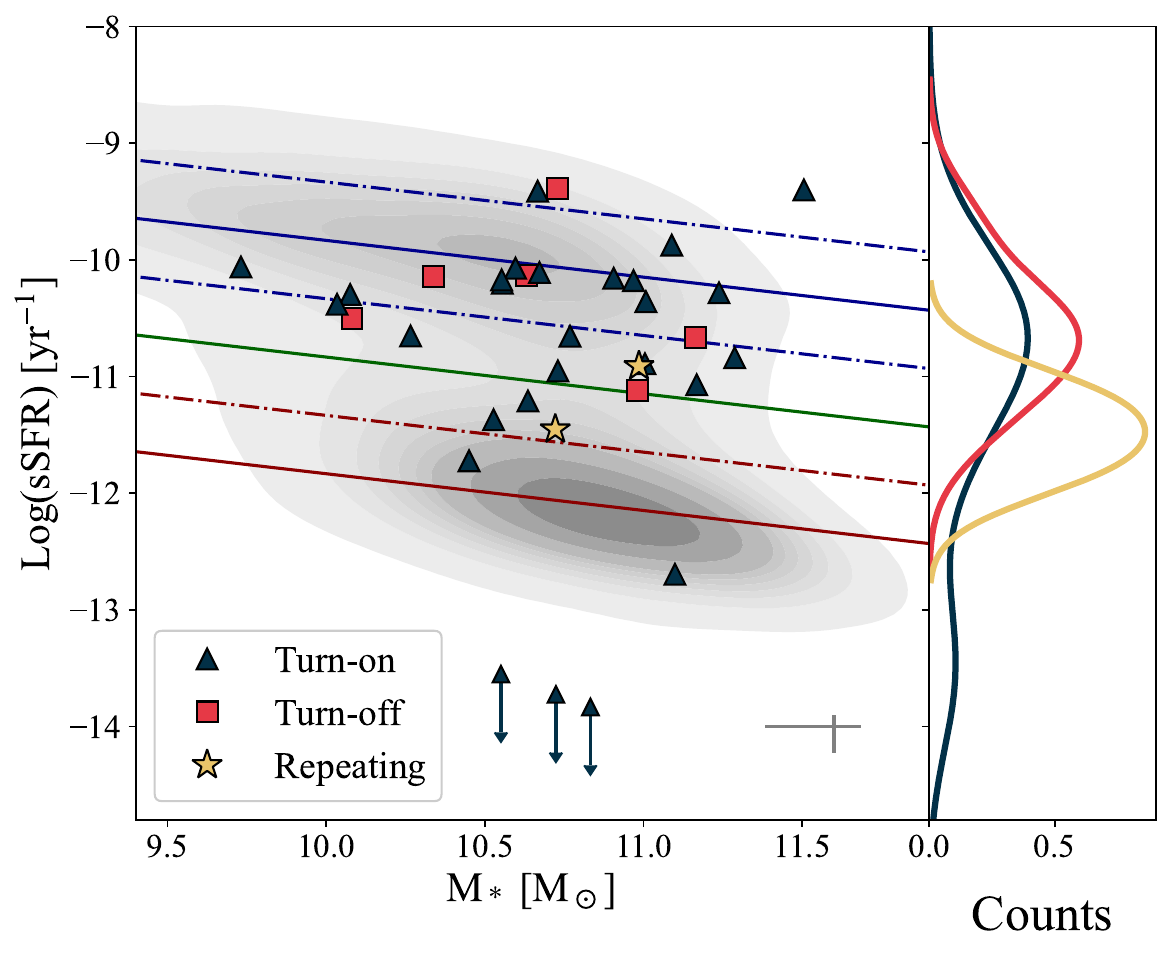}
    \caption{Mass-sSFR for the turn-on, turn-off, and repeating CL-AGN.  Gray contours represent the MPA-JHU sample; the gray error bars in the bottom right are median errors across the sample.  Most turn-on and turn-off CL-AGN hosts are on the star-forming main sequence, with statistically similar distributions in mass-sSFR space.  The two repeating CL-AGN hosts (yellow stars) have lower sSFRs than the turn-on and turn-off samples, though there are too few of these objects to make a statistical case about them.}
    \label{fig:mass-sfr-onoff}
\end{figure}

It is typically difficult to determine any statistical differences in turn-on and turn-off CL-AGN hosts, as most attempts to identify new CL-AGN are biased either towards turn-on CL-AGN (e.g. searches in transient surveys) or turn-off CL-AGN (e.g. monitoring known AGN for changes in luminosity; \citealt{ricci2022}).  We have attempted to circumvent this bias by combining samples which have a variety of CL-AGN selection methods, but our gold sample still has this bias; it contains only three turn-off CL-AGN as opposed to 18 turn-on CL-AGN and no repeating CL-AGN.  We therefore must turn to our combined gold and silver samples to provide a more balanced comparison, though we still have only seven turn-off CL-AGN compared to 30 turn-on CL-AGN.

Qualitatively, turn-on CL-AGN span the entire mass-sSFR parameter space, while turn-off CL-AGN fall entirely on the star forming main sequence.  However, this discrepancy is not statisically significant.  By the bootstrapped Kolmogorov-Smirnov test described above, turn-on and turn-off CL-AGN hosts have statistically indistinct distributions in mass-redshift space (see Figure \ref{fig:mass-sfr-onoff}).  They are also not statistically distinct under a chi-square test of homogeneity. 

We also have two repeating CL-AGN in our silver sample.  These may be triggered by a different physical process than other CL-AGN (e.g. an extreme mass ratio inspiral or partial TDE in an AGN disk), or it is possible that any turn-on and turn-off CL-AGN can repeat or revert at some point in the future. With such a small number of repeating CL-AGN, it is impossible to make statistical comments.  We do note that one of the repeating CL-AGN hosts falls in the Green Valley, while the other is quiescent.  Neither star formation history is reminiscent of a post-starburst galaxy as might be more likely for a TDE origin; however, with such a small number of repeating CL-AGN, further study would be needed to make a claim about their origin.

\subsection{Star formation histories of CL-AGN hosts} \label{sec:sfh}

\begin{figure*}
    \centering
    \includegraphics[width=\textwidth]{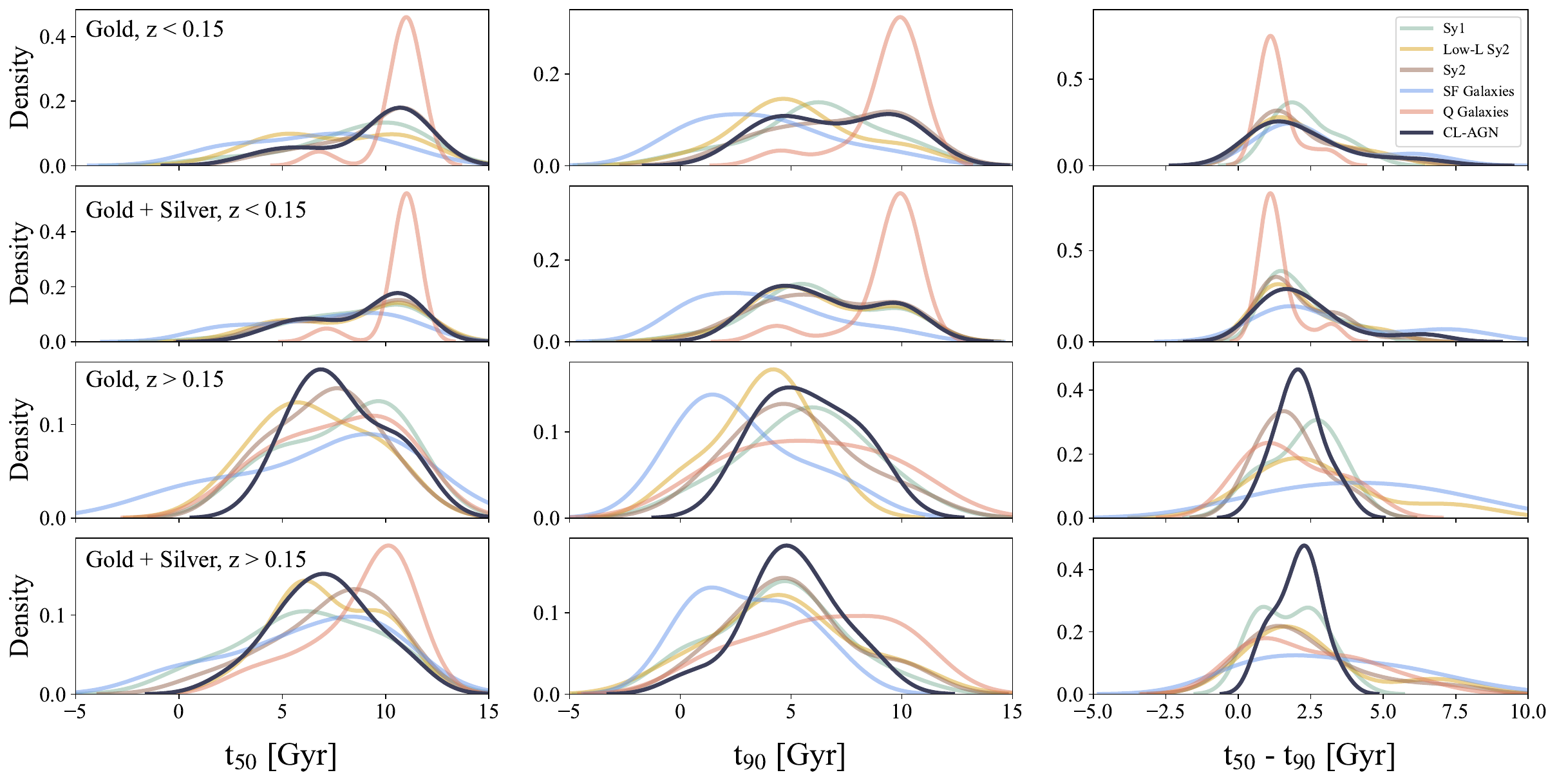}
    \caption{Kernel smoothed density distributions of time since 50 (left) and 90\% (center) of the mass formed in each galaxy and star formation duration for each combined sample (right).  Galaxies with higher $t_{50}$ and $t_{90}$ formed their mass earlier. Values under 0 are an artifact of the kernel smoothing. CL-AGN formed their mass at a similar time to other AGN types, and their star formation histories are only statistically distinct from those of star-forming galaxies at $z<0.15$.  There are no statistically significant differences in these parameters between CL-AGN hosts and other Seyfert hosts for the lower- or higher-redshift samples or for the gold or combined samples.}
    \label{fig:t50t90}
\end{figure*}

If we are to understand how the host galaxies of CL-AGN differ from our comparison samples, it is not enough to understand their current star-formation properties.  Several formation mechanisms exist to fuel AGN, including major mergers, disk instabilities, and mass loss from old stars \citep{wang2023, Cristello2024}.  These mechanisms are associated with different star formation histories for the AGN host galaxies, so we can use the star formation histories of CL-AGN host galaxies to understand which mechanisms may be at play in fueling the changing look phenomenon.

\begin{figure*}
    \centering
    \includegraphics[width=\textwidth]{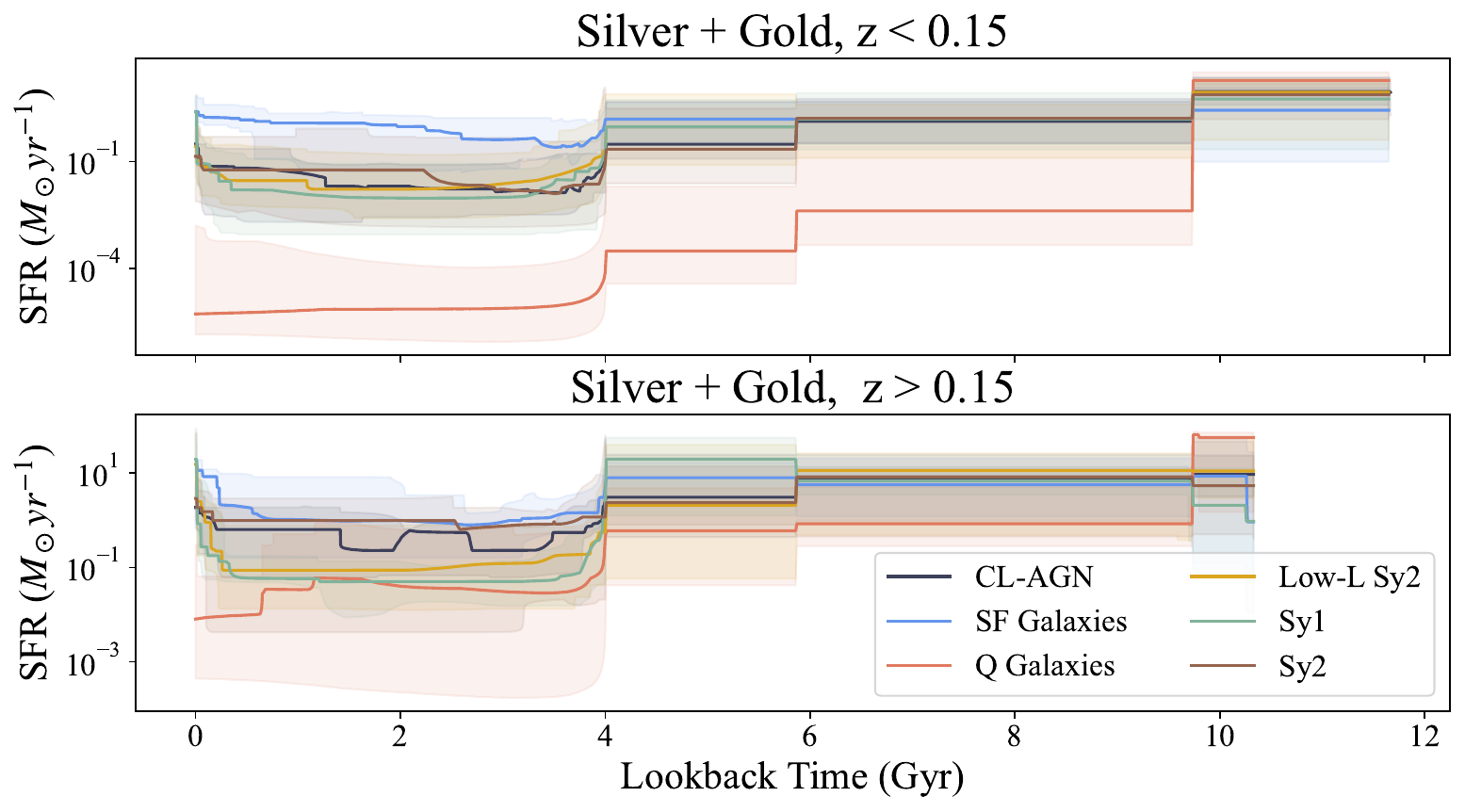}
    \caption{Combined star formation histories for each sample in this paper at z $< 0.15$ (top) and $z > 0.15$ (bottom).  The CL-AGN hosts have qualitatively similar star formation histories to other Seyfert hosts.  There is no evidence of significant starbursts in the combined CL-AGN star formation histories.  CL-AGN star formation histories, like those of other Seyfert hosts, have an uptick at late times, perhaps indicating rejuvenation or slow quenching as discussed in Section \ref{sec:sfh}.}
    \label{fig:combined_sfh}
\end{figure*}


First, we examine $t_{50}$ and $t_{90}$, two quantitative measures of star formation history.  $t_{50}$ and $t_{90}$ are the time since 50\% and 90\% of the mass in the galaxy formed, respectively.  This spans a significant period of star formation in the galaxy, and we therefore designate the value $t_{50}-t_{90}$ as ``star formation duration."  Galaxies with a shorter star formation duration are more similar to post-starburst galaxies at high redshift, while galaxies with a longer duration have experienced either stable star formation or are slowly quenching.

Distributions in $t_{50}$, $t_{90}$, and star formation duration for each sample are shown in Figure \ref{fig:t50t90}.  We find no significant difference in $t_{50}$, $t_{90}$, or star formation duration between the CL-AGN host galaxies and any of our AGN comparison samples using bootstrapped Kolmogorov-Smirnov tests.  This is true when we divide the sample by redshift and when we examine only the gold versus the combined samples.  We find that CL-AGN formed their stars statistically earlier than star forming galaxies at $z<0.15$ but not in the higher redshift bin.  This suggests that CL-AGN, Seyfert hosts, and quiescent galaxies experienced their primary epochs of star formation earlier than the local star forming galaxy population.

\begin{figure}
    \centering
    \includegraphics[width=\linewidth]{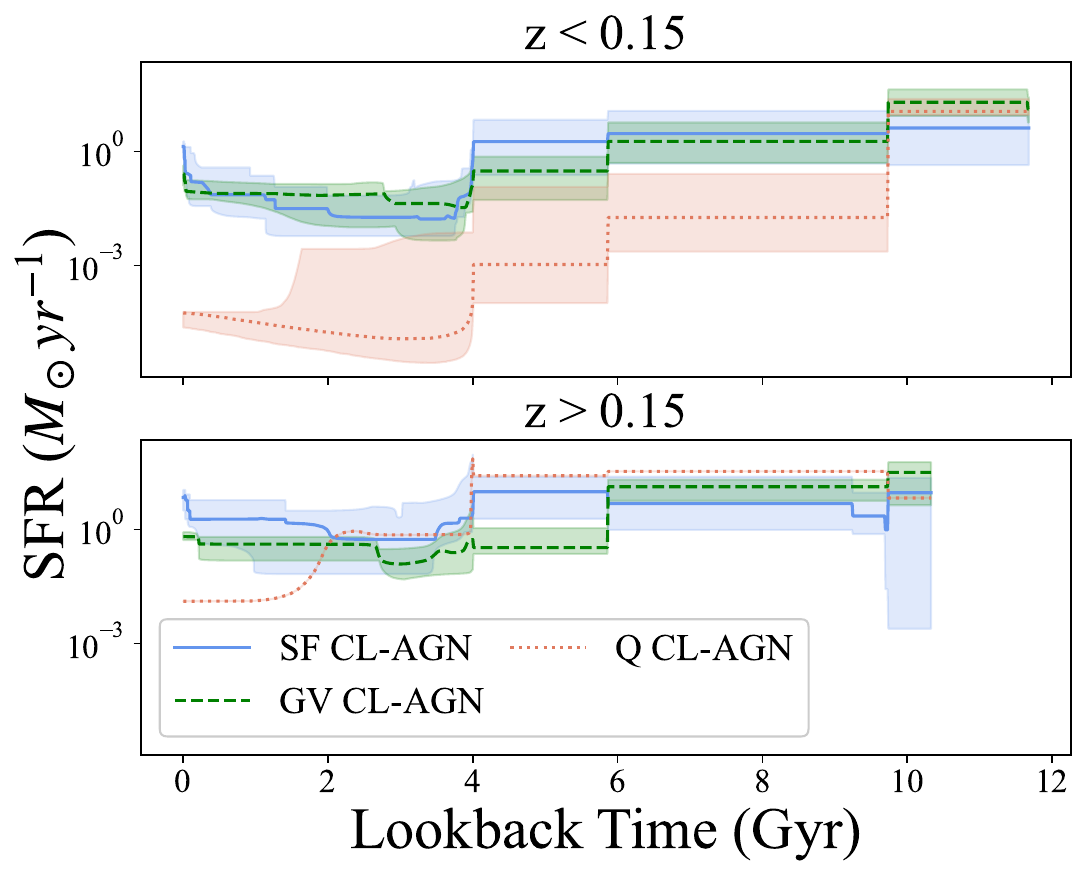} 
    \caption{Median (lines) and 25th–75th percentile (shaded region) star formation histories for star-forming (blue/solid), Green Valley (green/dashed), and quiescent (red/dotted) CL-AGN host galaxies. The top panel shows CL-AGN host star formation histories at z $<$ 0.15; the bottom panel shows CL-AGN host star formation histories at z $>$ 0.15. Even when we look at the star formation histories of only Green Valley CL-AGN hosts, we do not see evidence of rapid quenching or significant starbursts.}
    \label{fig:sfh_by_class}
\end{figure}

In Figure \ref{fig:sfh_by_class}, we show median star formation histories for our CL-AGN host galaxies, now separated by star formation rate classification (star-forming, Green Valley, or quiescent as described in Section \ref{sec:current_sfr}). Here, we see that CL-AGN hosts in the Green Valley do not show signs of rapid quenching as would be expected for the host galaxies of tidal disruption events. At z $<$ 0.15, CL-AGN hosts on the star forming main sequence and in the Green Valley have very similar star formation histories, and their final star formation rates are similar. This is likely because CL-AGN have lower SFRs in the local Universe than star-forming galaxies, indicating that they are located on the lower end of the star forming main sequence on average, consistent with the location of other Seyfert types. CL-AGN hosts which have quenched by z $<$ 0.15 tend to have been quiescent for at least 4 Gyr, with many experiencing their primary epoch of star formation more than 8 Gyr ago.

CL-AGN hosts at z $>$ 0.15 show slightly more distinct star formation history shapes by SFR class. Only one CL-AGN host in this redshift bin is fully quiescent. Here, Green Valley CL-AGN hosts have had a lower median SFR than star-forming CL-AGN hosts over the entire 4 Gyr flex-bin period of the \texttt{continuity\_psb\_sfh} template, though they still have some residual star formation. We discuss the impact of small amounts of residual star formation further in Section \ref{sec:discussion}.


Our results, like those from \citet{dodd2021}, indicate that CL-AGN host galaxies generally are not post-starburst.  Figure \ref{fig:combined_sfh} shows the non-parametric star formation histories for each population.  We find that CL-AGN hosts have a qualitatively flat star formation histories, with most CL-AGN not experiencing a starburst or rapid quenching within the past 4 Gyr. CL-AGN hosts formed most of their mass over the course of about 2 Gyr, and they completed their primary epoch of star formation on average 6-8 Gyr ago. This is similar to the star-forming and quiescent galaxy samples as well as the Seyfert 2 comparison samples. All AGN host galaxy samples, including CL-AGN hosts, have ongoing star formation with rising star formation histories over the past 0.25-0.5 Gyr. This is more dramatic in the higher-redshift bin, which also has generally higher star formation rates at the time of observation.  We observe the same behavior in the star-forming sample, indicating this is may be due to the relative brightness of stars formed in this epoch rather than a reflection of the true star formation history. 

 We discuss the star formation properties of CL-AGN hosts in Section \ref{sec:discussion} and possible interpretations regarding the physical drivers of CL-AGN in Section \ref{sec:clagn_theory}.

\section{The impact of broad lines on host galaxy analysis} \label{sec:sps_agn}

Gleaning AGN host galaxy properties from their spectra is a challenge.  AGN are bright and blue, mimicking the contribution from recent star formation.  Several stellar population synthesis codes include a prescription for AGN fitting, including CIGALE \citep[][UV through IR quasar models and AGN torus emission templates]{Boquien2019}, Prospector \citep[][AGN dust torus emission]{leja2018}, and GRAHSP \citep[][power-law, line, and dust emission]{Buchner2024}.  Properly accounting for AGN contamination in stellar population synthesis is crucial for extracting the physical parameters of both the AGN and its host galaxy.

CL-AGN provide a unique opportunity to understand the impact of broad-line AGN emission on SED modeling.  As previously stated, CL-AGN can vary on timescales of months to years, much shorter than the timescale on which the star-formation properties of a galaxy are expected to change.  We can therefore run the same pipeline on both the broad line and narrow line spectrum for each object and examine the differences in the output results.  Any difference here is expected to come from the difference in AGN emission.  

We select the only eight objects for which we have a broad-line spectrum in SDSS in addition to our original narrow-line spectrum.  We run each object through our pipeline for the \texttt{continuity\_psb\_sfh} template to extract features like the host galaxy mass and star formation rate.  For this test, the only difference in the input parameters is the width of the mask for the H$\alpha$, H$\beta$, and [OIII] $5007$ \si{\angstrom} emission lines (now 80, 50, and 70 \si{\angstrom} respectively); all other input parameters and priors are the same.

\begin{figure}[h]
    \centering
    \includegraphics[width=\columnwidth]{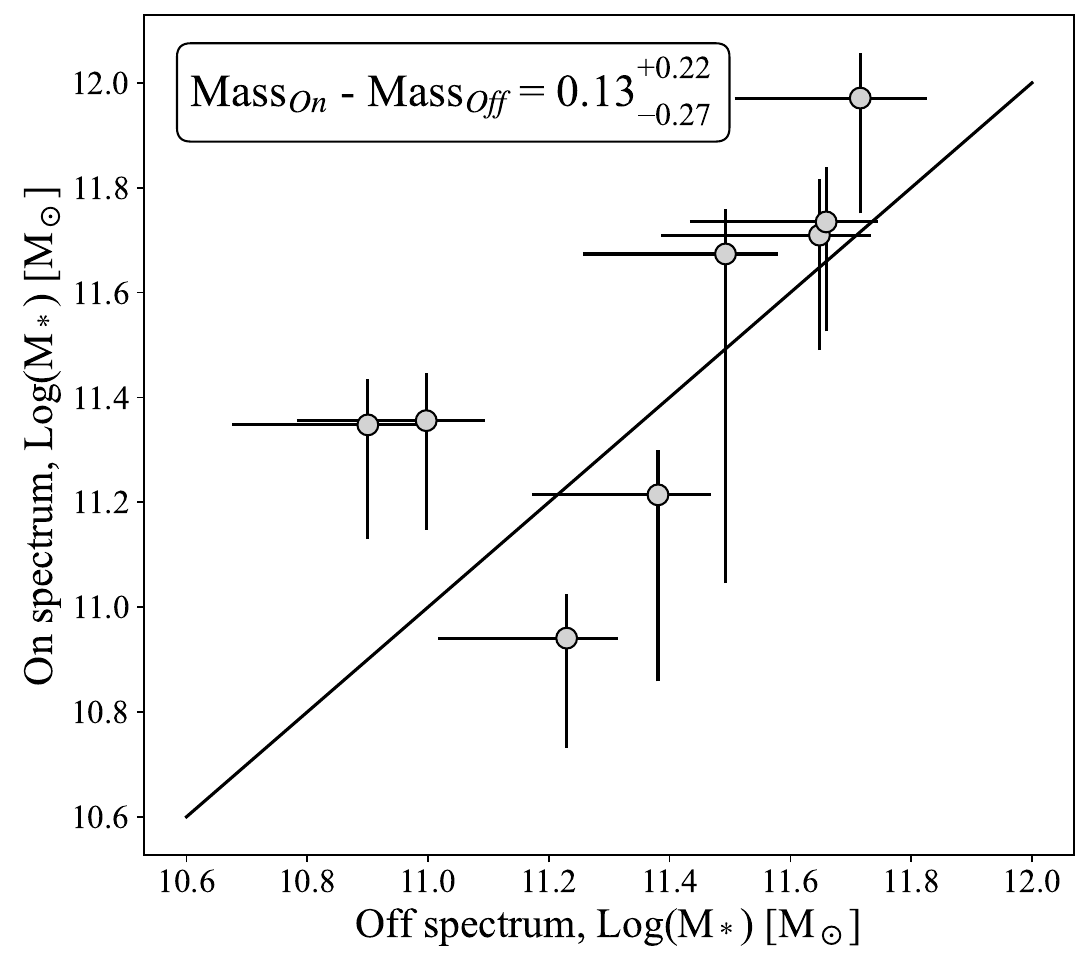}
    \includegraphics[width=\columnwidth]{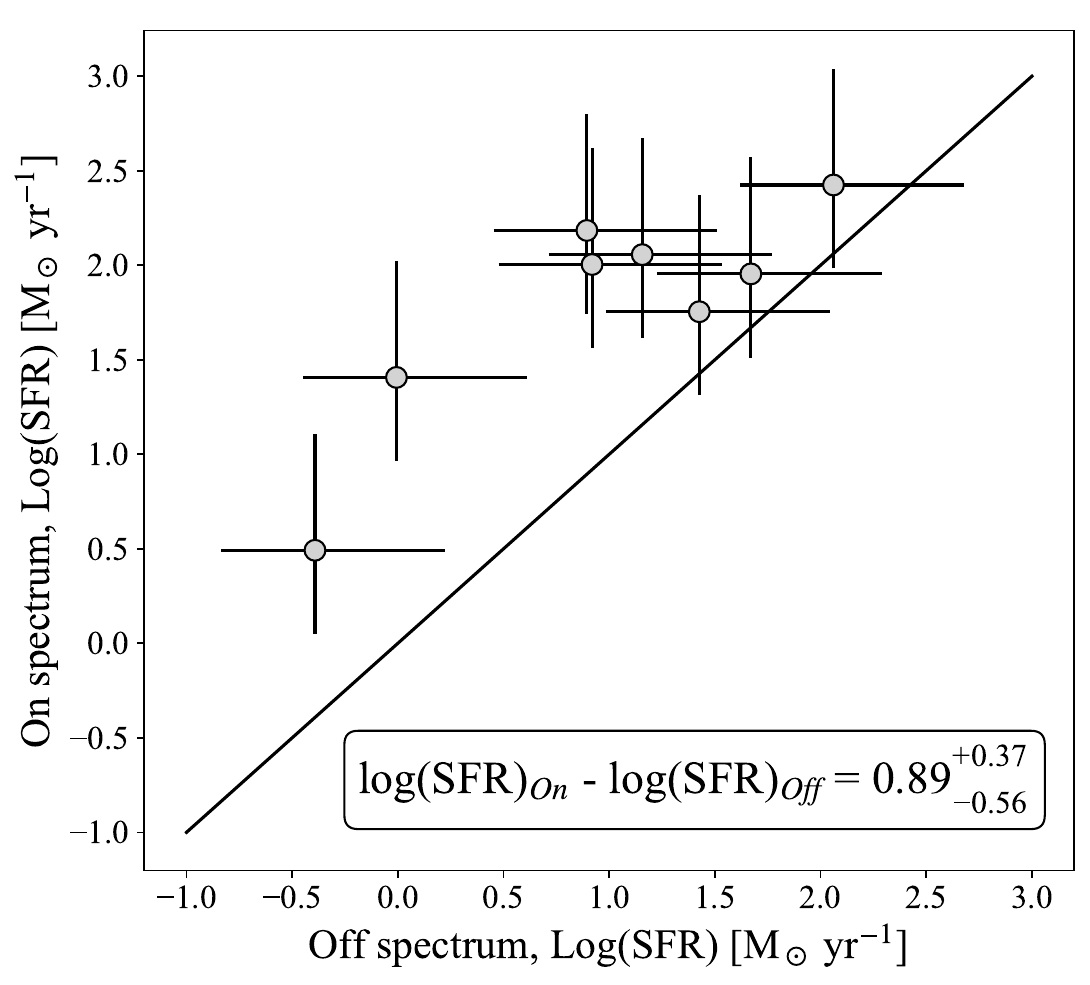}
    \caption{Stellar masses (top) and SFRs (bottom) recovered when fitting the broad-line versus narrow-line spectrum for each object.  We see some scatter but no significant offset in the recovered stellar mass for the same object ($0.13^{+0.22}_{-0.27}$ dex).  However, increased emission from the broad-line state of the CL-AGN significantly impacts recovered SFR, with an offset of $0.89^{+0.37}_{-0.56}$ dex.}
    \label{fig:onoffcompare}
\end{figure}

In Figure \ref{fig:onoffcompare}, we show a comparison between the output parameters from the broad line and narrow line spectra.  We find a high level of agreement in our stellar mass measurement between the broad line and narrow line spectrum. However, we find a $0.89^{+0.37}_{-0.56}$ dex increase in recovered SFR for the broad line spectrum, likely due to the blue excess introduced by the increased contribution from the AGN.  These objects are classified as ``QSO" in SDSS in their broad-line state \citep[M$_{i, z=2}< -20.5$][]{paris2012}, so the increase in luminosity from the AGN is significant---they are truly AGN dominated in the bright state.  

As seen with our mock spectra tests, as long as AGN emission lines are sufficiently masked, it is possible to recover AGN host galaxy information using non-parametric stellar population fits with a reasonable degree of accuracy for AGN hosts which are less luminous. For more luminous AGN, this test can be used to calibrate recovered SFRs from non-parametric fitting.



\section{Discussion} \label{sec:discussion}

We find that $60^{+9}_{-10}\%$ of z $<$ 0.15 and $50^{+13}_{-13}\%$ of of z $>$ 0.15 CL-AGN hosts fall in the color-magnitude Green Valley, while $32^{+9.9}_{-8.5}\%$ of z $<$ 0.15 and $33^{+14}_{-2}\%$ of z $>$ 0.15 CL-AGN hosts are in the star-formation Green Valley. These two Green Valleys trace different temporal and spatial scales of star formation history in a galaxy. In this section, we attempt to tie together the evidence gathered via CL-AGN colors, sSFRs, and star formation histories to understand the overall evolution of CL-AGN hosts in the context of other Seyfert host galaxies. 

As shown in Figure \ref{fig:colormag}, CL-AGN hosts are significantly redder than star-forming galaxies and Seyfert 1 hosts, and they are significantly bluer than quiescent galaxies.  They are similarly distributed in $g-r$ color to Seyfert 2 host galaxies.  The dust extinction in the CL-AGN hosts in our sample is low (median $\tau_{5500 \si{\angstrom}} \sim 0.9$) relative to the star-forming galaxies in our sample (median $\tau_{5500 \si{\angstrom}} \sim 1.6$), implying that dust reddening is not sufficient to cause this shift in color relative to the blue cloud.  Instead, the stellar populations of CL-AGN and Seyfert 2 host galaxies should be distinctly older than those of star-forming galaxies and Seyfert 1 hosts.

$z<0.15$ CL-AGN hosts are statistically indistinct from other AGN host galaxies in sSFR.  They are significantly less star forming than star-forming galaxies, though many of them fall on the star forming main sequence (Figure \ref{fig:samplefractions}). CL-AGN and Seyfert 2 AGN hosts are significantly less star forming than comparison Seyfert 1 and star-forming galaxies (Figure \ref{fig:mass-sfr}) at $z>0.15$, though again most CL-AGN and Seyfert 2 hosts still lie on the star forming main sequence.  This suggests a scenario in which CL-AGN and Seyfert 2 host galaxies still host a significant amount of star formation, though perhaps they are not among the most star-forming galaxies of these epochs.  The source of their red stellar populations is therefore not a lack of present-day star formation.

A clue may lie in the star formation histories of CL-AGN host galaxies.  Significant ongoing star formation along with a redder stellar population could indicate that CL-AGN and Seyfert 2 hosts are experiencing a period of rejuvenation atop an older stellar population.  In this scenario, their primary epoch of star formation would have already ended, but a recent inflow of gas to both the galaxy and the nuclear region could be responsible for triggering star formation and the onset of AGN activity. As shown in Figure \ref{fig:combined_sfh}, all AGN hosts in this sample show an increase in star formation within the past $\sim$ Gyr after a significant period of quiescence.  However, we caution that both these results could be impacted by AGN contamination, which would mimic recent star formation.  As discussed in Section \ref{sec:AGNimpact}, we do see a small increase in recovered late-time star formation in mock spectra with added AGN contamination, though the true late-time star formation history is captured by the \texttt{Prospector} uncertainties and therefore would not affect the results of the sSFR comparisons performed in Section \ref{sec:sfr}.

Another possibility is that CL-AGN hosts, along with Seyfert 2 host galaxies, are quenching slowly.  As discussed in \cite{schawinski2014}, galaxies with long quenching timescales ($\tau_{q} \gtrsim 1$ Gyr) would have enough residual star formation to contribute to the blue light of the galaxy for billions of years after the onset of galaxy quenching. This ongoing formation of O and B stars could outshine some of the intermediate-age stellar populations, creating an artificial rise in star formation in our \texttt{Prospector} star formation histories. If Seyfert 2 and CL-AGN hosts are quenching slowly, the slow removal of gas from the host galaxy would act to quench star formation while also ``starving" the central supermassive black hole, leading to a decrease in accretion rate and therefore a decrease in Eddington ratio associated with CL-AGN \citep{noda2018,guo2024b,jana2024}.  This lower accretion rate is also thought to drive ``true type 2" Seyferts, where the AGN is below the accretion rate required to maintain a broad-line region \citep{elitzur2014}.  Star-forming galaxies, possibly including Seyfert 1 hosts, would still retain significant molecular gas reservoirs and would still be in their primary epoch of star formation.  This is in agreement with our results for $t_{50}$, $t_{90}$, and star-formation duration (Figure \ref{fig:t50t90}), where most AGN host galaxies formed 90\% of their mass several billion years ago.  It could explain why CL-AGN hosts are more similar to Seyfert 2 hosts, rather than an intermediate population between Seyfert types 1 and 2. 

Finally, this scenario would be in agreement with the morphological studies of \cite{liu2021} and \cite{yu2020}, both of which find disky CL-AGN hosts with prominent bulges or pseudobulges rather than signatures of major mergers in CL-AGN hosts, as would be expected for rapid quenching. Indeed, we find that around half of the CL-AGN and AGN hosts in our sample are spiral or disk galaxies as classified by Galaxy Zoo, while around a third are elliptical and a final tenth have signatures of a recent merger (Agrawal et al. in prep), indicating that most CL-AGN either are transitioning slowly to quiescence or experienced any major mergers in the distant past. The conflicting results for CL-AGN star formation properties evaluated via different methods point towards a scenario in which CL-AGN hosts have already passed their star-formation prime, regardless of whether their current star formation is rejuvenated or residual.

\subsection{Comparison to the literature}

We find that most CL-AGN host galaxies at z $< 0.4$ lie on the star-forming main sequence, unlike the results of \cite{dodd2021,liu2021} and \cite{wang2023}, who found that CL-AGN hosts lie primarily in the Green Valley.  Our results agree with those of \cite{yu2020}, who found that CL-AGN in MaNGA were along the star-forming main sequence.

Each of these papers uses a different tracer for star formation. \citet{liu2021} and \citet{wang2023} use H$\delta_A$ and D$_n$4000 as tracers of stellar population. H$\delta_A$ absorption traces the A star population, which traces star formation on $\sim$Gyr timescales \citep{Worthey1997}, while D$_n$4000, or the strength of the 4000 \si{\angstrom} break, traces active star formation \citep[][]{Dressler1987,Kauffmann2003a}. We are unable to measure these quantities uniformly for our SDSS spectra; we see significant uncertainties on the equivalent width of H$\delta_A$, likely due to infilling from narrow-line AGN emission. The results of \citet{liu2021} and \citet{wang2023} show that CL-AGN occur preferentially at an intermediate D$_n$4000, likely indicating that CL-AGN hosts have ended their primary period of star formation. Their D$_n$4000 values are consistent with those of other AGN, though they are perhaps concentrated on the higher end of AGN D$_n$4000, which may indicate less star formation than other AGN hosts \citep{wang2023}. As we are unable to use H$\delta_A$ and D$_n$4000, we cannot compare directly to \citet{wang2023}. \citet{liu2021} uses more star formation tracers, and thus we are able to make a more direct comparison.

\citet{liu2021} finds that CL-AGN hosts are redder in \textit{g-i} color than the average Seyfert 2 host galaxy. We do not find the same trend in  \textit{g-r} color, though we do find (in agreement with \citet{liu2021}) that CL-AGN are primarily within the color-magnitude Green Valley. As discussed above, this may be due to slow quenching or a recent rejuvenation in SFR, either of which may be typical for AGN more broadly. 

The choice of star-forming main sequence can make a large difference in the results reported here. We use the redshift-dependent star forming main sequence from \citet{Whitaker2012b}, as it can be adjusted to span the full redshift range of our sample. \citet{Whitaker2012b} uses galaxies in NEWFIRM to create a star-forming main sequence. When selecting for only blue galaxies, the star-forming main sequence of \citet{Whitaker2012b} at low redshift is similar to that of \citet{peng2010}; however, \citet{Whitaker2012b} includes dusty star-forming galaxies at low redshift which impact the slope of the relation. We determine whether any individual object is star-forming, in the Green Valley, or quiescent by using its mass and SFR as well as its redshift. 

Other works use a variety of star-forming main sequences for their work. \citet{dodd2021} uses the star-forming main sequence from \citet{peng2010} and a width of 0.5 dex, while \citet{yu2020} uses the star-forming main sequence from \citet{Chang2015}. Both of these star-forming main sequences were constructed using SDSS. \citet{Chang2015} uses galaxies with SDSS and WISE galaxies with a median redshift of 0.1 for its SED modeling to obtain masses and SFRs.  \citet{peng2010} uses the SDSS DR7 sample (0.02 $<$ z $<$ 0.085) to derive masses using \textit{kcorrect} \citep{kcorrect} and SFRs from the \citet{Brinchmann2004} catalog. These different star-forming main sequences may explain some of the discrepancies in our absolute fractions of star-forming, Green Valley, and quiescent CL-AGN hosts. However, our relative results (namely, that CL-AGN hosts have similar star-formation properties to the hosts of Seyfert 2 AGN) should not be impacted by the change of star-forming main sequence.

The remaining differences in results may come from our attempt to construct a large sample from the literature, which may reduce selection bias in our CL-AGN. \citet{dodd2021} includes only two turn-off CL-AGN out of fifteen, while \citet{liu2021} includes no turn-off CL-AGN. These two papers have sample sizes of fifteen and twenty-six respectively.  \citet{wang2023} and \citet{yu2020} have nearly-equal numbers of turn-on and turn-off CL-AGN (5/9 turn-on and 3/5 turn-on, respectively). As shown in Figure \ref{fig:mass-sfr-onoff}, there is a slight (though not statistically significant) tendency for turn-on CL-AGN to be on the star-forming main sequence than turn-off CL-AGN; this may explain why \citet{yu2020} finds that CL-AGN hosts are primarily on the star-forming main sequence. It fails to explain the discrepancy between our results and those of \citet{dodd2021}, \citet{liu2021}, and \citet{wang2023}. 

In particular, it is interesting that we do not see a statistically significant difference between the host galaxies of CL-AGN and of low-luminosity Seyfert 2 from the MPA-JHU catalog, considering this is the exact comparison sample for \citet{dodd2021}. This may come down to the exact choice of statistical test used and to the fact that we compared to a mass- and redshift-matched comparison sample rather than the entire MPA-JHU catalog.

Finally, we can make limited comparisons to \citet{jin2021}, who found that CL-AGN hosts have a higher contribution from intermediate-age stellar populations (100 Myr--1.27 Gyr) than other AGN hosts. We consider only the most recent 10 Myr of star-formation when calculating our SFRs, which would be considered part of the young stellar population ($\leq55$ Myr old) by \citet{jin2021}. They find no statistically significant difference in the young stellar population between CL-AGN and Seyfert 2 hosts, consistent with our result that CL-AGN and Seyfert 2 host follow statistically indistinguishable sSFR distributions. They also find that Seyfert 1 hosts have a relatively significantly (P$_{KS} \leq 0.05$) younger stellar population than CL-AGN hosts, again consistent with our findings that CL-AGN have less ongoing star formation than Seyfert 1 hosts.

o\subsection{What drives CL-AGN?} \label{sec:clagn_theory}


\subsubsection{CL-AGN Triggered by Tidal Disruption Events}

One proposed mechanism for CL-AGN is tidal disruption events (TDEs) occuring in a pre-existing AGN disk.  Much like CL-AGN, TDEs are nuclear transients which occur on  timescales of several months and which are found predominantly in galaxies undergoing quenching.  Specifically, TDEs are overrepresented in E+A or ``post-starburst" galaxies, galaxies which have quenched rapidly within the past gigayear \citep{arcavi2014,french2016}.  Post-starburst TDE hosts in the local Universe have post-burst ages $\lesssim1$ Gyr, with burst mass fractions $>3\%$ \citep{french2016}.  Supermassive black hole binaries and AGN disks have both been invoked to explain this post-starburst preference \citep[e.g. ][]{chen2009,Kennedy2016}, both of which could trigger CL-AGN-like behavior.  AGN occur at a higher rate in younger or more star forming galaxies \citep{birchall2023,ni2023}, with the AGN fraction in post-starbursts peaking during the starburst or immediately after quenching \citep{greene2020,ellison2025}. Thus, AGN incidence should peak \textit{during} a starburst, while TDE incidence peaks \textit{after} a starburst. 

\begin{figure*}
    \centering
    \includegraphics[width=1\linewidth]{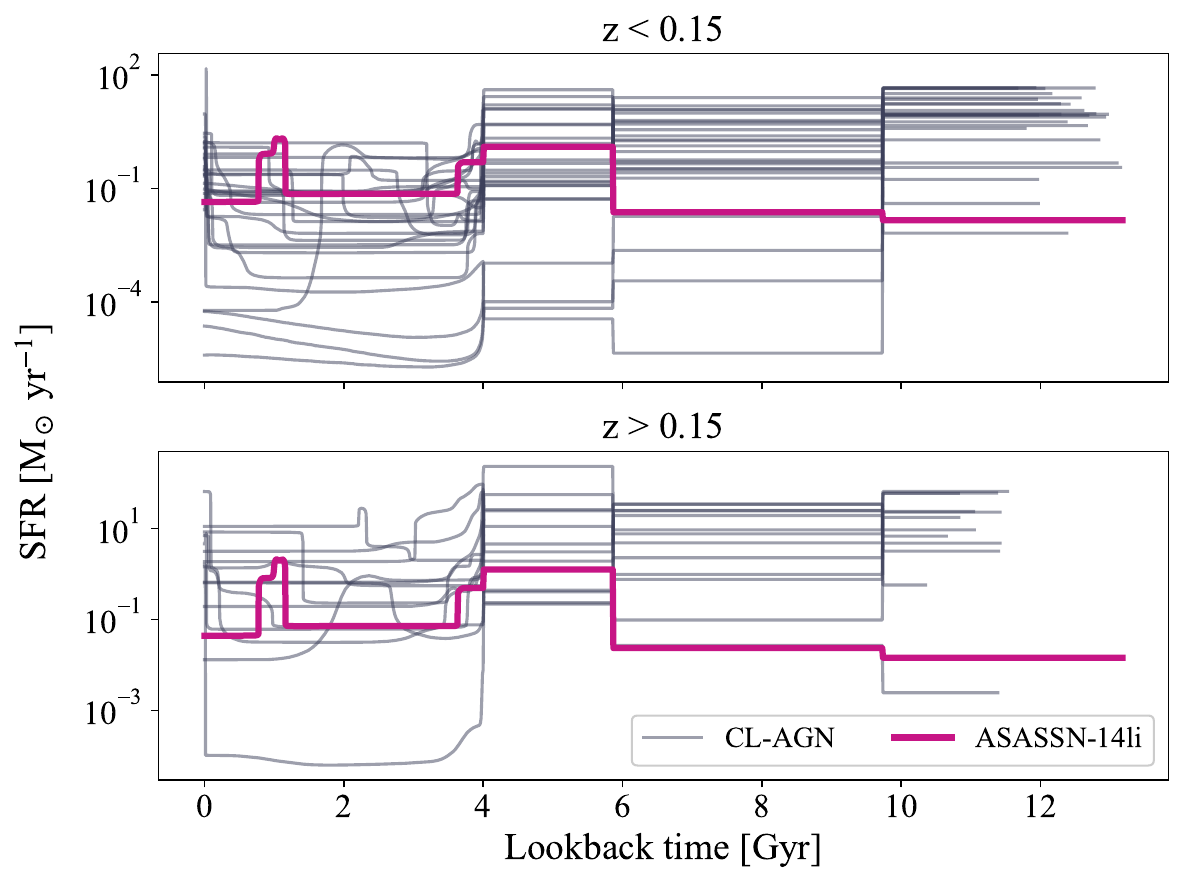}
    \caption{A comparison of CL-AGN host galaxy star formation histories at z $<$ 0.15 (top) and z $>$ 0.15 (bottom) with the star formation history of ASASSN-14li, a well-studied TDE in a post-starburst galaxy \citep[][]{Jose2014,french2016,holoien2016,Prieto2016}. While \texttt{Prospector} recovers a clear starburst in ASASSN-14li's host galaxy, most CL-AGN hosts do not show signs of a recent starburst, possibly indicating that most CL-AGN are not caused by TDEs.}
    \label{fig:clagntdesfh}
\end{figure*}

We compare our CL-AGN sample star formation histories to the star formation history of ASASSN-14li, a well-studied TDE in a post-starburst, post-merger galaxy \citep[][]{Jose2014,french2016,holoien2016,Prieto2016}. In Figure \ref{fig:clagntdesfh}, we demonstrate that the \texttt{continuity\_sfh\_psb} model recovers the signature of a starburst in the host galaxy of ASASSN-14li, a feature which is missing for most CL-AGN hosts. We find one case with a peak SFR $\sim3$ M$_\odot$ yr$^{-1}$ that rapidly declines at the same time as ASASSN-14li, but even if this galaxy is post-starburst, the post-starburst CL-AGN rate of 1/39 is much lower than the post-starburst rate in TDEs \citep[$\sim12-38\%$; see][and references therein]{french2020}. There is some indication that CL-AGN hosts have experienced a recent rise in star formation, perhaps indicating that they are experiencing a relative enhancement in star formation due to a recent merger (\citet{charlton2019} finds disturbed morphologies in three of four CL-AGN imaged with Gemini, perhaps pointing to an increased rate in post-mergers). However, considering that CL-AGN hosts in this study have a reduced star formation rate relative to Seyfert 1 hosts, and considering their relatively green colors, it does not seem likely that they are experiencing the elevated star formation rates typical of the \cite{hopkins2005a} merger evolutionary track expected for AGN. In either case, the lack of a clear preference for post-starburst hosts may point away from TDEs as an origin for CL-AGN.

On the other hand, it is possible that biases in optically-selected TDE samples are missing TDEs in pre-existing AGN, whose hosts may look more similar to CL-AGN. Nuclear dust can obscure optical transients, resulting in fewer observed TDEs in dusty galaxies \citep{Roth2021}. TDEs in these systems may appear as IR flares instead \citep{jiang2017,reynolds2022, masterson2024}. IR flares do not show an overenhancement in post-starburst host galaxies \citep{masterson2024}, though some IR flares may be driven by AGN instead \citep{dodd2023}. Several TDEs in dusty starburst galaxies have been seen as IR flares \citep{reynolds2022}, and while most nearby IR flares appear to be in non-starburst systems \citep{masterson2024}, the nuclear dust may be decoupled from what we observe in the larger host galaxy. 
%
Several studies \citep[e.g.][]{Kennedy2016,wangyihan2024a,kaur2025} have  proposed that AGN disks could be involved in inducing TDEs, producing a TDE host galaxy population more similar to the overall AGN host population.  \cite{ryu2024} suggest these TDEs would be more difficult to detect, as they would be ``muffled" by the mixing of tidal debris with the existing AGN disk, and that this interaction would lead to an accretion state change perhaps similar to CL-AGN.  Current TDE surveys also exclude AGN host galaxies due to the stochastic nature of AGN.  It is possible that, were AGN hosts included in TDE analyses, we would find that TDEs occur in a more similar host galaxy population to that of Seyferts, and therefore to CL-AGN host galaxies. On the other hand, this would again point toward a post-merger and/or post-starburst host galaxy preference for CL-AGN, which we do not see in our sample. 






\subsubsection{Feast versus famine fueling}

\cite{liu2021} proposed another link between galaxy quenching and changing-look behavior.  They suggest that luminous changing-look quasars occur when AGN are in ``feast" fueling mode, while CL-AGN occur in the ``famine" fueling.  These modes, first proposed by \cite{kauffmann2009}, distinguish between the high-Eddington ratio accretion fueled by the accretion of cold molecular gas (``feast") and low-Eddington ratio accretion of mass loss from aging stars in the galactic nucleus (``famine").  \cite{liu2021} restrict these feeding modes to the nuclear region of the galaxy, coining the phrases ``nuclear feasting" and ``nuclear famine" to describe the stage of galaxy evolution where gas in the nucleus is plentiful or depleted, respectively. Under this scenario, changing-look quasars are simply the most variable luminous quasars, while CL-AGN are triggered by stochastic accretion episodes from small clouds of molecular gas remaining in the gas-starved nucleus; \citet{veronese2024} uses this model to explain the variation of Mrk 1018, a well-studied local CL-AGN.

As explored by \cite{wangj2024}, instabilities in the accretion of cold molecular gas during this transition could trigger sudden changes in the accretion rate, leading to the observed changes in line width and continuum in CL-AGN spectra. \cite{liu2021} describes a scenario in which the stochastic fueling of AGN by the winds of aging stars (e.g. AGB, red giant) could lead to spikes in accretion rate in the outer disk, the observational signatures of which are not yet well predicted. This is consistent with work indicating that CL-AGN tend to have low Eddington ratios, with \cite{jana2024} finding that CL-AGN transitions occur around a median Eddington ratio of 0.01 and linking changing-look transitions in AGN with the transition in photon index-Eddington ratio correlation in black hole X-ray binaries.  As the amount of gas near the black hole varies, accretion rate changes could occur quickly, leading to a ``sputtering" behavior that manifests as a changing-look transition. The ``off" state for CL-AGN would then be a true Seyfert 2, without enough luminosity to support a broad-line region \citep{elitzur2014}.

In this case, a connection to the host galaxy becomes murky; the gas content of the circumnuclear region may be depleted by the AGN or by broader, galaxy-wide processes, perhaps with some time delay, which would wash out trends in SFR.  A major merger could drive chaotic accretion onto the central supermassive black hole and possibly rejuvenate star formation, which would lead to galaxies with similar sSFRs to other AGN hosts but perhaps irregular morphologies \citep{king2008}, consistent with initial high-resolution studies of CL-AGN hosts \citep{charlton2019}.  If CL-AGN are the result of a transition between ``feast" and ``famine" fueling, one would expect them to sit in a nuclear cluster filled with evolved stars.  This nuclear cluster would exist on scales of parsecs, far smaller than the area enclosed by the SDSS fiber at this redshift ($\sim 1$ kpc for our most nearby objects).  It would therefore be consistent with our result that CL-AGN hosts are similar in star formation rate and history to the host galaxies of other kinds of AGN at the galaxy-wide scale, though perhaps with distinctions at the nuclear scale.


\subsubsection{Disk instabilities}

As we can see, both the TDE and feast vs. famine theories for CL-AGN link their behavior to the host galaxy, either globally or on the nuclear scale.   Other proposed mechanisms for CL-AGN, such as accretion disk instabilities \citep[e.g.][]{elitzur2014,sniegowska2021} could potentially occur in any AGN, meaning the CL-AGN host galaxy population would be consistent with the overall AGN host galaxy population, consistent with our findings.  Several studies have found that changing-look transitions occur at an Eddington ratio $L_{\text{bol}}/L_{\text{Edd}} \approx 0.01$ \citep[e.g.][]{jana2024}, suggesting they occur at some critical accretion disk state transition associated with an increase or decrease in accretion rate.  \cite{noda2018} suggest CL-AGN may be triggered by similar state transitions to those which occur in black hole binaries at a critical luminosity Log($L_{\text{bol}}/L_{\text{Edd}}) = -2$, leading to a coincident change in UV/X-ray and broad-line emission. \cite{sniegowska2021} proposes radiation instability between the outer accretion disk and inner accretion flow could produce periodic outbursts resembling repeating CL-AGN. The repeating CL-AGN in our sample probe the highest and lowest ends of mass-SFR space, meaning they are not precluded from any point and may probe the entire AGN parameter space, and we therefore find no disagreement with this theory.

\subsection{Implications for AGN feedback}

Scaling relations between galaxies and their supermassive black holes point to coevolution, though the exact details of this coevolution are still an area of active research \citep{kormendyho2013}.  This coevolution is typically characterized as a cycle of feeding and feedback, where the energy emitted by an AGN disk blows gas away from the black hole, reducing the accretion rate, which then allows more gas to fall toward the black hole.  This duty cycle occurs on scales of $\sim10^5$ years \citep{hickox2014,Schawinski2015}, meaning it may occur many times throughout the lifespan of a galaxy or even within the SFR bins in our \texttt{Prospector} star formation history model.  Observationally, some studies find AGN at z $<0.35$ occur more frequently in young, star-forming galaxies than in older quiescent ones \citep{birchall2023}, and AGN in star-forming galaxies at z $<0.1$ have higher accretion rates than those in quiescent galaxies \citep{ni2023}, though this depends heavily on the AGN selection method. Other studies find little difference in the star-formation properties between high- and low-luminosity AGN \citep[e.g. ][]{Vijarnwannaluk2024}.  In simulations, AGN are also primarily hosted in gas-rich galaxies rather than quenching ones \citep[e.g. ][]{ward2022}, as quenching from AGN feedback occurs due to the integrated energy input from AGN over timescales much longer than AGN variability timescales.  The impact of AGN feedback can therefore be difficult to disentangle from other quenching mechanisms due to the spatial and time scales involved as well as the common fueling mechanism for AGN and star formation.

CL-AGN complicate this picture. If, as would be consistent with our results, all AGN are capable of ``changing look" throughout their lifetimes,  CL-AGN introduce another timescale on which AGN vary which is much shorter than timescales on which the star formation rate changes in a galaxy.  We caution that long-term tracers of AGN activity \citep[e.g.][]{lintott2009,keel2012,french2023} and statistical studies of AGN and their hosts \citep[e.g.][]{greene2011,hickox2014,Schawinski2015,terrazas2016} are necessary to trace the integrated AGN activity over timescales relevant to galactic star formation.




\subsection{The impact of model choices} \label{sec:modelchoices}

Star formation history modeling is sensitive to the choice of star formation history template and priors on galaxy properties \citep[e.g.][]{conroy2013,johnson2021,suess2022b}. To investigate the impact of our choices of metallicity, dust, and star formation history priors on our results, we run the gold, $z<0.15$ CL-AGN sample through our \texttt{Prospector} pipeline with different sets of priors to see whether our results change.

We first examine the impact of our metallicity prior, which is originally a Gaussian prior with a center and width taken from Table 2 of \citet{gallazzi2005}. In our test, we choose a less constraining uniform prior with $-1.7 \leq \mathrm{Log(Z/Z_\odot)} \leq 0.4$, the minimum and maximum observed values in the \citet{gallazzi2005} sample. We do not recover statistically significantly different values for the metallicity, dust extinction, mass, or SFR of our CL-AGN sample when we use a less constraining prior on metallicity.

We next test the impact of different dust laws on our results. We initially assume a fixed power-law absorption model with an index of -0.7 and an optical depth which is allowed to vary between 0.0 and 0.2. We test the impact of instead using two other common dust laws: the \citet{kriek2013} dust model, which ties the slope of the dust law to the strength of the UV bump, and the \citet{calzetti1994} dust law, which does not account for dust obscuration in star forming regions. Here, we do find some statistically significant differences in recovered values. With a \citet{kriek2013} dust law, the recovered metallicities for our sample are on average higher, and the dust extinction is on average significantly lower. This does not have a statistically significant impact on the recovered masses or SFRs of our CL-AGN sample, though it does lead to a slight decrease in the recovered stellar mass and a slight increase in the recovered SFR. We do not see any significant difference in recovered star formation history between the two models. The \citet{calzetti1994} dust law does not lead to any statistically significant changes in recovered dust extinction, metallicity, stellar mass, or SFR. Again, the recovered star formation histories do not change significantly with this dust model.

We then test the impact of a more physically-motivated prior on the SFR ratios in the \texttt{continuity\_psb\_sfh} template. \citet{suess2022b} uses a prior informed by simulations from UniverseMachine. In brief, they select a simulated galaxy from UniverseMachine for each object in their sample by matching on stellar mass. They determine the SFR ratios for the star formation history of the matched UniverseMachine galaxy, which they use to inform the SFR ratio prior on each object. This approach allows for more early-Universe star formation, which may be difficult to detect from observed spectra as older stellar populations are outshone by younger stars. We again do not find any statistically significantly different results in recovered dust extinction, metallicity, stellar mass, or SFR using this model. The recovered star formation history from this model does not differ significantly from the flat star formation history prior used in our analysis.

Finally, we address the impact of using the \texttt{continuity\_psb\_sfh} template rather than a parametric template. As discussed in Section \ref{sec:parametrictemplate}, we recover a slightly lower final stellar mass when using the \texttt{continuity\_psb\_sfh} template on mock spectra generated using a delayed-tau plus exponential starburst template. We do recover an accurate final star formation rate within the error bars returned by Prospector. This means we would be biased towards recovering a higher sSFR for galaxies which have experienced a delayed tau plus exponential burst star formation histories like post-starburst galaxies. As discussed in Suess et al. (2022b), the \texttt{continuity\_psb\_sfh} template is capable of recovering starbursts when tested against mock spectra. As such, we expect that for any post-starburst objects in our sample, we would see the signature of that starburst in the recovered star formation history using the \texttt{continuity\_psb\_sfh} template; an example of this is the starburst which is clearly recovered in the star formation history for ASASSN-14li (Figure 14). We do not see evidence of any significant starbursts in our CL-AGN sample. We do see evidence of rapid quenching of one object 1.7 Gyr before time of observation, with a small bump in its star formation beginning about 2.3 Gyr before observation. Assuming this one object is a post-starburst galaxy, its mass may be underestimated by a median value of 0.25 dex or an 84th percentile value of 0.42 dex, either of which would push this object onto the star-forming main sequence and would not change our finding that CL-AGN host galaxy star formation properties are not statistically distinct from other AGN hosts.

Overall, we find that our results are robust to the priors and model choices used in this analysis. 

\section{Conclusions}

In this work, we examine the star formation histories of 39 CL-AGN host galaxies.  We find:

\begin{enumerate}
    \item Despite AGN contamination, both stellar mass and star formation rate are recovered well by the \texttt{continuity\_psb\_sfh} template developed in \cite{suess2022c} for modeled galaxies with or without AGN. Non-parametric star formation history fitting with \texttt{Prospector} is capable of recovering the host galaxy parameters of low-luminosity AGN with optical spectroscopy, even without an incorporated AGN SED model in the optical bands.
    \item At z $<0.15$, $40-50\%$ of CL-AGN hosts are star forming, $\sim30\%$ are in the Green Valley, and $20-30\%$ are quiescent on a mass-sSFR diagram, consistent with the host galaxies of Type 1 and 2 Seyferts. 
    \item At z $>0.15$, $\sim60\%$ of CL-AGN are star forming, $\sim30\%$ are in the Green Valley, and $<15\%$ are quiescent, consistent with the host galaxies of Seyfert 2 hosts. They may be less star forming than Type 1 Seyfert hosts.
    \item The star formation histories of CL-AGN are similar to those of Seyfert 2 host galaxies and may indicate a recent ($\lesssim0.5$ Gyr) rejuvenation of star formation, possibly tied to the fueling of the central AGN. They may also indicate slow quenching mechanisms with some residual star formation outshining an older population.  CL-AGN hosts at these redshifts have not experienced significant starbursts in the past several Gyr.
    \item The \texttt{continuity\_psb\_sfh} template recovers similar masses for CL-AGN hosts in the broad- and narrow-line states, but returns a $0.89^{+0.37}_{-0.56}$ dex higher SFR for the broad-line spectrum.  This can be used to correct for the overestimation of SFR in broad-line AGN hosts.
\end{enumerate}

Finally, the similarity between CL-AGN host galaxies and the host galaxies of other Seyfert types suggests that all AGN may be capable of changing look at some point in their lives.  This short timescale of AGN variability allows us to test accretion physics with theories linking to X-ray binaries \citep[e.g.][]{noda2018}, extreme mass ratio inspirals \citep[e.g.][]{hernandezgarcia2025}, and rapid evolution of accretion disk instabilities \citep[e.g.][]{sniegowska2021}.  CL-AGN link accretion disk physics with galaxy-scale processes; their physics span timescales as short as hours and as long as ten thousand years.  By studying CL-AGN, we probe physics crucial for the fueling and growth of supermassive black holes on cosmic timescales.

\section*{Acknowledgements}
M.E.V., M.S., T.A., and K.D.F. \ acknowledge support from NSF grant AST 22-06164. M.E.V. and M.S.\ acknowledge support from the Illinois Space Grant Consortium. M.E.V. acknowledge support from the Center for Astrophysical Surveys Graduate Fellowship.

This work was in part supported by the NASA Astrophysics Data Analysis Program (ADAP) under grant 80NSSC23K0495.

M.E.V. would like to thank Vardha Bennert, Jordan Runco, Yue Shen, Qian Yang, and Grisha Zeltyn for providing data used in this paper. M.E.V. would also like to thank Nicholas Earl for providing programming support for this paper.

M.E.V. would like to thank Yue Shen, Gautham Narayan, Kirk Barrow, Katey Alatalo, Aidan Berres, Sierra Dodd, Nicholas Earl, Antoniu Fodor, Samaresh Mondal, Justin Otter, Haille Perkins, Sara Starecheski, David Setton, David Vizgan, Amanda Wasserman, and Grisha Zeltyn for conversations which improved this work.

Funding for the Sloan Digital Sky Survey IV has been provided by the Alfred P. Sloan Foundation, the U.S. Department of Energy Office of Science, and the Participating Institutions. SDSS acknowledges support and resources from the Center for High-Performance Computing at the University of Utah. The SDSS web site is www.sdss4.org.

SDSS is managed by the Astrophysical Research Consortium for the Participating Institutions of the SDSS Collaboration including the Brazilian Participation Group, the Carnegie Institution for Science, Carnegie Mellon University, Center for Astrophysics | Harvard \& Smithsonian (CfA), the Chilean Participation Group, the French Participation Group, Instituto de Astrofísica de Canarias, The Johns Hopkins University, Kavli Institute for the Physics and Mathematics of the Universe (IPMU) / University of Tokyo, the Korean Participation Group, Lawrence Berkeley National Laboratory, Leibniz Institut für Astrophysik Potsdam (AIP), Max-Planck-Institut für Astronomie (MPIA Heidelberg), Max-Planck-Institut für Astrophysik (MPA Garching), Max-Planck-Institut für Extraterrestrische Physik (MPE), National Astronomical Observatories of China, New Mexico State University, New York University, University of Notre Dame, Observatório Nacional / MCTI, The Ohio State University, Pennsylvania State University, Shanghai Astronomical Observatory, United Kingdom Participation Group, Universidad Nacional Autónoma de México, University of Arizona, University of Colorado Boulder, University of Oxford, University of Portsmouth, University of Utah, University of Virginia, University of Washington, University of Wisconsin, Vanderbilt University, and Yale University.

Funding for the Sloan Digital Sky Survey V has been provided by the Alfred P. Sloan Foundation, the Heising-Simons Foundation, the National Science Foundation, and the Participating Institutions. SDSS acknowledges support and resources from the Center for High-Performance Computing at the University of Utah. SDSS telescopes are located at Apache Point Observatory, funded by the Astrophysical Research Consortium and operated by New Mexico State University, and at Las Campanas Observatory, operated by the Carnegie Institution for Science. The SDSS web site is www.sdss.org.

SDSS is managed by the Astrophysical Research Consortium for the Participating Institutions of the SDSS Collaboration, including Caltech, the Carnegie Institution for Science, Chilean National Time Allocation Committee (CNTAC) ratified researchers, The Flatiron Institute, the Gotham Participation Group, Harvard University, Heidelberg University, The Johns Hopkins University, L’Ecole polytechnique fédérale de Lausanne (EPFL), Leibniz-Institut für Astrophysik Potsdam (AIP), Max-Planck-Institut für Astronomie (MPIA Heidelberg), Max-Planck-Institut für Extraterrestrische Physik (MPE), Nanjing University, National Astronomical Observatories of China (NAOC), New Mexico State University, The Ohio State University, Pennsylvania State University, Smithsonian Astrophysical Observatory, Space Telescope Science Institute (STScI), the Stellar Astrophysics Participation Group, Universidad Nacional Autónoma de México, University of Arizona, University of Colorado Boulder, University of Illinois at Urbana-Champaign, University of Toronto, University of Utah, University of Virginia, Yale University, and Yunnan University.

\subsection{Software}

Software used for this project includes Prospector \citep{leja17,johnson17,leja2018,johnson2021}, Python-FSPS \citep{pyfsps}, Corner \citep{corner}, \textit{kcorrect} \citep{kcorrect}, Astropy \citep{astropy:2013, astropy:2018, astropy:2022}, SedPy \citep{sedpy}, Scipy \citep{scipy}, Matplotlib \citep{matplotlib}, Numpy \citep{numpy}, Pandas \citep{pandas,pandas2} and Seaborn \citep{seaborn}.

\newpage 

\bibliography{clagn_prospector}{}

\begin{thebibliography}{}
\expandafter\ifx\csname natexlab\endcsname\relax\def\natexlab#1{#1}\fi
\providecommand{\url}[1]{\href{#1}{#1}}
\providecommand{\dodoi}[1]{doi:~\href{http://doi.org/#1}{\nolinkurl{#1}}}
\providecommand{\doeprint}[1]{\href{http://ascl.net/#1}{\nolinkurl{http://ascl.net/#1}}}
\providecommand{\doarXiv}[1]{\href{https://arxiv.org/abs/#1}{\nolinkurl{https://arxiv.org/abs/#1}}}

\bibitem[{{Aihara} {et~al.}(2011){Aihara}, {Allende Prieto}, {An}, {Anderson}, {Aubourg}, {Balbinot}, {Beers}, {Berlind}, {Bickerton}, {Bizyaev}, {Blanton}, {Bochanski}, {Bolton}, {Bovy}, {Brandt}, {Brinkmann}, {Brown}, {Brownstein}, {Busca}, {Campbell}, {Carr}, {Chen}, {Chiappini}, {Comparat}, {Connolly}, {Cortes}, {Croft}, {Cuesta}, {da Costa}, {Davenport}, {Dawson}, {Dhital}, {Ealet}, {Ebelke}, {Edmondson}, {Eisenstein}, {Escoffier}, {Esposito}, {Evans}, {Fan}, {Femen{\'\i}a Castell{\'a}}, {Font-Ribera}, {Frinchaboy}, {Ge}, {Gillespie}, {Gilmore}, {Gonz{\'a}lez Hern{\'a}ndez}, {Gott}, {Gould}, {Grebel}, {Gunn}, {Hamilton}, {Harding}, {Harris}, {Hawley}, {Hearty}, {Ho}, {Hogg}, {Holtzman}, {Honscheid}, {Inada}, {Ivans}, {Jiang}, {Johnson}, {Jordan}, {Jordan}, {Kazin}, {Kirkby}, {Klaene}, {Knapp}, {Kneib}, {Kochanek}, {Koesterke}, {Kollmeier}, {Kron}, {Lampeitl}, {Lang}, {Le Goff}, {Lee}, {Lin}, {Long}, {Loomis}, {Lucatello}, {Lundgren}, {Lupton}, {Ma}, {MacDonald}, {Mahadevan}, {Maia}, {Makler},
  {Malanushenko}, {Malanushenko}, {Mandelbaum}, {Maraston}, {Margala}, {Masters}, {McBride}, {McGehee}, {McGreer}, {M{\'e}nard}, {Miralda-Escud{\'e}}, {Morrison}, {Mullally}, {Muna}, {Munn}, {Murayama}, {Myers}, {Naugle}, {Neto}, {Nguyen}, {Nichol}, {O'Connell}, {Ogando}, {Olmstead}, {Oravetz}, {Padmanabhan}, {Palanque-Delabrouille}, {Pan}, {Pandey}, {P{\^a}ris}, {Percival}, {Petitjean}, {Pfaffenberger}, {Pforr}, {Phleps}, {Pichon}, {Pieri}, {Prada}, {Price-Whelan}, {Raddick}, {Ramos}, {Reyl{\'e}}, {Rich}, {Richards}, {Rix}, {Robin}, {Rocha-Pinto}, {Rockosi}, {Roe}, {Rollinde}, {Ross}, {Ross}, {Rossetto}, {S{\'a}nchez}, {Sayres}, {Schlegel}, {Schlesinger}, {Schmidt}, {Schneider}, {Sheldon}, {Shu}, {Simmerer}, {Simmons}, {Sivarani}, {Snedden}, {Sobeck}, {Steinmetz}, {Strauss}, {Szalay}, {Tanaka}, {Thakar}, {Thomas}, {Tinker}, {Tofflemire}, {Tojeiro}, {Tremonti}, {Vandenberg}, {Vargas Maga{\~n}a}, {Verde}, {Vogt}, {Wake}, {Wang}, {Weaver}, {Weinberg}, {White}, {White}, {Yanny}, {Yasuda}, {Yeche}, \&
  {Zehavi}}]{sdss8}
{Aihara}, H., {Allende Prieto}, C., {An}, D., {et~al.} 2011, \apjs, 193, 29, \dodoi{10.1088/0067-0049/193/2/29}

\bibitem[{{Alexander} \& {Hickox}(2012)}]{alexander2012}
{Alexander}, D.~M., \& {Hickox}, R.~C. 2012, \nar, 56, 93, \dodoi{10.1016/j.newar.2011.11.003}

\bibitem[{{Almeida} {et~al.}(2023){Almeida}, {Anderson}, {Argudo-Fern{\'a}ndez}, {Badenes}, {Barger}, {Barrera-Ballesteros}, {Bender}, {Benitez}, {Besser}, {Bird}, {Bizyaev}, {Blanton}, {Bochanski}, {Bovy}, {Brandt}, {Brownstein}, {Buchner}, {Bulbul}, {Burchett}, {Cano D{\'\i}az}, {Carlberg}, {Casey}, {Chandra}, {Cherinka}, {Chiappini}, {Coker}, {Comparat}, {Conroy}, {Contardo}, {Cortes}, {Covey}, {Crane}, {Cunha}, {Dabbieri}, {Davidson}, {Davis}, {de Andrade Queiroz}, {De Lee}, {M{\'e}ndez Delgado}, {Demasi}, {Di Mille}, {Donor}, {Dow}, {Dwelly}, {Eracleous}, {Eriksen}, {Fan}, {Farr}, {Frederick}, {Fries}, {Frinchaboy}, {G{\"a}nsicke}, {Ge}, {Gonz{\'a}lez {\'A}vila}, {Grabowski}, {Grier}, {Guiglion}, {Gupta}, {Hall}, {Hawkins}, {Hayes}, {Hermes}, {Hern{\'a}ndez-Garc{\'\i}a}, {Hogg}, {Holtzman}, {Ibarra-Medel}, {Ji}, {Jofre}, {Johnson}, {Jones}, {Kinemuchi}, {Kluge}, {Koekemoer}, {Kollmeier}, {Kounkel}, {Krishnarao}, {Krumpe}, {Lacerna}, {Lago}, {Laporte}, {Liu}, {Liu}, {Liu}, {Lopes}, {Macktoobian},
  {Majewski}, {Malanushenko}, {Maoz}, {Masseron}, {Masters}, {Matijevic}, {McBride}, {Medan}, {Merloni}, {Morrison}, {Myers}, {M{\'e}sz{\'a}ros}, {Negrete}, {Nidever}, {Nitschelm}, {Oravetz}, {Oravetz}, {Pan}, {Peng}, {Pinsonneault}, {Pogge}, {Qiu}, {Ramirez}, {Rix}, {Fern{\'a}ndez Rosso}, {Runnoe}, {Salvato}, {Sanchez}, {Santana}, {Saydjari}, {Sayres}, {Schlaufman}, {Schneider}, {Schwope}, {Serna}, {Shen}, {Sobeck}, {Song}, {Souto}, {Spoo}, {Stassun}, {Steinmetz}, {Straumit}, {Stringfellow}, {S{\'a}nchez-Gallego}, {Taghizadeh-Popp}, {Tayar}, {Thakar}, {Tissera}, {Tkachenko}, {Hernandez Toledo}, {Trakhtenbrot}, {Fern{\'a}ndez-Trincado}, {Troup}, {Trump}, {Tuttle}, {Ulloa}, {Vazquez-Mata}, {Vera Alfaro}, {Villanova}, {Wachter}, {Weijmans}, {Wheeler}, {Wilson}, {Wojno}, {Wolf}, {Xue}, {Ybarra}, {Zari}, \& {Zasowski}}]{sdss18}
{Almeida}, A., {Anderson}, S.~F., {Argudo-Fern{\'a}ndez}, M., {et~al.} 2023, \apjs, 267, 44, \dodoi{10.3847/1538-4365/acda98}

\bibitem[{{Antonucci}(1993)}]{Antonucci1993}
{Antonucci}, R. 1993, \araa, 31, 473, \dodoi{10.1146/annurev.aa.31.090193.002353}

\bibitem[{{Arcavi} {et~al.}(2014){Arcavi}, {Gal-Yam}, {Sullivan}, {Pan}, {Cenko}, {Horesh}, {Ofek}, {De Cia}, {Yan}, {Yang}, {Howell}, {Tal}, {Kulkarni}, {Tendulkar}, {Tang}, {Xu}, {Sternberg}, {Cohen}, {Bloom}, {Nugent}, {Kasliwal}, {Perley}, {Quimby}, {Miller}, {Theissen}, \& {Laher}}]{arcavi2014}
{Arcavi}, I., {Gal-Yam}, A., {Sullivan}, M., {et~al.} 2014, \apj, 793, 38, \dodoi{10.1088/0004-637X/793/1/38}

\bibitem[{{Astropy Collaboration} {et~al.}(2013){Astropy Collaboration}, {Robitaille}, {Tollerud}, {Greenfield}, {Droettboom}, {Bray}, {Aldcroft}, {Davis}, {Ginsburg}, {Price-Whelan}, {Kerzendorf}, {Conley}, {Crighton}, {Barbary}, {Muna}, {Ferguson}, {Grollier}, {Parikh}, {Nair}, {Unther}, {Deil}, {Woillez}, {Conseil}, {Kramer}, {Turner}, {Singer}, {Fox}, {Weaver}, {Zabalza}, {Edwards}, {Azalee Bostroem}, {Burke}, {Casey}, {Crawford}, {Dencheva}, {Ely}, {Jenness}, {Labrie}, {Lim}, {Pierfederici}, {Pontzen}, {Ptak}, {Refsdal}, {Servillat}, \& {Streicher}}]{astropy:2013}
{Astropy Collaboration}, {Robitaille}, T.~P., {Tollerud}, E.~J., {et~al.} 2013, \aap, 558, A33, \dodoi{10.1051/0004-6361/201322068}

\bibitem[{{Astropy Collaboration} {et~al.}(2018){Astropy Collaboration}, {Price-Whelan}, {Sip{\H{o}}cz}, {G{\"u}nther}, {Lim}, {Crawford}, {Conseil}, {Shupe}, {Craig}, {Dencheva}, {Ginsburg}, {Vand erPlas}, {Bradley}, {P{\'e}rez-Su{\'a}rez}, {de Val-Borro}, {Aldcroft}, {Cruz}, {Robitaille}, {Tollerud}, {Ardelean}, {Babej}, {Bach}, {Bachetti}, {Bakanov}, {Bamford}, {Barentsen}, {Barmby}, {Baumbach}, {Berry}, {Biscani}, {Boquien}, {Bostroem}, {Bouma}, {Brammer}, {Bray}, {Breytenbach}, {Buddelmeijer}, {Burke}, {Calderone}, {Cano Rodr{\'\i}guez}, {Cara}, {Cardoso}, {Cheedella}, {Copin}, {Corrales}, {Crichton}, {D'Avella}, {Deil}, {Depagne}, {Dietrich}, {Donath}, {Droettboom}, {Earl}, {Erben}, {Fabbro}, {Ferreira}, {Finethy}, {Fox}, {Garrison}, {Gibbons}, {Goldstein}, {Gommers}, {Greco}, {Greenfield}, {Groener}, {Grollier}, {Hagen}, {Hirst}, {Homeier}, {Horton}, {Hosseinzadeh}, {Hu}, {Hunkeler}, {Ivezi{\'c}}, {Jain}, {Jenness}, {Kanarek}, {Kendrew}, {Kern}, {Kerzendorf}, {Khvalko}, {King}, {Kirkby}, {Kulkarni},
  {Kumar}, {Lee}, {Lenz}, {Littlefair}, {Ma}, {Macleod}, {Mastropietro}, {McCully}, {Montagnac}, {Morris}, {Mueller}, {Mumford}, {Muna}, {Murphy}, {Nelson}, {Nguyen}, {Ninan}, {N{\"o}the}, {Ogaz}, {Oh}, {Parejko}, {Parley}, {Pascual}, {Patil}, {Patil}, {Plunkett}, {Prochaska}, {Rastogi}, {Reddy Janga}, {Sabater}, {Sakurikar}, {Seifert}, {Sherbert}, {Sherwood-Taylor}, {Shih}, {Sick}, {Silbiger}, {Singanamalla}, {Singer}, {Sladen}, {Sooley}, {Sornarajah}, {Streicher}, {Teuben}, {Thomas}, {Tremblay}, {Turner}, {Terr{\'o}n}, {van Kerkwijk}, {de la Vega}, {Watkins}, {Weaver}, {Whitmore}, {Woillez}, {Zabalza}, \& {Astropy Contributors}}]{astropy:2018}
{Astropy Collaboration}, {Price-Whelan}, A.~M., {Sip{\H{o}}cz}, B.~M., {et~al.} 2018, \aj, 156, 123, \dodoi{10.3847/1538-3881/aabc4f}

\bibitem[{{Astropy Collaboration} {et~al.}(2022){Astropy Collaboration}, {Price-Whelan}, {Lim}, {Earl}, {Starkman}, {Bradley}, {Shupe}, {Patil}, {Corrales}, {Brasseur}, {N{"o}the}, {Donath}, {Tollerud}, {Morris}, {Ginsburg}, {Vaher}, {Weaver}, {Tocknell}, {Jamieson}, {van Kerkwijk}, {Robitaille}, {Merry}, {Bachetti}, {G{"u}nther}, {Aldcroft}, {Alvarado-Montes}, {Archibald}, {B{'o}di}, {Bapat}, {Barentsen}, {Baz{'a}n}, {Biswas}, {Boquien}, {Burke}, {Cara}, {Cara}, {Conroy}, {Conseil}, {Craig}, {Cross}, {Cruz}, {D'Eugenio}, {Dencheva}, {Devillepoix}, {Dietrich}, {Eigenbrot}, {Erben}, {Ferreira}, {Foreman-Mackey}, {Fox}, {Freij}, {Garg}, {Geda}, {Glattly}, {Gondhalekar}, {Gordon}, {Grant}, {Greenfield}, {Groener}, {Guest}, {Gurovich}, {Handberg}, {Hart}, {Hatfield-Dodds}, {Homeier}, {Hosseinzadeh}, {Jenness}, {Jones}, {Joseph}, {Kalmbach}, {Karamehmetoglu}, {Ka{l}uszy{'n}ski}, {Kelley}, {Kern}, {Kerzendorf}, {Koch}, {Kulumani}, {Lee}, {Ly}, {Ma}, {MacBride}, {Maljaars}, {Muna}, {Murphy}, {Norman}, {O'Steen},
  {Oman}, {Pacifici}, {Pascual}, {Pascual-Granado}, {Patil}, {Perren}, {Pickering}, {Rastogi}, {Roulston}, {Ryan}, {Rykoff}, {Sabater}, {Sakurikar}, {Salgado}, {Sanghi}, {Saunders}, {Savchenko}, {Schwardt}, {Seifert-Eckert}, {Shih}, {Jain}, {Shukla}, {Sick}, {Simpson}, {Singanamalla}, {Singer}, {Singhal}, {Sinha}, {Sip{H{o}}cz}, {Spitler}, {Stansby}, {Streicher}, {{{S}}umak}, {Swinbank}, {Taranu}, {Tewary}, {Tremblay}, {Val-Borro}, {Van Kooten}, {Vasovi{'c}}, {Verma}, {de Miranda Cardoso}, {Williams}, {Wilson}, {Winkel}, {Wood-Vasey}, {Xue}, {Yoachim}, {Zhang}, {Zonca}, \& {Astropy Project Contributors}}]{astropy:2022}
{Astropy Collaboration}, {Price-Whelan}, A.~M., {Lim}, P.~L., {et~al.} 2022, \apj, 935, 167, \dodoi{10.3847/1538-4357/ac7c74}

\bibitem[{{Baldwin} {et~al.}(1981){Baldwin}, {Phillips}, \& {Terlevich}}]{baldwin1981}
{Baldwin}, J.~A., {Phillips}, M.~M., \& {Terlevich}, R. 1981, \pasp, 93, 5, \dodoi{10.1086/130766}

\bibitem[{{Behroozi} {et~al.}(2019){Behroozi}, {Wechsler}, {Hearin}, \& {Conroy}}]{behroozi2019}
{Behroozi}, P., {Wechsler}, R.~H., {Hearin}, A.~P., \& {Conroy}, C. 2019, \mnras, 488, 3143, \dodoi{10.1093/mnras/stz1182}

\bibitem[{{Bell} {et~al.}(2003){Bell}, {McIntosh}, {Katz}, \& {Weinberg}}]{bell2003}
{Bell}, E.~F., {McIntosh}, D.~H., {Katz}, N., \& {Weinberg}, M.~D. 2003, \apjs, 149, 289, \dodoi{10.1086/378847}

\bibitem[{{Birchall} {et~al.}(2023){Birchall}, {Watson}, {Aird}, \& {Starling}}]{birchall2023}
{Birchall}, K.~L., {Watson}, M.~G., {Aird}, J., \& {Starling}, R.~L.~C. 2023, \mnras, 523, 4756, \dodoi{10.1093/mnras/stad1723}

\bibitem[{{Blanton} \& {Roweis}(2007)}]{kcorrect}
{Blanton}, M.~R., \& {Roweis}, S. 2007, \aj, 133, 734, \dodoi{10.1086/510127}

\bibitem[{{Boquien} {et~al.}(2019){Boquien}, {Burgarella}, {Roehlly}, {Buat}, {Ciesla}, {Corre}, {Inoue}, \& {Salas}}]{Boquien2019}
{Boquien}, M., {Burgarella}, D., {Roehlly}, Y., {et~al.} 2019, \aap, 622, A103, \dodoi{10.1051/0004-6361/201834156}

\bibitem[{{Brinchmann} {et~al.}(2004){Brinchmann}, {Charlot}, {White}, {Tremonti}, {Kauffmann}, {Heckman}, \& {Brinkmann}}]{Brinchmann2004}
{Brinchmann}, J., {Charlot}, S., {White}, S.~D.~M., {et~al.} 2004, \mnras, 351, 1151, \dodoi{10.1111/j.1365-2966.2004.07881.x}

\bibitem[{{Buchner} {et~al.}(2024){Buchner}, {Starck}, {Salvato}, {Netzer}, {Igo}, {Laloux}, {Georgakakis}, {Gauger}, {Olechowska}, {Lopez}, {Shankar}, {Li}, {Nandra}, \& {Merloni}}]{Buchner2024}
{Buchner}, J., {Starck}, H., {Salvato}, M., {et~al.} 2024, arXiv e-prints, arXiv:2405.19297, \dodoi{10.48550/arXiv.2405.19297}

\bibitem[{{Calzetti} {et~al.}(1994){Calzetti}, {Kinney}, \& {Storchi-Bergmann}}]{calzetti1994}
{Calzetti}, D., {Kinney}, A.~L., \& {Storchi-Bergmann}, T. 1994, \apj, 429, 582, \dodoi{10.1086/174346}

\bibitem[{{Chabrier}(2003)}]{chabrier2003}
{Chabrier}, G. 2003, \pasp, 115, 763, \dodoi{10.1086/376392}

\bibitem[{{Chang} {et~al.}(2015){Chang}, {van der Wel}, {da Cunha}, \& {Rix}}]{Chang2015}
{Chang}, Y.-Y., {van der Wel}, A., {da Cunha}, E., \& {Rix}, H.-W. 2015, \apjs, 219, 8, \dodoi{10.1088/0067-0049/219/1/8}

\bibitem[{{Charlton} {et~al.}(2019){Charlton}, {Ruan}, {Haggard}, {Anderson}, {Eracleous}, {MacLeod}, \& {Runnoe}}]{charlton2019}
{Charlton}, P. J.~L., {Ruan}, J.~J., {Haggard}, D., {et~al.} 2019, \apj, 876, 75, \dodoi{10.3847/1538-4357/ab0ec1}

\bibitem[{{Chen} {et~al.}(2009){Chen}, {Madau}, {Sesana}, \& {Liu}}]{chen2009}
{Chen}, X., {Madau}, P., {Sesana}, A., \& {Liu}, F.~K. 2009, \apjl, 697, L149, \dodoi{10.1088/0004-637X/697/2/L149}

\bibitem[{{Choi} {et~al.}(2016){Choi}, {Dotter}, {Conroy}, {Cantiello}, {Paxton}, \& {Johnson}}]{choi2016}
{Choi}, J., {Dotter}, A., {Conroy}, C., {et~al.} 2016, \apj, 823, 102, \dodoi{10.3847/0004-637X/823/2/102}

\bibitem[{Conroy(2013)}]{conroy2013}
Conroy, C. 2013, Annual Review of Astronomy and Astrophysics, 51, 393, \dodoi{10.1146/annurev-astro-082812-141017}

\bibitem[{{Conroy} \& {Gunn}(2010)}]{conroy2010}
{Conroy}, C., \& {Gunn}, J.~E. 2010, \apj, 712, 833, \dodoi{10.1088/0004-637X/712/2/833}

\bibitem[{{Conroy} {et~al.}(2009){Conroy}, {Gunn}, \& {White}}]{conroy2009}
{Conroy}, C., {Gunn}, J.~E., \& {White}, M. 2009, \apj, 699, 486, \dodoi{10.1088/0004-637X/699/1/486}

\bibitem[{{Cristello} {et~al.}(2024){Cristello}, {Zou}, {Brandt}, {Chen}, {Leja}, {Ni}, \& {Yang}}]{Cristello2024}
{Cristello}, N., {Zou}, F., {Brandt}, W.~N., {et~al.} 2024, \apj, 962, 156, \dodoi{10.3847/1538-4357/ad2177}

\bibitem[{{Dodd} {et~al.}(2021){Dodd}, {Law-Smith}, {Auchettl}, {Ramirez-Ruiz}, \& {Foley}}]{dodd2021}
{Dodd}, S.~A., {Law-Smith}, J. A.~P., {Auchettl}, K., {Ramirez-Ruiz}, E., \& {Foley}, R.~J. 2021, ApJl, 907, L21, \dodoi{10.3847/2041-8213/abd852}

\bibitem[{{Dodd} {et~al.}(2023){Dodd}, {Nukala}, {Connor}, {Auchettl}, {French}, {Law-Smith}, {Hammerstein}, \& {Ramirez-Ruiz}}]{dodd2023}
{Dodd}, S.~A., {Nukala}, A., {Connor}, I., {et~al.} 2023, \apjl, 959, L19, \dodoi{10.3847/2041-8213/ad1112}

\bibitem[{{Dong} {et~al.}(2024){Dong}, {Zhang}, {Gu}, {Sun}, \& {Zheng}}]{dong2024}
{Dong}, Q., {Zhang}, Z.-X., {Gu}, W.-M., {Sun}, M., \& {Zheng}, Y.-G. 2024, arXiv e-prints, arXiv:2408.07335, \dodoi{10.48550/arXiv.2408.07335}

\bibitem[{{Dotter}(2016)}]{dotter2016}
{Dotter}, A. 2016, \apjs, 222, 8, \dodoi{10.3847/0067-0049/222/1/8}

\bibitem[{{Dressler} \& {Shectman}(1987)}]{Dressler1987}
{Dressler}, A., \& {Shectman}, S.~A. 1987, \aj, 94, 899, \dodoi{10.1086/114524}

\bibitem[{{Elitzur} {et~al.}(2014){Elitzur}, {Ho}, \& {Trump}}]{elitzur2014}
{Elitzur}, M., {Ho}, L.~C., \& {Trump}, J.~R. 2014, MNRAS, 438, 3340, \dodoi{10.1093/mnras/stt2445}

\bibitem[{{Ellison} {et~al.}(2025){Ellison}, {Ferreira}, {Bickley}, {Grindlay}, {Salim}, {Byrne-Mamahit}, {Satyapal}, {Patton}, \& {Scudder}}]{ellison2025}
{Ellison}, S., {Ferreira}, L., {Bickley}, R., {et~al.} 2025, The Open Journal of Astrophysics, 8, 12, \dodoi{10.33232/001c.129235}

\bibitem[{{Feroz} {et~al.}(2009){Feroz}, {Hobson}, \& {Bridges}}]{feroz2009}
{Feroz}, F., {Hobson}, M.~P., \& {Bridges}, M. 2009, \mnras, 398, 1601, \dodoi{10.1111/j.1365-2966.2009.14548.x}

\bibitem[{Foreman-Mackey(2016)}]{corner}
Foreman-Mackey, D. 2016, The Journal of Open Source Software, 1, 24, \dodoi{10.21105/joss.00024}

\bibitem[{{Frederick} {et~al.}(2019){Frederick}, {Gezari}, {Graham}, {Cenko}, {van Velzen}, {Stern}, {Blagorodnova}, {Kulkarni}, {Yan}, {De}, {Fremling}, {Hung}, {Kara}, {Shupe}, {Ward}, {Bellm}, {Dekany}, {Duev}, {Feindt}, {Giomi}, {Kupfer}, {Laher}, {Masci}, {Miller}, {Neill}, {Ngeow}, {Patterson}, {Porter}, {Rusholme}, {Sollerman}, \& {Walters}}]{Frederick2019}
{Frederick}, S., {Gezari}, S., {Graham}, M.~J., {et~al.} 2019, \apj, 883, 31, \dodoi{10.3847/1538-4357/ab3a38}

\bibitem[{{French} {et~al.}(2016){French}, {Arcavi}, \& {Zabludoff}}]{french2016}
{French}, K.~D., {Arcavi}, I., \& {Zabludoff}, A. 2016, \apjl, 818, L21, \dodoi{10.3847/2041-8205/818/1/L21}

\bibitem[{{French} {et~al.}(2017){French}, {Arcavi}, \& {Zabludoff}}]{French2017}
---. 2017, \apj, 835, 176, \dodoi{10.3847/1538-4357/835/2/176}

\bibitem[{{French} {et~al.}(2023){French}, {Earl}, {Novack}, {Pardasani}, {Pillai}, {Tripathi}, \& {Verrico}}]{french2023}
{French}, K.~D., {Earl}, N., {Novack}, A.~B., {et~al.} 2023, \apj, 950, 153, \dodoi{10.3847/1538-4357/acd249}

\bibitem[{{French} {et~al.}(2020){French}, {Wevers}, {Law-Smith}, {Graur}, \& {Zabludoff}}]{french2020}
{French}, K.~D., {Wevers}, T., {Law-Smith}, J., {Graur}, O., \& {Zabludoff}, A.~I. 2020, \ssr, 216, 32, \dodoi{10.1007/s11214-020-00657-y}

\bibitem[{{Gallazzi} {et~al.}(2005){Gallazzi}, {Charlot}, {Brinchmann}, {White}, \& {Tremonti}}]{gallazzi2005}
{Gallazzi}, A., {Charlot}, S., {Brinchmann}, J., {White}, S. D.~M., \& {Tremonti}, C.~A. 2005, \mnras, 362, 41, \dodoi{10.1111/j.1365-2966.2005.09321.x}

\bibitem[{{Gezari} {et~al.}(2017){Gezari}, {Hung}, {Cenko}, {Blagorodnova}, {Yan}, {Kulkarni}, {Mooley}, {Kong}, {Cantwell}, {Yu}, {Cao}, {Fremling}, {Neill}, {Ngeow}, {Nugent}, \& {Wozniak}}]{gezari2017}
{Gezari}, S., {Hung}, T., {Cenko}, S.~B., {et~al.} 2017, \apj, 835, 144, \dodoi{10.3847/1538-4357/835/2/144}

\bibitem[{{Green} {et~al.}(2022){Green}, {Pulgarin-Duque}, {Anderson}, {MacLeod}, {Eracleous}, {Ruan}, {Runnoe}, {Graham}, {Roulston}, {Schneider}, {Ahlf}, {Bizyaev}, {Brownstein}, {del Casal}, {Dodd}, {Hoover}, {Matt}, {Merloni}, {Pan}, {Ramirez}, {Ridder}, \& {Moseley}}]{green2022}
{Green}, P.~J., {Pulgarin-Duque}, L., {Anderson}, S.~F., {et~al.} 2022, \apj, 933, 180, \dodoi{10.3847/1538-4357/ac743f}

\bibitem[{{Greene} {et~al.}(2020){Greene}, {Setton}, {Bezanson}, {Suess}, {Kriek}, {Spilker}, {Goulding}, \& {Feldmann}}]{greene2020}
{Greene}, J.~E., {Setton}, D., {Bezanson}, R., {et~al.} 2020, ApJl, 899, L9, \dodoi{10.3847/2041-8213/aba534}

\bibitem[{{Greene} {et~al.}(2011){Greene}, {Zakamska}, {Ho}, \& {Barth}}]{greene2011}
{Greene}, J.~E., {Zakamska}, N.~L., {Ho}, L.~C., \& {Barth}, A.~J. 2011, \apj, 732, 9, \dodoi{10.1088/0004-637X/732/1/9}

\bibitem[{{Guo} {et~al.}(2024{\natexlab{a}}){Guo}, {Zou}, {Fawcett}, {Canning}, {Juneau}, {Davis}, {Alexander}, {Jiang}, {Aguilar}, {Ahlen}, {Brooks}, {Claybaugh}, {de la Macorra}, {Doel}, {Fanning}, {Forero-Romero}, {Gontcho A Gontcho}, {Honscheid}, {Kisner}, {Kremin}, {Landriau}, {Meisner}, {Miquel}, {Moustakas}, {Nie}, {Pan}, {Poppett}, {Prada}, {Rezaie}, {Rossi}, {Siudek}, {Sanchez}, {Schubnell}, {Seo}, {Sui}, {Tarl{\'e}}, \& {Zhou}}]{guo2024a}
{Guo}, W.-J., {Zou}, H., {Fawcett}, V.~A., {et~al.} 2024{\natexlab{a}}, \apjs, 270, 26, \dodoi{10.3847/1538-4365/ad118a}

\bibitem[{{Guo} {et~al.}(2024{\natexlab{b}}){Guo}, {Zou}, {Greenwell}, {Alexander}, {Fawcett}, {Pan}, {Siudek}, {Aguilar}, {Ahlen}, {Brooks}, {Claybaugh}, {Dawson}, {De La Macorra}, {Doel}, {Font-Ribera}, {Gaztanaga}, {Gontcho}, {Gutierrez}, {Kehoe}, {Kisner}, {Landriau}, {Le Guillou}, {Manera}, {Meisner}, {Mique}, {Moustakas}, {Prada}, {Rossi}, {Sanchez}, {Schubnell}, {Sprayberry}, {Sui}, {Tarle}, {Weaver}, {Xiao}, \& {Zou}}]{guo2024b}
{Guo}, W.-J., {Zou}, H., {Greenwell}, C.~L., {et~al.} 2024{\natexlab{b}}, arXiv e-prints, arXiv:2408.00402, \dodoi{10.48550/arXiv.2408.00402}

\bibitem[{Harris {et~al.}(2020)Harris, Millman, van~der Walt, Gommers, Virtanen, Cournapeau, Wieser, Taylor, Berg, Smith, Kern, Picus, Hoyer, van Kerkwijk, Brett, Haldane, del R{\'{i}}o, Wiebe, Peterson, G{\'{e}}rard-Marchant, Sheppard, Reddy, Weckesser, Abbasi, Gohlke, \& Oliphant}]{numpy}
Harris, C.~R., Millman, K.~J., van~der Walt, S.~J., {et~al.} 2020, Nature, 585, 357, \dodoi{10.1038/s41586-020-2649-2}

\bibitem[{{Heckman}(1980)}]{heckman1980}
{Heckman}, T.~M. 1980, \aap, 87, 152

\bibitem[{{Hern{\'a}ndez-Garc{\'\i}a} {et~al.}(2025){Hern{\'a}ndez-Garc{\'\i}a}, {Chakraborty}, {S{\'a}nchez-S{\'a}ez}, {Ricci}, {Cuadra}, {McKernan}, {Ford}, {Ar{\'e}valo}, {Rau}, {Arcodia}, {Kara}, {Liu}, {Merloni}, {Bruni}, {Goodwin}, {Arzoumanian}, {Assef}, {Baldini}, {Bayo}, {Bauer}, {Bernal}, {Brightman}, {Calistro Rivera}, {Gendreau}, {Homan}, {Krumpe}, {Lira}, {Mart{\'\i}nez-Aldama}, {Salvato}, \& {Sotomayor}}]{hernandezgarcia2025}
{Hern{\'a}ndez-Garc{\'\i}a}, L., {Chakraborty}, J., {S{\'a}nchez-S{\'a}ez}, P., {et~al.} 2025, arXiv e-prints, arXiv:2504.07169, \dodoi{10.48550/arXiv.2504.07169}

\bibitem[{{Hickox} {et~al.}(2014){Hickox}, {Mullaney}, {Alexander}, {Chen}, {Civano}, {Goulding}, \& {Hainline}}]{hickox2014}
{Hickox}, R.~C., {Mullaney}, J.~R., {Alexander}, D.~M., {et~al.} 2014, \apj, 782, 9, \dodoi{10.1088/0004-637X/782/1/9}

\bibitem[{{Higson} {et~al.}(2019){Higson}, {Handley}, {Hobson}, \& {Lasenby}}]{higson2019}
{Higson}, E., {Handley}, W., {Hobson}, M., \& {Lasenby}, A. 2019, Statistics and Computing, 29, 891, \dodoi{10.1007/s11222-018-9844-0}

\bibitem[{{Holoien} {et~al.}(2016){Holoien}, {Kochanek}, {Prieto}, {Stanek}, {Dong}, {Shappee}, {Grupe}, {Brown}, {Basu}, {Beacom}, {Bersier}, {Brimacombe}, {Danilet}, {Falco}, {Guo}, {Jose}, {Herczeg}, {Long}, {Pojmanski}, {Simonian}, {Szczygie{\l}}, {Thompson}, {Thorstensen}, {Wagner}, \& {Wo{\'z}niak}}]{holoien2016}
{Holoien}, T.~W.~S., {Kochanek}, C.~S., {Prieto}, J.~L., {et~al.} 2016, \mnras, 455, 2918, \dodoi{10.1093/mnras/stv2486}

\bibitem[{{Hon} {et~al.}(2022){Hon}, {Wolf}, {Onken}, {Webster}, \& {Auchettl}}]{hon2022}
{Hon}, W.~J., {Wolf}, C., {Onken}, C.~A., {Webster}, R., \& {Auchettl}, K. 2022, \mnras, 511, 54, \dodoi{10.1093/mnras/stab3694}

\bibitem[{{Hopkins} {et~al.}(2005){Hopkins}, {Hernquist}, {Martini}, {Cox}, {Robertson}, {Di Matteo}, \& {Springel}}]{hopkins2005a}
{Hopkins}, P.~F., {Hernquist}, L., {Martini}, P., {et~al.} 2005, \apjl, 625, L71, \dodoi{10.1086/431146}

\bibitem[{Hunter(2007)}]{matplotlib}
Hunter, J.~D. 2007, Computing in Science \& Engineering, 9, 90, \dodoi{10.1109/MCSE.2007.55}

\bibitem[{{Jana} {et~al.}(2024){Jana}, {Ricci}, {Temple}, {Chang}, {Shablovinskaya}, {Trakhtenbrot}, {Diaz}, {Ilic}, {Nandi}, \& {Koss}}]{jana2024}
{Jana}, A., {Ricci}, C., {Temple}, M.~J., {et~al.} 2024, arXiv e-prints, arXiv:2411.08676.
\newblock \doarXiv{2411.08676}

\bibitem[{{Jiang} {et~al.}(2017){Jiang}, {Wang}, {Yan}, {Xiao}, {Yang}, {Dou}, {Wang}, {Cutri}, \& {Mainzer}}]{jiang2017}
{Jiang}, N., {Wang}, T., {Yan}, L., {et~al.} 2017, \apj, 850, 63, \dodoi{10.3847/1538-4357/aa93f5}

\bibitem[{{Jin} {et~al.}(2021){Jin}, {Ruan}, {Haggard}, {Gingras}, {Hountalas}, {MacLeod}, {Anderson}, {Doan}, {Eracleous}, {Green}, \& {Runnoe}}]{jin2021}
{Jin}, X., {Ruan}, J.~J., {Haggard}, D., {et~al.} 2021, ApJ, 912, 20, \dodoi{10.3847/1538-4357/abeb17}

\bibitem[{{Johnson} \& {Leja}(2017)}]{johnson17}
{Johnson}, B., \& {Leja}, J. 2017, {Bd-J/Prospector: Initial Release}, v0.1,  Zenodo, \dodoi{10.5281/zenodo.1116491}

\bibitem[{{Johnson} {et~al.}(2024){Johnson}, {Foreman-Mackey}, {Sick}, {Leja}, {Walmsley}, {Tollerud}, {Leung}, {Scott}, \& {Park}}]{pyfsps}
{Johnson}, B., {Foreman-Mackey}, D., {Sick}, J., {et~al.} 2024, {dfm/python-fsps: v0.4.7}, v0.4.7,  Zenodo, \dodoi{10.5281/zenodo.12447779}

\bibitem[{{Johnson}(2021)}]{sedpy}
{Johnson}, B.~D. 2021, {bd-j/sedpy: sedpy v0.2.0 (v0.2.0)}, Zenodo

\bibitem[{Johnson {et~al.}(2021)Johnson, Leja, Conroy, \& Speagle}]{johnson2021}
Johnson, B.~D., Leja, J., Conroy, C., \& Speagle, J.~S. 2021, The Astrophysical Journal Supplement Series, 254, 22, \dodoi{10.3847/1538-4365/abef67}

\bibitem[{{Jose} {et~al.}(2014){Jose}, {Guo}, {Long}, {Herczeg}, {Dong}, {Holoien}, {Prieto}, {Grupe}, {Shappee}, {Stanek}, {Kochanek}, {Davis}, {Simonian}, {Basu}, {Beacom}, {Bersier}, {Brimacombe}, {Szczygiel}, \& {Pojmanski}}]{Jose2014}
{Jose}, J., {Guo}, Z., {Long}, F., {et~al.} 2014, The Astronomer's Telegram, 6777, 1

\bibitem[{Kauffmann \& Heckman(2009)}]{kauffmann2009}
Kauffmann, G., \& Heckman, T.~M. 2009, MNRAS, 397, 135, \dodoi{10.1111/j.1365-2966.2009.14960.x}

\bibitem[{{Kauffmann} {et~al.}(2003){Kauffmann}, {Heckman}, {White}, {Charlot}, {Tremonti}, {Brinchmann}, {Bruzual}, {Peng}, {Seibert}, {Bernardi}, {Blanton}, {Brinkmann}, {Castander}, {Cs{\'a}bai}, {Fukugita}, {Ivezic}, {Munn}, {Nichol}, {Padmanabhan}, {Thakar}, {Weinberg}, \& {York}}]{Kauffmann2003a}
{Kauffmann}, G., {Heckman}, T.~M., {White}, S. D.~M., {et~al.} 2003, \mnras, 341, 33, \dodoi{10.1046/j.1365-8711.2003.06291.x}

\bibitem[{{Kaur} \& {Stone}(2025)}]{kaur2025}
{Kaur}, K., \& {Stone}, N.~C. 2025, \apj, 979, 172, \dodoi{10.3847/1538-4357/ad9b86}

\bibitem[{{Keel} {et~al.}(2012){Keel}, {Chojnowski}, {Bennert}, {Schawinski}, {Lintott}, {Lynn}, {Pancoast}, {Harris}, {Nierenberg}, {Sonnenfeld}, \& {Proctor}}]{keel2012}
{Keel}, W.~C., {Chojnowski}, S.~D., {Bennert}, V.~N., {et~al.} 2012, \mnras, 420, 878, \dodoi{10.1111/j.1365-2966.2011.20101.x}

\bibitem[{{Kennedy} {et~al.}(2016){Kennedy}, {Meiron}, {Shukirgaliyev}, {Panamarev}, {Berczik}, {Just}, \& {Spurzem}}]{Kennedy2016}
{Kennedy}, G.~F., {Meiron}, Y., {Shukirgaliyev}, B., {et~al.} 2016, \mnras, 460, 240, \dodoi{10.1093/mnras/stw908}

\bibitem[{{Kennicutt} \& {Evans}(2012)}]{kennicutt2012}
{Kennicutt}, R.~C., \& {Evans}, N.~J. 2012, \araa, 50, 531, \dodoi{10.1146/annurev-astro-081811-125610}

\bibitem[{{King} {et~al.}(2008){King}, {Pringle}, \& {Hofmann}}]{king2008}
{King}, A.~R., {Pringle}, J.~E., \& {Hofmann}, J.~A. 2008, \mnras, 385, 1621, \dodoi{10.1111/j.1365-2966.2008.12943.x}

\bibitem[{Kokubo \& Minezaki(2019)}]{kokubo2019}
Kokubo, M., \& Minezaki, T. 2019, MNRAS, 491, 4615, \dodoi{10.1093/mnras/stz3397}

\bibitem[{{Kormendy} \& {Ho}(2013)}]{kormendyho2013}
{Kormendy}, J., \& {Ho}, L.~C. 2013, \araa, 51, 511, \dodoi{10.1146/annurev-astro-082708-101811}

\bibitem[{{Kriek} \& {Conroy}(2013)}]{kriek2013}
{Kriek}, M., \& {Conroy}, C. 2013, \apjl, 775, L16, \dodoi{10.1088/2041-8205/775/1/L16}

\bibitem[{{LaMassa} {et~al.}(2015){LaMassa}, {Cales}, {Moran}, {Myers}, {Richards}, {Eracleous}, {Heckman}, {Gallo}, \& {Urry}}]{lamassa2015}
{LaMassa}, S.~M., {Cales}, S., {Moran}, E.~C., {et~al.} 2015, ApJ, 800, 144, \dodoi{10.1088/0004-637X/800/2/144}

\bibitem[{{Leja} {et~al.}(2019{\natexlab{a}}){Leja}, {Carnall}, {Johnson}, {Conroy}, \& {Speagle}}]{leja2019}
{Leja}, J., {Carnall}, A.~C., {Johnson}, B.~D., {Conroy}, C., \& {Speagle}, J.~S. 2019{\natexlab{a}}, \apj, 876, 3, \dodoi{10.3847/1538-4357/ab133c}

\bibitem[{{Leja} {et~al.}(2018){Leja}, {Johnson}, {Conroy}, \& {van Dokkum}}]{leja2018}
{Leja}, J., {Johnson}, B.~D., {Conroy}, C., \& {van Dokkum}, P. 2018, \apj, 854, 62, \dodoi{10.3847/1538-4357/aaa8db}

\bibitem[{{Leja} {et~al.}(2020){Leja}, {Johnson}, {Conroy}, {van Dokkum}, {Speagle}, \& {the 3D-HST Team}}]{leja2021_prospectormasses}
{Leja}, J., {Johnson}, B.~D., {Conroy}, C., {et~al.} 2020, in IAU Symposium, Vol. 352, Uncovering Early Galaxy Evolution in the ALMA and JWST Era, ed. E.~{da Cunha}, J.~{Hodge}, J.~{Afonso}, L.~{Pentericci}, \& D.~{Sobral}, 99--102, \dodoi{10.1017/S1743921319009025}

\bibitem[{{Leja} {et~al.}(2017){Leja}, {Johnson}, {Conroy}, {van Dokkum}, \& {Byler}}]{leja17}
{Leja}, J., {Johnson}, B.~D., {Conroy}, C., {van Dokkum}, P.~G., \& {Byler}, N. 2017, ApJ, 837, 170, \dodoi{10.3847/1538-4357/aa5ffe}

\bibitem[{{Leja} {et~al.}(2019{\natexlab{b}}){Leja}, {Johnson}, {Conroy}, {van Dokkum}, {Speagle}, {Brammer}, {Momcheva}, {Skelton}, {Whitaker}, {Franx}, \& {Nelson}}]{leja2019_3dhst}
{Leja}, J., {Johnson}, B.~D., {Conroy}, C., {et~al.} 2019{\natexlab{b}}, \apj, 877, 140, \dodoi{10.3847/1538-4357/ab1d5a}

\bibitem[{{Lintott} {et~al.}(2009){Lintott}, {Schawinski}, {Keel}, {van Arkel}, {Bennert}, {Edmondson}, {Thomas}, {Smith}, {Herbert}, {Jarvis}, {Virani}, {Andreescu}, {Bamford}, {Land}, {Murray}, {Nichol}, {Raddick}, {Slosar}, {Szalay}, \& {Vandenberg}}]{lintott2009}
{Lintott}, C.~J., {Schawinski}, K., {Keel}, W., {et~al.} 2009, \mnras, 399, 129, \dodoi{10.1111/j.1365-2966.2009.15299.x}

\bibitem[{{Liu} {et~al.}(2021){Liu}, {Lira}, {Yao}, {Xu}, {Wang}, {Dong}, \& {Mart{\'\i}nez-Palomera}}]{liu2021}
{Liu}, W.-J., {Lira}, P., {Yao}, S., {et~al.} 2021, \apj, 915, 63, \dodoi{10.3847/1538-4357/abf82c}

\bibitem[{{L{\'o}pez-Navas} {et~al.}(2022){L{\'o}pez-Navas}, {Mart{\'\i}nez-Aldama}, {Bernal}, {S{\'a}nchez-S{\'a}ez}, {Ar{\'e}valo}, {Graham}, {Hern{\'a}ndez-Garc{\'\i}a}, {Lira}, \& {Rojas Lobos}}]{lopeznavas2022}
{L{\'o}pez-Navas}, E., {Mart{\'\i}nez-Aldama}, M.~L., {Bernal}, S., {et~al.} 2022, \mnras, 513, L57, \dodoi{10.1093/mnrasl/slac033}

\bibitem[{{Madau} \& {Dickinson}(2014)}]{madau2014}
{Madau}, P., \& {Dickinson}, M. 2014, \araa, 52, 415, \dodoi{10.1146/annurev-astro-081811-125615}

\bibitem[{{Malizia} {et~al.}(1997){Malizia}, {Bassani}, {Stephen}, {Malaguti}, \& {Palumbo}}]{Malizia1997}
{Malizia}, A., {Bassani}, L., {Stephen}, J.~B., {Malaguti}, G., \& {Palumbo}, G.~G.~C. 1997, \apjs, 113, 311, \dodoi{10.1086/313057}

\bibitem[{{Martig} {et~al.}(2009){Martig}, {Bournaud}, {Teyssier}, \& {Dekel}}]{Martig2009}
{Martig}, M., {Bournaud}, F., {Teyssier}, R., \& {Dekel}, A. 2009, \apj, 707, 250, \dodoi{10.1088/0004-637X/707/1/250}

\bibitem[{{Martig} {et~al.}(2013){Martig}, {Crocker}, {Bournaud}, {Emsellem}, {Gabor}, {Alatalo}, {Blitz}, {Bois}, {Bureau}, {Cappellari}, {Davies}, {Davis}, {Dekel}, {de Zeeuw}, {Duc}, {Falc{\'o}n-Barroso}, {Khochfar}, {Krajnovi{\'c}}, {Kuntschner}, {Morganti}, {McDermid}, {Naab}, {Oosterloo}, {Sarzi}, {Scott}, {Serra}, {Griffin}, {Teyssier}, {Weijmans}, \& {Young}}]{Martig2013}
{Martig}, M., {Crocker}, A.~F., {Bournaud}, F., {et~al.} 2013, \mnras, 432, 1914, \dodoi{10.1093/mnras/sts594}

\bibitem[{{Masterson} {et~al.}(2024){Masterson}, {De}, {Panagiotou}, {Kara}, {Arcavi}, {Eilers}, {Frostig}, {Gezari}, {Grotova}, {Liu}, {Malyali}, {Meisner}, {Merloni}, {Newsome}, {Rau}, {Simcoe}, \& {van Velzen}}]{masterson2024}
{Masterson}, M., {De}, K., {Panagiotou}, C., {et~al.} 2024, \apj, 961, 211, \dodoi{10.3847/1538-4357/ad18bb}

\bibitem[{Matt {et~al.}(2003)Matt, Guainazzi, \& Maiolino}]{matt2003}
Matt, G., Guainazzi, M., \& Maiolino, R. 2003, MNRAS, 342, 422, \dodoi{10.1046/j.1365-8711.2003.06539.x}

\bibitem[{{Melchor} {et~al.}(2024){Melchor}, {Mockler}, {Naoz}, {Rose}, \& {Ramirez-Ruiz}}]{melchor2024}
{Melchor}, D., {Mockler}, B., {Naoz}, S., {Rose}, S.~C., \& {Ramirez-Ruiz}, E. 2024, \apj, 960, 39, \dodoi{10.3847/1538-4357/acfee0}

\bibitem[{{Ni} {et~al.}(2023){Ni}, {Aird}, {Merloni}, {Birchall}, {Buchner}, {Salvato}, \& {Yang}}]{ni2023}
{Ni}, Q., {Aird}, J., {Merloni}, A., {et~al.} 2023, \mnras, 524, 4778, \dodoi{10.1093/mnras/stad2070}

\bibitem[{{Noda} \& {Done}(2018)}]{noda2018}
{Noda}, H., \& {Done}, C. 2018, \mnras, 480, 3898, \dodoi{10.1093/mnras/sty2032}

\bibitem[{pandas~development team(2020)}]{pandas}
pandas~development team, T. 2020, pandas-dev/pandas: Pandas, latest,  Zenodo, \dodoi{10.5281/zenodo.3509134}

\bibitem[{{P{\^a}ris} {et~al.}(2012){P{\^a}ris}, {Petitjean}, {Aubourg}, {Bailey}, {Ross}, {Myers}, {Strauss}, {Anderson}, {Arnau}, {Bautista}, {Bizyaev}, {Bolton}, {Bovy}, {Brandt}, {Brewington}, {Browstein}, {Busca}, {Capellupo}, {Carithers}, {Croft}, {Dawson}, {Delubac}, {Ebelke}, {Eisenstein}, {Engelke}, {Fan}, {Filiz Ak}, {Finley}, {Font-Ribera}, {Ge}, {Gibson}, {Hall}, {Hamann}, {Hennawi}, {Ho}, {Hogg}, {Ivezi{\'c}}, {Jiang}, {Kimball}, {Kirkby}, {Kirkpatrick}, {Lee}, {Le Goff}, {Lundgren}, {MacLeod}, {Malanushenko}, {Malanushenko}, {Maraston}, {McGreer}, {McMahon}, {Miralda-Escud{\'e}}, {Muna}, {Noterdaeme}, {Oravetz}, {Palanque-Delabrouille}, {Pan}, {Perez-Fournon}, {Pieri}, {Richards}, {Rollinde}, {Sheldon}, {Schlegel}, {Schneider}, {Slosar}, {Shelden}, {Shen}, {Simmons}, {Snedden}, {Suzuki}, {Tinker}, {Viel}, {Weaver}, {Weinberg}, {White}, {Wood-Vasey}, \& {Y{\`e}che}}]{paris2012}
{P{\^a}ris}, I., {Petitjean}, P., {Aubourg}, {\'E}., {et~al.} 2012, \aap, 548, A66, \dodoi{10.1051/0004-6361/201220142}

\bibitem[{{Paxton} {et~al.}(2011){Paxton}, {Bildsten}, {Dotter}, {Herwig}, {Lesaffre}, \& {Timmes}}]{paxton2011}
{Paxton}, B., {Bildsten}, L., {Dotter}, A., {et~al.} 2011, \apjs, 192, 3, \dodoi{10.1088/0067-0049/192/1/3}

\bibitem[{{Paxton} {et~al.}(2013){Paxton}, {Cantiello}, {Arras}, {Bildsten}, {Brown}, {Dotter}, {Mankovich}, {Montgomery}, {Stello}, {Timmes}, \& {Townsend}}]{paxton2013}
{Paxton}, B., {Cantiello}, M., {Arras}, P., {et~al.} 2013, \apjs, 208, 4, \dodoi{10.1088/0067-0049/208/1/4}

\bibitem[{{Paxton} {et~al.}(2015){Paxton}, {Marchant}, {Schwab}, {Bauer}, {Bildsten}, {Cantiello}, {Dessart}, {Farmer}, {Hu}, {Langer}, {Townsend}, {Townsley}, \& {Timmes}}]{paxton2015}
{Paxton}, B., {Marchant}, P., {Schwab}, J., {et~al.} 2015, \apjs, 220, 15, \dodoi{10.1088/0067-0049/220/1/15}

\bibitem[{{Peng} {et~al.}(2010){Peng}, {Lilly}, {Kova{\v{c}}}, {Bolzonella}, {Pozzetti}, {Renzini}, {Zamorani}, {Ilbert}, {Knobel}, {Iovino}, {Maier}, {Cucciati}, {Tasca}, {Carollo}, {Silverman}, {Kampczyk}, {de Ravel}, {Sanders}, {Scoville}, {Contini}, {Mainieri}, {Scodeggio}, {Kneib}, {Le F{\`e}vre}, {Bardelli}, {Bongiorno}, {Caputi}, {Coppa}, {de la Torre}, {Franzetti}, {Garilli}, {Lamareille}, {Le Borgne}, {Le Brun}, {Mignoli}, {Perez Montero}, {Pello}, {Ricciardelli}, {Tanaka}, {Tresse}, {Vergani}, {Welikala}, {Zucca}, {Oesch}, {Abbas}, {Barnes}, {Bordoloi}, {Bottini}, {Cappi}, {Cassata}, {Cimatti}, {Fumana}, {Hasinger}, {Koekemoer}, {Leauthaud}, {Maccagni}, {Marinoni}, {McCracken}, {Memeo}, {Meneux}, {Nair}, {Porciani}, {Presotto}, \& {Scaramella}}]{peng2010}
{Peng}, Y.-j., {Lilly}, S.~J., {Kova{\v{c}}}, K., {et~al.} 2010, \apj, 721, 193, \dodoi{10.1088/0004-637X/721/1/193}

\bibitem[{{Prieto} {et~al.}(2016){Prieto}, {Kr{\"u}hler}, {Anderson}, {Galbany}, {Kochanek}, {Aquino}, {Brown}, {Dong}, {F{\"o}rster}, {Holoien}, {Kuncarayakti}, {Maureira}, {Rosales-Ortega}, {S{\'a}nchez}, {Shappee}, \& {Stanek}}]{Prieto2016}
{Prieto}, J.~L., {Kr{\"u}hler}, T., {Anderson}, J.~P., {et~al.} 2016, \apjl, 830, L32, \dodoi{10.3847/2041-8205/830/2/L32}

\bibitem[{{Reynolds} {et~al.}(2022){Reynolds}, {Mattila}, {Efstathiou}, {Kankare}, {Kool}, {Ryder}, {Pe{\~n}a-Mo{\~n}ino}, \& {P{\'e}rez-Torres}}]{reynolds2022}
{Reynolds}, T.~M., {Mattila}, S., {Efstathiou}, A., {et~al.} 2022, \aap, 664, A158, \dodoi{10.1051/0004-6361/202243289}

\bibitem[{Ricci \& Trakhtenbrot(2022)}]{ricci2022}
Ricci, C., \& Trakhtenbrot, B. 2022, Changing-look Active Galactic Nuclei, \dodoi{10.48550/ARXIV.2211.05132}

\bibitem[{{Risaliti} {et~al.}(2007){Risaliti}, {Elvis}, {Fabbiano}, {Baldi}, {Zezas}, \& {Salvati}}]{Risaliti2007}
{Risaliti}, G., {Elvis}, M., {Fabbiano}, G., {et~al.} 2007, \apjl, 659, L111, \dodoi{10.1086/517884}

\bibitem[{{Roth} {et~al.}(2021){Roth}, {van Velzen}, {Cenko}, \& {Mushotzky}}]{Roth2021}
{Roth}, N., {van Velzen}, S., {Cenko}, S.~B., \& {Mushotzky}, R.~F. 2021, \apj, 910, 93, \dodoi{10.3847/1538-4357/abdf50}

\bibitem[{{Ruan} {et~al.}(2016){Ruan}, {Anderson}, {Cales}, {Eracleous}, {Green}, {Morganson}, {Runnoe}, {Shen}, {Wilkinson}, {Blanton}, {Dwelly}, {Georgakakis}, {Greene}, {LaMassa}, {Merloni}, \& {Schneider}}]{ruan2016}
{Ruan}, J.~J., {Anderson}, S.~F., {Cales}, S.~L., {et~al.} 2016, \apj, 826, 188, \dodoi{10.3847/0004-637X/826/2/188}

\bibitem[{{Ryu} {et~al.}(2024){Ryu}, {McKernan}, {Ford}, {Cantiello}, {Graham}, {Stern}, \& {Leigh}}]{ryu2024}
{Ryu}, T., {McKernan}, B., {Ford}, K.~E.~S., {et~al.} 2024, \mnras, 527, 8103, \dodoi{10.1093/mnras/stad3487}

\bibitem[{{Salim}(2014)}]{salim2014}
{Salim}, S. 2014, Serbian Astronomical Journal, 189, 1, \dodoi{10.2298/SAJ1489001S}

\bibitem[{{S{\'a}nchez-Bl{\'a}zquez} {et~al.}(2006){S{\'a}nchez-Bl{\'a}zquez}, {Gorgas}, {Cardiel}, \& {Gonz{\'a}lez}}]{sanchez-blazquez2006}
{S{\'a}nchez-Bl{\'a}zquez}, P., {Gorgas}, J., {Cardiel}, N., \& {Gonz{\'a}lez}, J.~J. 2006, \aap, 457, 809, \dodoi{10.1051/0004-6361:20064845}

\bibitem[{{Schawinski} {et~al.}(2015){Schawinski}, {Koss}, {Berney}, \& {Sartori}}]{Schawinski2015}
{Schawinski}, K., {Koss}, M., {Berney}, S., \& {Sartori}, L.~F. 2015, \mnras, 451, 2517, \dodoi{10.1093/mnras/stv1136}

\bibitem[{Schawinski {et~al.}(2014)Schawinski, Urry, Simmons, Fortson, Kaviraj, Keel, Lintott, Masters, Nichol, Sarzi, Skibba, Treister, Willett, Wong, \& Yi}]{schawinski2014}
Schawinski, K., Urry, C.~M., Simmons, B.~D., {et~al.} 2014, MNRAS, 440, 889, \dodoi{10.1093/mnras/stu327}

\bibitem[{{Schneider} {et~al.}(2010){Schneider}, {Richards}, {Hall}, {Strauss}, {Anderson}, {Boroson}, {Ross}, {Shen}, {Brandt}, {Fan}, {Inada}, {Jester}, {Knapp}, {Krawczyk}, {Thakar}, {Vanden Berk}, {Voges}, {Yanny}, {York}, {Bahcall}, {Bizyaev}, {Blanton}, {Brewington}, {Brinkmann}, {Eisenstein}, {Frieman}, {Fukugita}, {Gray}, {Gunn}, {Hibon}, {Ivezi{\'c}}, {Kent}, {Kron}, {Lee}, {Lupton}, {Malanushenko}, {Malanushenko}, {Oravetz}, {Pan}, {Pier}, {Price}, {Saxe}, {Schlegel}, {Simmons}, {Snedden}, {SubbaRao}, {Szalay}, \& {Weinberg}}]{sdssqso}
{Schneider}, D.~P., {Richards}, G.~T., {Hall}, P.~B., {et~al.} 2010, \aj, 139, 2360, \dodoi{10.1088/0004-6256/139/6/2360}

\bibitem[{{Shen} {et~al.}(2011){Shen}, {Richards}, {Strauss}, {Hall}, {Schneider}, {Snedden}, {Bizyaev}, {Brewington}, {Malanushenko}, {Malanushenko}, {Oravetz}, {Pan}, \& {Simmons}}]{shen2011}
{Shen}, Y., {Richards}, G.~T., {Strauss}, M.~A., {et~al.} 2011, \apjs, 194, 45, \dodoi{10.1088/0067-0049/194/2/45}

\bibitem[{Skilling(2006)}]{Skilling2006}
Skilling, J. 2006, Bayesian Analysis, 1, 833 , \dodoi{10.1214/06-BA127}

\bibitem[{{S}niegowska {et~al.}(2021){S}niegowska, Grzkedzielski, Czerny, \& Janiuk}]{sniegowska2021}
{S}niegowska, M., Grzkedzielski, M., Czerny, B., \& Janiuk, A. 2021, Astronomische Nachrichten, 343, \dodoi{10.1002/asna.20210065}

\bibitem[{{Speagle}(2020)}]{speagle2020}
{Speagle}, J.~S. 2020, \mnras, 493, 3132, \dodoi{10.1093/mnras/staa278}

\bibitem[{{Stone} {et~al.}(2018){Stone}, {Generozov}, {Vasiliev}, \& {Metzger}}]{stone2018}
{Stone}, N.~C., {Generozov}, A., {Vasiliev}, E., \& {Metzger}, B.~D. 2018, \mnras, 480, 5060, \dodoi{10.1093/mnras/sty2045}

\bibitem[{{Stone} {et~al.}(2020){Stone}, {Vasiliev}, {Kesden}, {Rossi}, {Perets}, \& {Amaro-Seoane}}]{stone2020}
{Stone}, N.~C., {Vasiliev}, E., {Kesden}, M., {et~al.} 2020, \ssr, 216, 35, \dodoi{10.1007/s11214-020-00651-4}

\bibitem[{{Suess} {et~al.}(2022{\natexlab{a}}){Suess}, {Leja}, {Johnson}, {Bezanson}, {Greene}, {Kriek}, {Lower}, {Narayanan}, {Setton}, \& {Spilker}}]{suess2022c}
{Suess}, K.~A., {Leja}, J., {Johnson}, B.~D., {et~al.} 2022{\natexlab{a}}, \apj, 935, 146, \dodoi{10.3847/1538-4357/ac82b0}

\bibitem[{{Suess} {et~al.}(2022{\natexlab{b}}){Suess}, {Leja}, {Johnson}, {Bezanson}, {Greene}, {Kriek}, {Lower}, {Narayanan}, {Setton}, \& {Spilker}}]{suess2022b}
---. 2022{\natexlab{b}}, ApJ, 935, 146, \dodoi{10.3847/1538-4357/ac82b0}

\bibitem[{{Teboul} \& {Perets}(2025)}]{teboul2025}
{Teboul}, O., \& {Perets}, H.~B. 2025, \apj, 984, 12, \dodoi{10.3847/1538-4357/adc09f}

\bibitem[{{Terrazas} {et~al.}(2016){Terrazas}, {Bell}, {Henriques}, {White}, {Cattaneo}, \& {Woo}}]{terrazas2016}
{Terrazas}, B.~A., {Bell}, E.~F., {Henriques}, B. M.~B., {et~al.} 2016, \apjl, 830, L12, \dodoi{10.3847/2041-8205/830/1/L12}

\bibitem[{{Tozzi} {et~al.}(2022){Tozzi}, {Lusso}, {Casetti}, {Romoli}, {Andreuzzi}, {Montoya Arroyave}, {Nardini}, {Cresci}, {Middei}, {Bertolini}, {Calabretto}, {Cammelli}, {Cuadra}, {Dalla Ragione}, {Marconcini}, {Miceli}, {Mini}, {Palazzini}, {Rotellini}, {Saccardi}, {Sam{\`a}}, {Sangalli}, {Serafini}, \& {Spaccino}}]{tozzi2022}
{Tozzi}, G., {Lusso}, E., {Casetti}, L., {et~al.} 2022, \aap, 667, L12, \dodoi{10.1051/0004-6361/202244987}

\bibitem[{{Trakhtenbrot} {et~al.}(2019){Trakhtenbrot}, {Arcavi}, {MacLeod}, {Ricci}, {Kara}, {Graham}, {Stern}, {Harrison}, {Burke}, {Hiramatsu}, {Hosseinzadeh}, {Howell}, {Smartt}, {Rest}, {Prieto}, {Shappee}, {Holoien}, {Bersier}, {Filippenko}, {Brink}, {Zheng}, {Li}, {Remillard}, \& {Loewenstein}}]{Trakhtenbrot2019}
{Trakhtenbrot}, B., {Arcavi}, I., {MacLeod}, C.~L., {et~al.} 2019, \apj, 883, 94, \dodoi{10.3847/1538-4357/ab39e4}

\bibitem[{{Tremonti} {et~al.}(2004){Tremonti}, {Heckman}, {Kauffmann}, {Brinchmann}, {Charlot}, {White}, {Seibert}, {Peng}, {Schlegel}, {Uomoto}, {Fukugita}, \& {Brinkmann}}]{tremonti2004}
{Tremonti}, C.~A., {Heckman}, T.~M., {Kauffmann}, G., {et~al.} 2004, \apj, 613, 898, \dodoi{10.1086/423264}

\bibitem[{{Urry} \& {Padovani}(1995)}]{urry1995}
{Urry}, C.~M., \& {Padovani}, P. 1995, \pasp, 107, 803, \dodoi{10.1086/133630}

\bibitem[{{V{\'e}ron-Cetty} \& {V{\'e}ron}(2010)}]{veroncetty2010}
{V{\'e}ron-Cetty}, M.~P., \& {V{\'e}ron}, P. 2010, \aap, 518, A10, \dodoi{10.1051/0004-6361/201014188}

\bibitem[{{Veronese} {et~al.}(2024){Veronese}, {Vignali}, {Severgnini}, {Matzeu}, \& {Cignoni}}]{veronese2024}
{Veronese}, S., {Vignali}, C., {Severgnini}, P., {Matzeu}, G.~A., \& {Cignoni}, M. 2024, \aap, 683, A131, \dodoi{10.1051/0004-6361/202348098}

\bibitem[{{Vijarnwannaluk} {et~al.}(2024){Vijarnwannaluk}, {Akiyama}, {Schramm}, {Ueda}, {Matsuoka}, {Toba}, {Matsumoto}, {Ruiz}, {Georgantopoulos}, {Pouliasis}, {Koulouridis}, {Ichikawa}, {Sawicki}, \& {Gwyn}}]{Vijarnwannaluk2024}
{Vijarnwannaluk}, B., {Akiyama}, M., {Schramm}, M., {et~al.} 2024, \mnras, 529, 3610, \dodoi{10.1093/mnras/stae728}

\bibitem[{Virtanen {et~al.}(2020)Virtanen, Gommers, Oliphant, Haberland, Reddy, Cournapeau, Burovski, Peterson, Weckesser, Bright, {van der Walt}, Brett, Wilson, Millman, Mayorov, Nelson, Jones, Kern, Larson, Carey, Polat, Feng, Moore, {VanderPlas}, Laxalde, Perktold, Cimrman, Henriksen, Quintero, Harris, Archibald, Ribeiro, Pedregosa, {van Mulbregt}, \& {SciPy 1.0 Contributors}}]{scipy}
Virtanen, P., Gommers, R., Oliphant, T.~E., {et~al.} 2020, Nature Methods, 17, 261, \dodoi{10.1038/s41592-019-0686-2}

\bibitem[{{Wang} {et~al.}(2024{\natexlab{a}}){Wang}, {Xu}, {Cao}, {Gao}, {Xie}, \& {Wei}}]{wangj2024}
{Wang}, J., {Xu}, D.~W., {Cao}, X., {et~al.} 2024{\natexlab{a}}, arXiv e-prints, arXiv:2405.10663, \dodoi{10.48550/arXiv.2405.10663}

\bibitem[{{Wang} {et~al.}(2023){Wang}, {Zheng}, {Brink}, {Xu}, {Filippenko}, {Gao}, {Xie}, \& {Wei}}]{wang2023}
{Wang}, J., {Zheng}, W.~K., {Brink}, T.~G., {et~al.} 2023, arXiv e-prints, arXiv:2308.16521, \dodoi{10.48550/arXiv.2308.16521}

\bibitem[{{Wang} {et~al.}(2024{\natexlab{b}}){Wang}, {Ma}, {Wu}, \& {Jiang}}]{wangm2024}
{Wang}, M., {Ma}, Y., {Wu}, Q., \& {Jiang}, N. 2024{\natexlab{b}}, \apj, 960, 69, \dodoi{10.3847/1538-4357/ad0bfb}

\bibitem[{{Wang} {et~al.}(2024{\natexlab{c}}){Wang}, {Woo}, {Gallo}, {Guo}, {Son}, {Kong}, {Mandal}, {Cho}, {Kim}, \& {Shin}}]{wang2024}
{Wang}, S., {Woo}, J.-H., {Gallo}, E., {et~al.} 2024{\natexlab{c}}, arXiv e-prints, arXiv:2402.18131, \dodoi{10.48550/arXiv.2402.18131}

\bibitem[{{Wang} {et~al.}(2024{\natexlab{d}}){Wang}, {Lin}, {Zhang}, \& {Zhu}}]{wangyihan2024a}
{Wang}, Y., {Lin}, D. N.~C., {Zhang}, B., \& {Zhu}, Z. 2024{\natexlab{d}}, \apjl, 962, L7, \dodoi{10.3847/2041-8213/ad20e5}

\bibitem[{{Ward} {et~al.}(2022){Ward}, {Harrison}, {Costa}, \& {Mainieri}}]{ward2022}
{Ward}, S.~R., {Harrison}, C.~M., {Costa}, T., \& {Mainieri}, V. 2022, \mnras, 514, 2936, \dodoi{10.1093/mnras/stac1219}

\bibitem[{Waskom(2021)}]{seaborn}
Waskom, M.~L. 2021, Journal of Open Source Software, 6, 3021, \dodoi{10.21105/joss.03021}

\bibitem[{{Weinmann} {et~al.}(2006){Weinmann}, {van den Bosch}, {Yang}, \& {Mo}}]{Weinmann2006}
{Weinmann}, S.~M., {van den Bosch}, F.~C., {Yang}, X., \& {Mo}, H.~J. 2006, \mnras, 366, 2, \dodoi{10.1111/j.1365-2966.2005.09865.x}

\bibitem[{{W}es {M}c{K}inney(2010)}]{pandas2}
{W}es {M}c{K}inney. 2010, in {P}roceedings of the 9th {P}ython in {S}cience {C}onference, ed. {S}t\'efan van~der {W}alt \& {J}arrod {M}illman, 56 -- 61, \dodoi{10.25080/Majora-92bf1922-00a}

\bibitem[{{Wevers} {et~al.}(2024){Wevers}, {French}, {Zabludoff}, {Fischer}, {Rowlands}, {Guolo}, {Dalla Barba}, {Arcodia}, {Berton}, {Bian}, {Linial}, {Miniutti}, \& {Pasham}}]{wevers2024}
{Wevers}, T., {French}, K.~D., {Zabludoff}, A.~I., {et~al.} 2024, \apjl, 970, L23, \dodoi{10.3847/2041-8213/ad5f1b}

\bibitem[{{Whitaker} {et~al.}(2012){Whitaker}, {van Dokkum}, {Brammer}, \& {Franx}}]{Whitaker2012b}
{Whitaker}, K.~E., {van Dokkum}, P.~G., {Brammer}, G., \& {Franx}, M. 2012, \apjl, 754, L29, \dodoi{10.1088/2041-8205/754/2/L29}

\bibitem[{{Worthey} \& {Ottaviani}(1997)}]{Worthey1997}
{Worthey}, G., \& {Ottaviani}, D.~L. 1997, \apjs, 111, 377, \dodoi{10.1086/313021}

\bibitem[{{Yang} {et~al.}(2025){Yang}, {Green}, {Wu}, {Eracleous}, {Jiang}, \& {Fu}}]{yang2024}
{Yang}, Q., {Green}, P.~J., {Wu}, X.-B., {et~al.} 2025, \apj, 980, 91, \dodoi{10.3847/1538-4357/ad94ed}

\bibitem[{{Yang} {et~al.}(2018){Yang}, {Wu}, {Fan}, {Jiang}, {McGreer}, {Shangguan}, {Yao}, {Wang}, {Joshi}, {Green}, {Wang}, {Feng}, {Fu}, {Yang}, \& {Liu}}]{yang2018}
{Yang}, Q., {Wu}, X.-B., {Fan}, X., {et~al.} 2018, ApJ, 862, 109, \dodoi{10.3847/1538-4357/aaca3a}

\bibitem[{{Yu} {et~al.}(2020){Yu}, {Shi}, {Chen}, {Chen}, {Li}, {Bing}, {Ge}, {Riffel}, \& {Riffel}}]{yu2020}
{Yu}, X., {Shi}, Y., {Chen}, Y., {et~al.} 2020, MNRAS, 498, 3985, \dodoi{10.1093/mnras/staa2627}

\bibitem[{{Zeltyn} {et~al.}(2022){Zeltyn}, {Trakhtenbrot}, {Eracleous}, {Runnoe}, {Trump}, {Stern}, {Shen}, {Hern{\'a}ndez-Garc{\'\i}a}, {Bauer}, {Yang}, {Dwelly}, {Ricci}, {Green}, {Anderson}, {Assef}, {Guolo}, {MacLeod}, {Davis}, {Fries}, {Gezari}, {Grogin}, {Homan}, {Koekemoer}, {Krumpe}, {LaMassa}, {Liu}, {Merloni}, {Mart{\'\i}nez-Aldama}, {Schneider}, {Temple}, {Brownstein}, {Ibarra-Medel}, {Burke}, {Pellegrino}, \& {Kollmeier}}]{zeltyn2022}
{Zeltyn}, G., {Trakhtenbrot}, B., {Eracleous}, M., {et~al.} 2022, ApJl, 939, L16, \dodoi{10.3847/2041-8213/ac9a47}

\bibitem[{{Zeltyn} {et~al.}(2024){Zeltyn}, {Trakhtenbrot}, {Eracleous}, {Yang}, {Green}, {Anderson}, {LaMassa}, {Runnoe}, {Assef}, {Bauer}, {Brandt}, {Davis}, {Frederick}, {Fries}, {Graham}, {Grogin}, {Guolo}, {Hern{\'a}ndez-Garc{\'\i}a}, {Koekemoer}, {Krumpe}, {Liu}, {Mart{\'\i}nez-Aldama}, {Ricci}, {Schneider}, {Shen}, {{\'S}niegowska}, {Temple}, {Trump}, {Xue}, {Brownstein}, {Dwelly}, {Morrison}, {Bizyaev}, {Pan}, \& {Kollmeier}}]{zeltyn2024}
---. 2024, \apj, 966, 85, \dodoi{10.3847/1538-4357/ad2f30}

\bibitem[{{Zhu} {et~al.}(2024){Zhu}, {Li}, {Wang}, \& {Zhang}}]{zhu2024}
{Zhu}, L.-T., {Li}, J., {Wang}, Z., \& {Zhang}, J.-J. 2024, \mnras, 530, 3538, \dodoi{10.1093/mnras/stae1044}

\end{thebibliography}
\bibliographystyle{aasjournal}

\newpage

\appendix

\section{Star formation rate calibration} \label{sec:sfrcalib}

Due to the excessively variable nature of CL-AGN, we cannot confirm that the SDSS photometry and spectroscopy for any CL-AGN host should have the same colors and brightnesses unless the data are taken quasi-simultaneously.  We therefore cannot calibrate individual spectra used in our \texttt{Prospector} fitting without potentially recovering inaccurate host galaxy properties due to varying levels of AGN contamination.

We instead obtain a median correction factor on \texttt{Prospector} output parameters using our comparison galaxy samples.  We run an additional fit on the gold star forming and quiescent galaxy comparison samples at all redshifts, this time incorporating the built-in PolySpecModel object from \texttt{Prospector}, which optimizes out a polynomial to calibrate the spectrum and photometry for a given object.  As these galaxies samples do not include AGN, these calibrations should hold regardless of the time between observations.  

\begin{figure}[ht!]
    \centering
    \includegraphics[width=.5\textwidth]{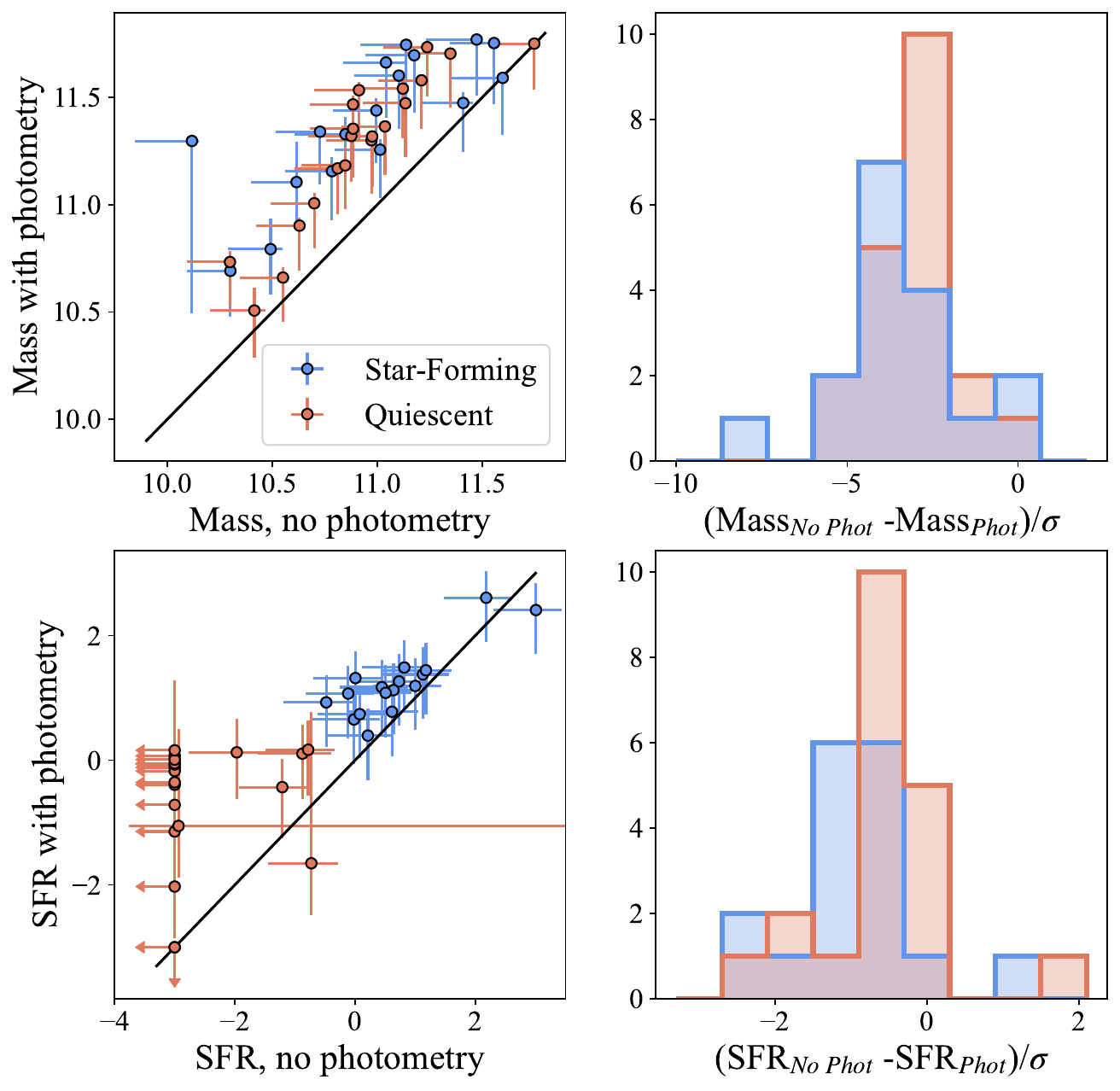}
    \caption{Results of our fitting with and without calibrating a polynomial fit to the photometry. (Top) Our recovered surviving stellar masses are higher by a median value of 0.37 dex when we include photometry in the fit.  (Bottom)  Our recovered star formation rates are higher by a median value of 0.66 dex when we include photometry in the fit.}
    \label{fig:sfrcalib}
\end{figure}

We find that the recovered stellar masses for the calibrated spectra are a median of 0.37 dex higher than the recovered stellar masses for the uncalibrated spectra.  The recovered star formation rates are a median of 0.66 dex higher for the calibrated spectra than the uncalibrated spectra (Figure \ref{fig:sfrcalib}).  We therefore apply an across-the-board 0.37 dex increase for all masses and 0.66 dex increase for all SFRs stated in this paper.

This calibration effect does not impact the relative position of CL-AGN to the comparison samples on a mass-sSFR diagram, and therefore does not impact our finding that CL-AGN are statistically indistinct from other AGN host galaxies in sSFR.  However, it does impact the classification of CL-AGN and our comparison samples relative to the star-forming main sequence. To recover the uncalibrated \texttt{Prospector} SFRs from the tables in Section \ref{sec:fittable}, subtract 0.37 from the given value for Log(M$_*$) and 0.66 from the given value for Log(SFR).  


\section{Testing our comparison sample selection} \label{sec:compare}

Three of the comparison samples analyzed in this work (star-forming and quiescent galaxies and low-luminosity Seyfert 2 hosts) are matched to CL-AGN hosts in mass and redshift. We use the outputted surviving mass from \texttt{Prospector} and the \texttt{lgm\_tot\_p50} mass from the MPA-JHU catalog \citep{Brinchmann2004} for this matching. However, the outputted mass values from \texttt{Prospector} have on average a 0.30 dex uncertainty. This uncertainty comes from both the width of the posterior distribution of recovered stellar masses and from the systematic error calculated as described in Section \ref{sec:AGNimpact}. This error on the recovered stellar mass may impact the comparison sample matching process and therefore our results.

Ideally, we would consider a large number of comparison objects to mitigate the effects of this error. Unfortunately, stellar population synthesis with \texttt{Prospector} and \texttt{Dynesty} dynamic nested sampling for stellar population synthesis, while efficient relative to other SPS methods \citep{leja2019_3dhst}, is computationally expensive. We therefore want to minimize the number of comparison objects to keep computation time reasonable. To that end, we test the impact of including a larger comparison sample on our gold, $\textrm{z}<0.15$ CL-AGN sample, which includes 14 objects.

To perform this test, we first remove the star-forming, quiescent, and low-luminosity Seyfert 2 comparison objects from Figure \ref{fig:compare} from our parent MPA-JHU catalog. We then re-perform a Euclidean distance match as described in Section \ref{sec:comparisonsample} to find the second-closest match in mass and redshift space to each CL-AGN. We use our \texttt{Prospector} pipeline to recover stellar masses, star formation rates, and star formation histories for these samples. We combine those results with the results of the original runs such that each comparison sample now contains two comparison objects per CL-AGN. This means that for 14 gold-sample, z$<$0.15 CL-AGN, we now have 28 star-forming galaxies, 28 quiescent galaxies, and 28 low-luminosity Seyfert 2 hosts for our comparison samples.

We use the bootstrapped Kolmogorov-Smirnov test described in Section \ref{sec:sfr} to determine whether CL-AGN are statistically distinct from star-forming or quiescent galaxies or low-luminosity Seyfert 2 hosts in sSFR, time since half of the mass formed ($t_{50}$) and time since 90\% of the mass formed ($t_{90}$). Under these tests, our results do not change: Gold-sample CL-AGN at $z < 0.15$ are still statistically distinct from star-forming galaxies in sSFR, $t_{50}$ and $t_{90}$. They are still statistically distinct from quiescent galaxies in sSFR and are not statistically distinct from quiescent galaxies in $t_{50}$ and $t_{90}$. They are not statistically distinct from low-luminosity Seyfert 2 hosts in any of these quantities. We conclude that adding more comparison galaxies to our mass-matched star-forming, quiescent, and low-luminosity Seyfert 2 host galaxy samples do not change our results.


\pagestyle{empty}

\section{Results from \texttt{Prospector} fits} \label{sec:fittable}

\renewcommand{\tabcolsep}{4pt}
\begin{sidewaystable}
\centering

\begin{tabular}{cccccccccccccccccc}
R.A. & Dec. & Redshift & Source (a) & On/Off (b & Epoch & Log(M$_*$) & $\sigma_{Mass}$ & Log(SFR) & $\sigma_{SFR}$ & $t_{50}$ & $\sigma_{t_{50}}$ & $t_{90}$ & $\sigma_{t_{90}}$ & Dust  & $\sigma_{Dust}$ & Log(Z) & $\sigma_Z$ \\
\hline \\
(J200) & (J200) & (J200) &  &  &   [MJD] & [M$\odot$] & [M$\odot$] & [M$_\odot$ yr$^{-1}$] & [M$_\odot$ yr$^{-1}$] & [Gyr] & [Gyr] &  [Gyr] & [Gyr] & [$\tau_{5500 \si{\angstrom}}$] & [$\tau_{5500 \si{\angstrom}}$] & [Z$_\odot$] & [Z$_\odot$] \\
\hline \\
12:10:49.60 & +39:28:22.1 & 0.0226 & 7 & 1 & 53415 & 10.81 & 0.15 & 0.15 & 0.53 & 4.93 & 3.99 & 4.17 & 2.43 & 0.97 & 0.58 & -0.46 & 0.69 \\
15:33:08.01 & +44:32:08.2 & 0.0367 & 4 & 1 & 52446 & 11.03 & 0.15 & -3.59 & 3.76 & 11.34 & 3.99 & 10.05 & 2.43 & 0.58 & 0.58 & -0.7 & 0.74 \\
11:33:55.83 & +67:01:08.0 & 0.0397 & 4 & 1 & 51955 & 11.0 & 0.15 & -0.02 & 0.53 & 11.17 & 3.99 & 9.78 & 2.44 & 0.59 & 0.58 & -0.57 & 0.69 \\
08:17:26.42 & +10:12:10.1 & 0.0458 & 4 & 1 & 54149 & 10.47 & 0.15 & 0.04 & 0.53 & 10.73 & 3.99 & 5.78 & 2.69 & 0.49 & 0.59 & -0.59 & 0.7 \\
08:37:16.47 & +03:56:06.1 & 0.0635 & 12 & 1 & 52646 & 11.19 & 0.15 & -0.9 & 0.86 & 11.14 & 3.99 & 10.01 & 2.43 & 0.67 & 0.58 & -0.66 & 0.69 \\
12:54:03.80 & +49:14:52.9 & 0.067 & 4 & 1 & 52736 & 11.46 & 0.15 & -4.74 & 12.36 & 10.93 & 3.99 & 9.96 & 2.43 & 0.62 & 0.58 & -0.55 & 0.71 \\
13:03:39.71 & +19:01:20.9 & 0.0822 & 8 & 1 & 54499 & 11.34 & 0.16 & 0.89 & 0.53 & 10.28 & 4.48 & 7.26 & 3.04 & 1.08 & 0.58 & -0.51 & 0.7 \\
14:05:15.59 & +54:24:58.0 & 0.0833 & 12 & 1 & 53088 & 11.29 & 0.16 & 0.72 & 0.53 & 7.09 & 4.1 & 4.49 & 2.5 & 0.9 & 0.58 & -0.55 & 0.7 \\
11:15:36.57 & +05:44:49.7 & 0.09 & 3 & 1 & 52326 & 11.27 & 0.15 & -0.47 & 0.54 & 10.77 & 3.99 & 9.86 & 2.43 & 0.58 & 0.58 & -0.54 & 0.7 \\
09:15:31.06 & +48:14:08.0 & 0.1005 & 4 & 1 & 52637 & 11.57 & 0.15 & -3.97 & 14.44 & 6.02 & 4.1 & 4.27 & 2.43 & 0.74 & 0.58 & -0.69 & 0.73 \\
10:03:23.47 & +35:25:03.8 & 0.1189 & 3 & 1 & 53389 & 11.47 & 0.15 & 0.15 & 0.53 & 9.73 & 4.14 & 6.53 & 2.46 & 0.56 & 0.58 & -0.65 & 0.71 \\
22:10:44.76 & +24:59:58.0 & 0.1199 & 9 & 1 & 56213 & 11.75 & 0.15 & 1.02 & 0.53 & 10.65 & 3.99 & 8.53 & 3.12 & 1.46 & 0.58 & -0.58 & 0.7 \\
01:42:09.02 & -00:50:50.0 & 0.1325 & 12 & 1 & 51788 & 11.56 & 0.15 & 0.84 & 0.53 & 10.37 & 3.99 & 3.93 & 2.53 & 0.64 & 0.58 & -0.53 & 0.72 \\
09:14:59.16 & +05:02:43.4 & 0.1425 & 12 & 1 & 52652 & 11.74 & 0.16 & 0.48 & 0.53 & 3.79 & 4.0 & 3.1 & 2.45 & 1.09 & 0.59 & -0.61 & 0.75 \\
08:45:28.72 & -00:27:23.3 & 0.1544 & 12 & 2 & 59222 & 11.9 & 0.15 & 0.87 & 0.53 & 7.6 & 3.99 & 5.88 & 2.45 & 0.0 & 0.57 & 0.8 & 0.69 \\
11:04:23.21 & +63:43:05.3 & 0.1643 & 3 & 2 & 54498 & 10.82 & 0.15 & -0.05 & 0.53 & 5.06 & 4.0 & 4.14 & 2.43 & 0.2 & 0.61 & -0.85 & 0.87 \\
13:19:30.75 & +67:53:55.4 & 0.1664 & 3 & 1 & 51988 & 11.41 & 0.15 & 0.93 & 0.53 & 6.42 & 4.27 & 2.99 & 2.47 & 0.68 & 0.59 & -0.51 & 0.73 \\
12:25:50.31 & +51:08:46.5 & 0.168 & 4 & 1 & 52664 & 11.91 & 0.15 & 0.47 & 0.62 & 10.93 & 4.0 & 8.53 & 3.15 & 0.79 & 0.59 & -0.57 & 0.77 \\
08:31:32.25 & +36:46:17.2 & 0.195 & 3 & 1 & 52312 & 11.71 & 0.16 & 1.16 & 0.53 & 6.41 & 4.3 & 4.31 & 2.43 & 0.92 & 0.63 & -0.5 & 0.71 \\
15:54:40.25 & +36:29:52.0 & 0.2368 & 2 & 1 & 53172 & 11.84 & 0.15 & -1.23 & 1.71 & 7.88 & 4.0 & 6.17 & 2.44 & 0.6 & 0.59 & -0.48 & 0.75 \\
16:10:03.12 & +54:36:27.9 & 0.2678 & 12 & 2 & 56430 & 11.54 & 0.15 & 1.51 & 0.53 & 10.54 & 4.0 & 8.09 & 3.86 & 0.53 & 0.65 & -0.3 & 0.7 \\

\end{tabular}
\caption{Results of the \texttt{Prospector} fits for our gold CL-AGN host sample. (a) Sources are (1) \cite{ruan2016}, (2) \cite{gezari2017}, (3) \cite{yang2018}, (4) \cite{Frederick2019}, (5) \cite{yu2020}, (6) \cite{green2022}, (7) \cite{tozzi2022}, (8) \cite{dong2024}, (9) \cite{guo2024a}, (10) \cite{wang2023}, (11) \cite{yang2024}, and (12) \cite{zeltyn2024}.  (b) Turn-on (1), turn-off (2), or repeating (3).}
\end{sidewaystable}

\renewcommand{\tabcolsep}{4pt}
\begin{sidewaystable}
\centering
\begin{tabular}{cccccccccccccccccc}
R.A. & Dec. & Redshift & Source (a) & On/Off (b) & Epoch & Log(M$_*$) & $\sigma_{Mass}$ & Log(SFR) & $\sigma_{SFR}$ & $t_{50}$ & $\sigma_{t_{50}}$ & $t_{90}$ & $\sigma_{t_{90}}$ & Dust  & $\sigma_{Dust}$ & Log(Z) & $\sigma_Z$ \\
\hline \\
(J200) & (J200) & (J200) &  &  &   [MJD] & [M$\odot$] & [M$\odot$] & [M$_\odot$ yr$^{-1}$] & [M$_\odot$ yr$^{-1}$] & [Gyr] & [Gyr] &  [Gyr] & [Gyr] & [$\tau_{5500 \si{\angstrom}}$] & [$\tau_{5500 \si{\angstrom}}$] & [Z$_\odot$] & [Z$_\odot$] \\
\hline \\
16:25:01.43 & +24:15:47.3 & 0.0503 & 5 & 1 & 53476 & 11.47 & 0.15 & 1.71 & 0.53 & 2.63 & 4.02 & 0.54 & 2.43 & 1.38 & 0.6 & -0.39 & 0.76 \\
07:52:44.20 & +45:56:57.4 & 0.0517 & 10 & 3 & 53055 & 11.72 & 0.15 & 0.44 & 0.53 & 11.24 & 3.99 & 10.03 & 2.43 & 1.13 & 0.58 & -0.63 & 0.7 \\
08:53:37.26 & +01:43:03.0 & 0.0582 & 12 & 1 & 51913 & 11.08 & 0.16 & 0.57 & 0.53 & 8.99 & 4.02 & 6.27 & 2.43 & 0.71 & 0.58 & -0.6 & 0.71 \\
14:10:20.59 & +13:08:29.3 & 0.0593 & 10 & 2 & 53442 & 11.37 & 0.15 & 0.86 & 0.53 & 10.74 & 3.99 & 4.21 & 2.44 & 0.97 & 0.58 & -0.57 & 0.7 \\
10:25:30.44 & +46:28:08.6 & 0.0796 & 10 & 2 & 52615 & 11.72 & 0.16 & 0.23 & 0.6 & 10.85 & 3.99 & 9.89 & 2.43 & 0.82 & 0.58 & -0.5 & 0.69 \\
11:32:29.14 & +03:57:29.0 & 0.0909 & 3, 11 & 3 & 52642 & 10.77 & 0.18 & 0.02 & 0.54 & 4.92 & 5.09 & 2.79 & 4.52 & 0.2 & 0.6 & -0.59 & 0.71 \\
15:02:40.63 & +61:28:51.4 & 0.1093 & 10 & 2 & 52055 & 11.37 & 0.15 & -0.2 & 0.59 & 6.42 & 4.28 & 4.39 & 2.45 & 0.99 & 0.59 & -0.63 & 0.73 \\
13:58:55.82 & +49:34:14.1 & 0.1159 & 3 & 1 & 53438 & 11.64 & 0.16 & 1.12 & 0.53 & 6.46 & 4.18 & 4.38 & 2.44 & 0.92 & 0.58 & -0.5 & 0.7 \\
09:09:32.02 & +47:47:30.6 & 0.1169 & 3 & 1 & 52620 & 11.83 & 0.15 & 1.58 & 0.53 & 5.18 & 3.99 & 4.04 & 2.43 & 1.16 & 0.59 & -0.65 & 0.76 \\
15:45:29.64 & +25:11:27.9 & 0.117 & 3 & 1 & 53846 & 11.98 & 0.15 & 1.33 & 0.58 & 7.6 & 4.28 & 4.84 & 2.88 & 1.22 & 0.6 & -0.34 & 0.72 \\
14:47:54.23 & +28:33:24.1 & 0.1634 & 3 & 1 & 53764 & 12.03 & 0.15 & 0.82 & 0.53 & 9.06 & 4.16 & 6.44 & 2.44 & 0.71 & 0.59 & -0.52 & 0.78 \\
09:02:54.12 & +01:54:29.1 & 0.1661 & 12 & 1 & 51924 & 11.46 & 0.15 & -0.36 & 0.58 & 10.26 & 4.09 & 6.55 & 2.81 & 0.86 & 0.58 & -0.58 & 0.71 \\
01:26:48.08 & -08:39:48.0 & 0.198 & 1 & 2 & 54465 & 11.29 & 0.15 & 0.75 & 0.53 & 7.78 & 3.99 & 6.15 & 2.43 & 1.41 & 0.59 & -0.61 & 0.7 \\
12:59:16.74 & +55:15:07.2 & 0.1987 & 3 & 1 & 52707 & 11.41 & 0.2 & 1.62 & 0.53 & 8.97 & 4.03 & 6.43 & 2.43 & 1.51 & 0.59 & -0.9 & 0.75 \\
15:50:17.24 & +41:39:02.2 & 0.2201 & 3 & 1 & 52468 & 11.29 & 0.15 & -3.56 & 4.79 & 5.78 & 5.74 & 4.05 & 2.44 & 1.04 & 0.59 & -0.82 & 0.72 \\
21:51:30.35 & +00:35:22.26 & 0.34 & 6 & 1 & 52078 & 11.51 & 0.15 & 0.48 & 0.53 & 7.66 & 4.01 & 5.32 & 2.62 & 0.77 & 0.58 & -0.52 & 0.76 \\
11:34:57.21 & +29:25:02.4 & 0.3922 & 12 & 1 & 53795 & 12.44 & 0.16 & 2.66 & 0.53 & 6.81 & 4.05 & 3.08 & 2.48 & 1.14 & 0.61 & 0.1 & 0.7 \\
\end{tabular}
\caption{Results of the \texttt{Prospector} fits for our silver CL-AGN host sample. (a) Sources are (1) \cite{ruan2016}, (2) \cite{gezari2017}, (3) \cite{yang2018}, (4) \cite{Frederick2019}, (5) \cite{yu2020}, (6) \cite{green2022}, (7) \cite{tozzi2022}, (8) \cite{dong2024}, (9) \cite{guo2024a}, (10) \cite{wang2023}, (11) \cite{yang2024}, and (12) \cite{zeltyn2024}.  (b) Turn-on (1), turn-off (2), or repeating (3). (c) This object was originally identified in \cite{yang2018} and was found to re-brighten in \cite{yang2024}.}
\end{sidewaystable}
 



\end{document}